\documentclass[letter,11pt]{article}
 \pdfoutput=1
 \usepackage{jheppub}
 
 \usepackage{graphicx}
 \usepackage{hyperref}
 \usepackage[utf8]{inputenc}
 \usepackage{amssymb}
\usepackage{multirow}
\usepackage{soul}
 
\usepackage{dcolumn}
\usepackage{bm}
\usepackage{color}
 \newcommand{\ud}{\mathrm{d}}

\allowdisplaybreaks

 \newcommand{\lsim}{{\;\raise0.3ex\hbox{$<$\kern-0.75em\raise-1.1ex\hbox{$\sim$}}\;}}
\newcommand{\gsim}{{\;\raise0.3ex\hbox{$>$\kern-0.75em\raise-1.1ex\hbox{$\sim$}}\;}}
\newcommand{\beq}{\begin{equation}}
\newcommand{\eeq}{\end{equation}}
\newcommand{\bea}{\begin{eqnarray}}
\newcommand{\eea}{\end{eqnarray}}
\newcommand{\MJ}{\textrm{MJ}}

\def\baa{\begin{array}}
\def\eaa{\end{array}}

\mathchardef\minus="002D

\newcommand{\GeV}{\mathrm{\;GeV}}

\newcommand{\TeV}{\mathrm{\;TeV}}

\preprint{
\begin{minipage}[b]{0.75\linewidth}
\begin{flushright}
CTPU-16-19 \\
IPMU16-0133 \\
PITT-PACC-1608 \\
CERN-TH-2016-218
 \end{flushright}
\end{minipage}
}

\title{Identifying a New Particle with Jet Substructures}
\author[a]{Chengcheng Han} 
\author[b,c]{Doojin Kim}
\author[d,e]{Minho Kim}
\author[f,g]{Kyoungchul Kong}
\author[d,h,1]{Sung Hak Lim}
\author[d,1]{Myeonghun Park\note{Corresponding author.}}

\affiliation[a]{Kavli IPMU (WPI), The University of Tokyo, Kashiwa, Chiba 277-8583, Japan}
\affiliation[b]{Department of Physics, University of Florida, Gainesville, FL 32611, USA}
\affiliation[c]{CERN, Theory Division, CH-1211 Geneva 23, Switzerland}
\affiliation[d]{Center for Theoretical Physics of the Universe, Institute for Basic Science (IBS), Daejeon, 34051, Korea} 
\affiliation[e]{Department of Physics, Postech, Pohang 790-784, Korea}
\affiliation[f]{Department of Physics and Astronomy, University of Kansas, Lawrence, KS 66045, USA}
\affiliation[g]{Pittsburgh Particle physics, Astrophysics, and Cosmology Center, Department of Physics and Astronomy, University of Pittsburgh, Pittsburgh, PA 15260, USA}
\affiliation[h]{Department of Physics, KAIST, 291 Daehak-ro, Yuseong-gu, Daejeon, 34141, Korea}

\emailAdd{chengcheng.han@ipmu.jp}
\emailAdd{doojin.kim@cern.ch}
\emailAdd{kmhmon@postech.ac.kr}
\emailAdd{kckong@ku.edu}
\emailAdd{sunghak.lim@kaist.ac.kr}
\emailAdd{parc@ibs.re.kr}

\abstract{
We investigate a potential of determining properties of a new heavy resonance of mass $\mathcal{O}(1)$ TeV which decays to collimated jets via heavy Standard Model intermediary states, exploiting jet substructure techniques.
Employing the $Z$ gauge boson as a concrete example for the intermediary state, we utilize a ``merged jet'' defined by a large jet size to capture the two quarks from its decay.
The use of the merged jet benefits the identification of a $Z$-induced jet as a single, reconstructed object without any combinatorial ambiguity. 
We find that jet substructure procedures may enhance features in some kinematic observables formed with subjet four-momenta extracted from a merged jet. 
This observation motivates us to feed subjet momenta into the matrix elements associated with plausible hypotheses on the nature of the heavy resonance, which are further processed to construct a matrix element method (MEM)-based observable. 
For both moderately and highly boosted $Z$ bosons, we demonstrate that the MEM in combination with jet substructure techniques can be a very powerful tool for identifying its physical properties.
We also discuss effects from choosing different jet sizes for merged jets and jet-grooming parameters upon the MEM analyses. 
}

\makeatletter
\def\@fpheader{\relax}
\makeatother

\date{Sept. 20, 2016}

\begin{document} 
\maketitle
\flushbottom

\section{Introduction}
\label{sec:introduction}

The Large hadron collider (LHC) has played an important role in deepening our understanding of electroweak symmetry breaking by discovering a Higgs particle. 
As the LHC experiment 
reaches the energy scale of tera electronvolt (TeV), 
it is of paramount importance to study potential new physics such as
various extended Higgs sectors, existence of other fundamental scalars~\cite{Lee:1973iz,Gunion:2002zf,Branco:2011iw}, vector resonances under the set-up of composite models~\cite{Contino:2011np,Bellazzini:2012tv,Pappadopulo:2014qza,Greco:2014aza,Low:2015uha,Franzosi:2016aoo}, and so on.  
We remark that resonances in those new physics models often have sizable branching fractions to heavy SM particles including the weak gauge bosons, the Higgs, and the top quark,
if kinematically allowed.
As increased center-of-mass energy at the LHC enables us to probe heavier new particles of $\mathcal{O}(1)\TeV$, a substantial mass gap between a new particle and a heavy SM state would result in a large boost of the latter, accompanying highly collimated objects along the boost direction of the latter in the final state. 
While the leptonic decay products of the above-listed heavy SM particles often carry advantages in conducting data analyses thanks to their cleanness, hadronic decay products are expected to play an important role in not only discovery opportunity but property measurement at the {\it early} stage due to their larger branching fractions. 
However, their jetty nature at the detection level renders associated analyses challenging because of significant overlaps between the final state jets, requiring robust analysis tools to deal with such hadronic objects reliably.
A promising venue in developing relevant techniques is the field of jet substructure~\cite{Altheimer:2013yza}.

A successful application of the jet substructure techniques is to tag single-jet-looking objects from decays of boosted, heavy SM states (e.g., $t/W/Z/H$) against structureless or single-prong QCD jets~\cite{CMS-PAS-JME-14-002}.
The idea is that one can capture hadrons from the decay of a heavy SM particle, using a single ``merged'' jet which is defined by a proper choice of the jet size. 
An expected benefit from utilizing a resultant (massive) merged jet is mitigation of the systematics which often arises in considering multi-particle final states (e.g., combinatorial ambiguity), by reducing the number of reconstructed objects.
The price for it is the possibility that even a normal QCD jet may acquire a sizable mass in combination with underlying QCD activities including pile-ups.\footnote{See Ref.~\cite{Aad:2013gja} for the jet substructure techniques alleviating the pile-up contamination.} 
In this regard, there are dedicated studies
\begin{itemize}
\item to reduce corruptions from irrelevant hadrons for a given jet~\cite{Butterworth:2008iy,Ellis:2009su,Krohn:2009th,Larkoski:2014wba}, and
\item to differentiate a jet resulting from a boosted heavy SM state from an ordinary QCD jet by looking into its substructure~\cite{Butterworth:2008iy,Kaplan:2008ie,CMS:2009lxa,Plehn:2010st,Kasieczka:2015jma,Almeida:2010pa,Almeida:2011aa,Thaler:2010tr,Thaler:2011gf,Kim:2016plm}.
\end{itemize}  

Many proposed methods along the line have been successfully implemented for analyzing the LHC data, 
and they concurrently improve the sensitivities for the high mass region by reducing relevant SM backgrounds efficiently.
While tagging a boosted jet by jet substructure techniques is useful for discovery opportunities e.g., heavy resonance searches, the constituent-jet information itself allows to construct various experimental observables for further data analyses.
In this context, it is interesting to question
how far characteristic features in kinematic distributions are preserved after subjet isolations, if included are various realistic effects such as parton shower, hadronization/fragmentation, detector response, and jet clustering.
We {\it first} point out that rather precise identification of the features is viable in some controlled environment, despite the presence of realistic effects. 
Motivated by the spin-parity determination of the SM Higgs boson\,\cite{Chatrchyan:2012jja} and the diboson resonance\,\cite{Kim:2015vba, Buschmann:2016kwr} through massive bosonic intermediary states in relevant decay processes, we focus on the analysis of $W/Z$-induced two-prong jets and examine well-motivated angular variables formed with reconstructed subjets.
In the case of production of a new, bosonic heavy resonance, the jet substructure techniques are relevant to the channels of $WW$, $ZZ$, and $Z\gamma$ in which
the associated final state is, at least, partially hadronic. 

For a sufficiently boosted, heavy state $V$, the angular separation $\Delta R$ between its two decay products is given by
\bea
\Delta R \approx \frac{2m_V}{P_T^V}\,,
\eea
where $m_V$ and $P_T^V$ denote the mass and the transverse momentum of particle $V$. 
Since usual jet substructure techniques begin with identifying a ``merged'' jet by a fairly large fixed cone size to capture all constituent jets followed by a declustering procedure to find subjets, the hardness of $P_T^V$ is crucial in choosing a reasonable cone size, hence too a successful subjet analysis. 
Moreover, considering the fact that the generic shape analysis demands global information, we see that a proper definition of merged jets is a key component for posterior analyses.
In particular, the phase-space reduction induced by fixing a cone size for merged jets would cause adverse distortions of the kinematic distributions of interest, becoming an obstacle in decoding the physics behind signals. 
To illustrate these points, we employ two benchmark points for a heavy resonance decaying into a $ZZ$ final state in order to cover kinematically distinctive regions, one for the moderately boosted $Z$ case and the other for the highly boosted one. We contrast/compare them in terms of the angle particularly sensitive to the CP state of the resonances. We there explicitly show that {\it remarkably}, jet substructure techniques preserve useful information quite well. 

Being confident of the above single-variable analysis, we then move our focus onto matrix element method (MEM)-based observables which allow us to make full use of all available information encrypted in four-momenta of final state particles\,\cite{Gao:2010qx, DeRujula:2010ys, Gainer:2011xz, Campbell:2012cz, Campbell:2012ct, Bolognesi:2012mm, Avery:2012um, Chen:2012jy,Soper:2014rya,Englert:2015dlp}. 
Unlike other statistical methods based on distributions of multiple observables, the MEM is predicated on a straightforward and elegant interpretation on the probability measure $\mathcal{P}$, that is, the quantum amplitude of a given process with hypothesis $\alpha$ is schematically given as follows:
\beq
\mathcal{P}\left(\{{\bf p}^{\textrm{reco}} \}| \alpha \right) \propto  \int\ud \Pi_{q_i} \, \mathcal{W}\left( {\bf q_i} ,\{ \bf{p}^{\textrm{reco}}\}\right) \, 
\Big|\mathcal{M} \left({\bf q}_i \,; \alpha\right) \Big|^2  \, ,
\label{eq:MEM}
\eeq
where $\mathcal{M}$ is the matrix element for hypothesis $\alpha$ and $\mathcal{W}$ is the transfer function introduced to map parton-level momentum vectors $(\{{\bf q}\})$ to reconstruction-level ones $(\{{\bf p^\textrm{reco}}\})$.
Markedly, the usefulness of the MEM has been proven in discriminating different spin/CP state hypotheses~\cite{Chatrchyan:2012jja, Aad:2013xqa, Gao:2010qx, DeRujula:2010ys, Bolognesi:2012mm, Avery:2012um}. 
In particular, the MEM was a driving force to determine various properties of the SM Higgs particle in the four-lepton channel, which has been considered as one of the most exciting achievements at the LHC. 
In more detail, by identifying the interaction between the Higgs boson and a $Z$-boson pair, it has been shown that the Higgs boson is indeed related to the $SU(2)_L \times U(1)_Y$ gauge symmetry breaking mechanism. 
We note that this channel comes with ten degrees of freedom compared to its competing diphoton channel with only four degrees of freedom although the former involves smaller statistics than the latter. 
Therefore, given low statistics, it is imperative to combine different information from various degrees of freedom in an optimized way, for which the MEM is well-suited. 

We remind that many of the collider studies for the decay of a heavy resonance into the final state particles via massive SM states often advocate fully leptonic channels in not only search for new particles but measurement of their properties, due to the clean nature of leptonic final states even at the reconstruction level.
While it is challenging to extract useful information from hadronic decay products unlike leptonic ones, the remarkable discriminating power of the MEM motivates us to construct an MEM-based kinematic discriminant (KD) using four-momenta of subjets.
We then investigate how much the discrimination potential is retained in the context of jet substructure techniques again employing the benchmark scalar resonances.

To convey our main ideas coherently, we organize this paper as follows. In Section~\ref{sec:mergedJet}, we begin with the discussion on the phase-space reduction occurred by the introduction of a fixed cone size. In Section~\ref{sec:angular}, we provide a brief review on various angular variables for discriminating the spin and the CP states of heavy resonances, and discuss the impact of the phase-space reduction upon kinematic observables, in particular, CP-sensitive ones.
In Section~\ref{sec:results}, we confirm the observations made in the two previous sections, using detector-level Monte Carlo simulation. 
We then, in Section~\ref{sec:MEManal}, present our main results obtained from the MEM-based analyses under the circumstance of negligible background contamination, in conjunction with the jet substructure techniques.   
Our concluding remarks and outlook appear in Section~\ref{sec:conclusion}.
Finally, Appendices~\ref{sec:BKG} and~\ref{sec:appB} are reserved for the discussion on the MEM-based analyses including backgrounds and the phase-space reduction in other jet substructure techniques, respectively.

\section{Phase-space reduction \label{sec:mergedJet}}
We begin this section by estimating the cone size $R$ for ``Merged Jets'' (MJ) to capture both of the {\it two} visible particles $v_1$ and $v_2$ emitted from a highly boosted massive particle (e.g., $W/Z/H\rightarrow v_1v_2$). 
For simplicity, we assume that the two partonic decay products are massless and well-approximated to two subjets $j_1$ and $j_2$ which are {\it the} constituents of a merged jet.
We define $P_{T(\MJ)}$ and $m_{\MJ}$ as the laboratory-frame transverse momentum and the mass of a merged jet, respectively.
With the assumption of $P_{T(\MJ)} \gg m_\MJ$, simple kinematics in leading-order QCD leads to  
\beq
R\simeq \frac{1}{\sqrt{z(1-z)}}\frac{m_{\MJ}}{P_{T(\MJ)}} 
\geq
 \frac{2 m_{\MJ}}{P_{T(\MJ)}}\, , 
\label{eq:R}
\eeq
where $z$ is defined as $\frac{\min\left(P_{T(j_1)}, P_{T(j_2)}\right)}{P_{T(\MJ)}}$, i.e., the fractional transverse momentum of the leading subjet (say, $j_1$) with respect to the total transverse momentum. 
Here the equality is obtained in the limit of $z=1/2$.

We then closely look at the relation between $R$ and the angular separation $\Delta R_{12}$ of two subjets which is defined as 
\bea
\Delta R_{12} \equiv \sqrt{\Delta \eta_{12}^2+\Delta \phi_{12}^2}\,,
\eea
where $\Delta \eta_{12}$ and $\Delta \phi_{12}$ denote the differences between the two subjets in pseudorapidity and azimuthal angle in the laboratory frame, respectively. 
The angular distance between $j_1$ and $j_2$ in the laboratory frame can be expressed in terms of the polar angle $\theta$ and the azimuthal angle $\phi$ of the leading subjet in the heavy particle rest frame relative to the boost direction to the laboratory frame~\cite{Ellis:2009me}: 
\bea
\Delta R_{12}^2 \hspace{-0.1cm}&=& \hspace{-0.1cm}\left[\tanh^{-1}\left(\frac{2 \cosh\eta \sin\theta \sin\phi}{\sin^2\theta (\sinh^2\eta+\sin^2\phi)+1}\right)\right]^2 \hspace{-0.1cm}+\hspace{-0.1cm}\left[\tan^{-1}\left(\frac{2 \sinh\eta \sin\theta \cos\phi}{\sin^2\theta(\sinh^2\eta +\sin^2\phi)-1}\right)\right]^2 , \nonumber \\
&&
\eea
where $\cosh\eta =E_{\MJ}/m_{\MJ}$ is a Lorentz boost factor 
of the MJ.
One can show that $\Delta R_{12}$ has a minimum at $\theta=\pi/2$ and $\phi=0$ for any fixed $\eta$~\cite{Ellis:2009me}.
Therefore, a necessary condition to capture the two subjets for a given $\eta$ is that the cone size $R$ should be greater than the lower limit of $\Delta R_{12}$:
\bea
R\geq \Delta R_{12}^{\min} = 2\csc^{-1}(\cosh \eta) \xrightarrow[P_{T(\MJ)} \gg m_\MJ ]{} \frac{2\,m_\MJ}{P_{T(\MJ)}}\, ,
\label{eq:Rsize}
\eea 
where the last step is done by setting $\cosh\eta$ in the transverse plane and taking a large transverse momentum limit. 
Note that this asymptotic behavior is identical to the estimate in eq.~\eqref{eq:R}. 
Now if we set the cone size to be $R_{\MJ}$, all events with $R<R_{\MJ}$ are accepted. 
We then translate this inequality to the upper bound for the polar angle $\theta$:
\beq
|\cos\theta| \le  \sqrt{1- \frac{1}{\sinh^2\eta}\cot^2\left(\frac{R_\MJ}{2}\right)} =  \sqrt{1- \left(\frac{m_\MJ}{P_{T(\MJ)}}\right)^2\cot^2\left(\frac{R_\MJ}{2}\right)} \, .
\label{eq:cosUpper}
\eeq
This inequality implies that fixing the cone size for MJs confines the polar angle to a certain range, resulting in a reduction of the accessible phase space. 

 \begin{figure*}[t!]
 \centering
 \includegraphics[width=7.2cm]{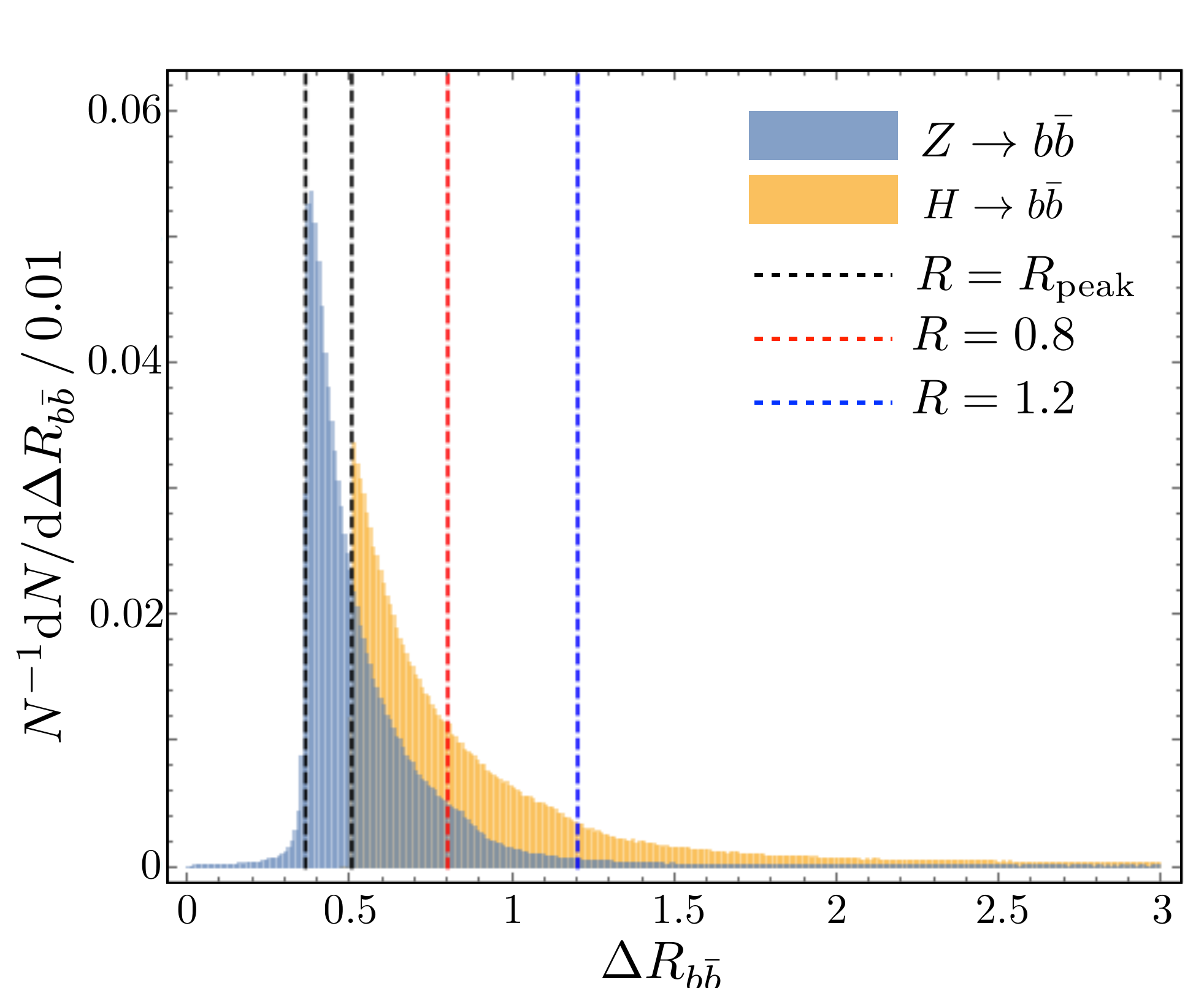} \hspace{0.3cm}
 \includegraphics[width=7.2cm]{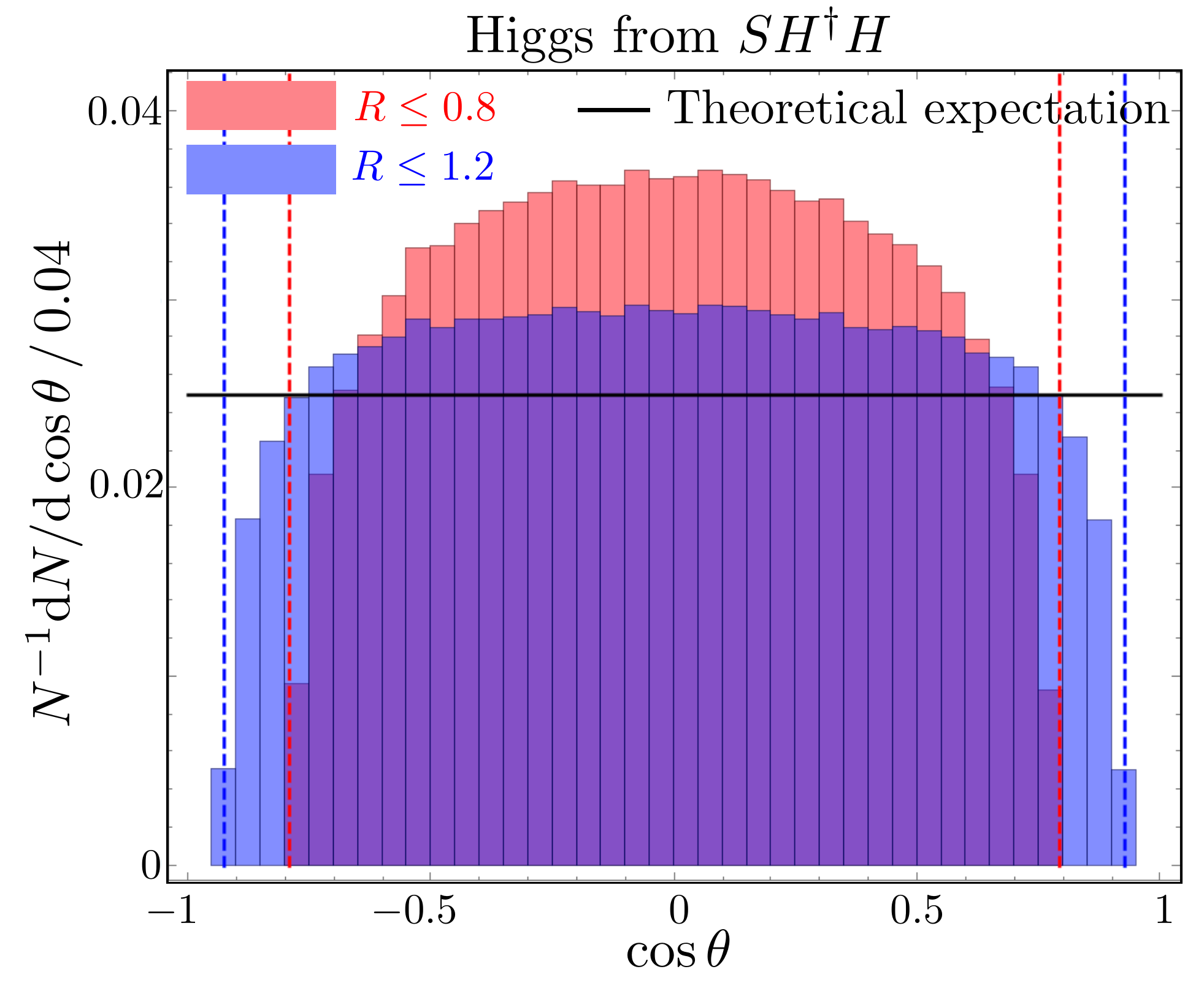}  \\
 \vspace{0.2cm}
  \includegraphics[width=7.2cm]{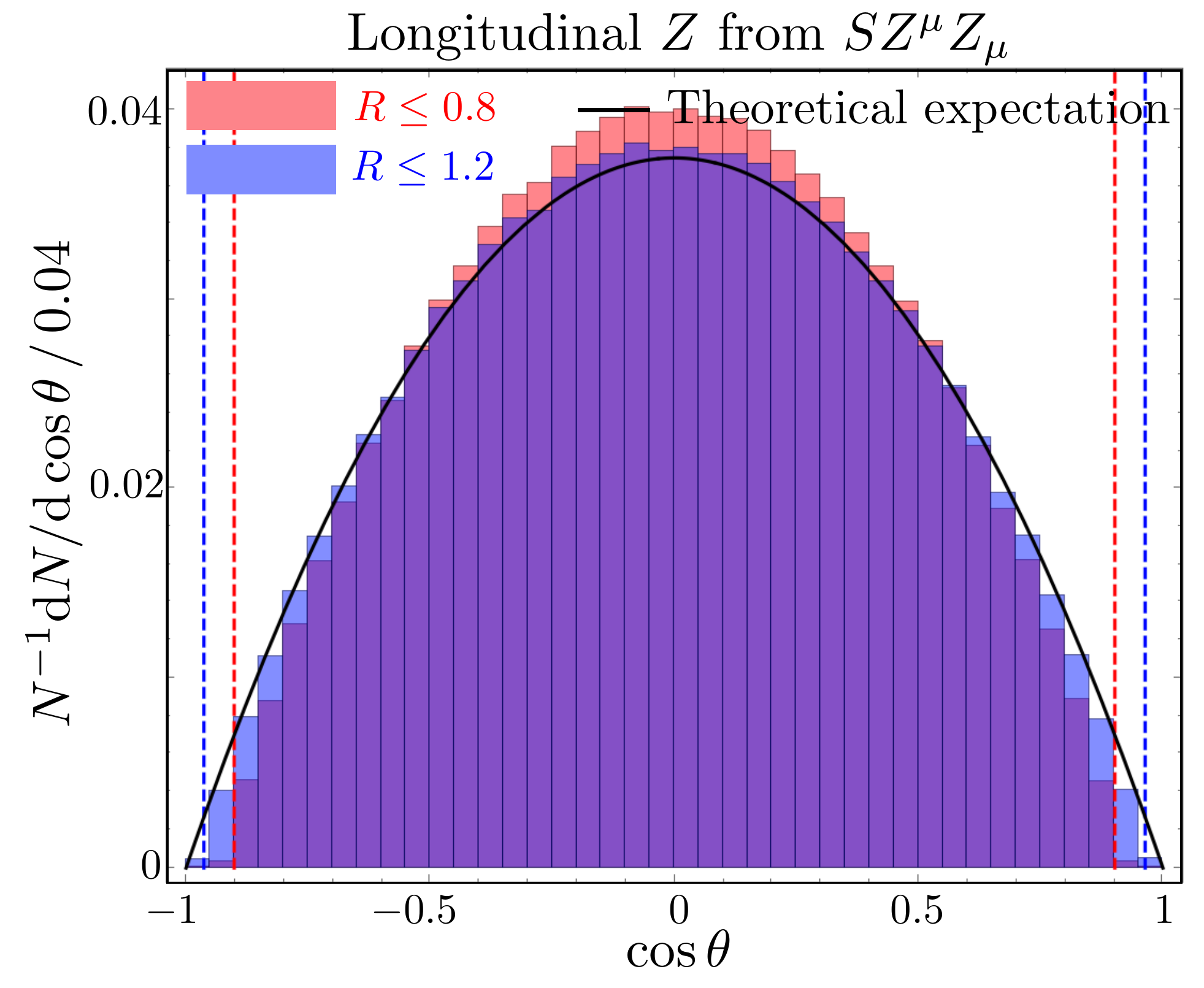}\hspace{0.3cm}
 \includegraphics[width=7.2cm]{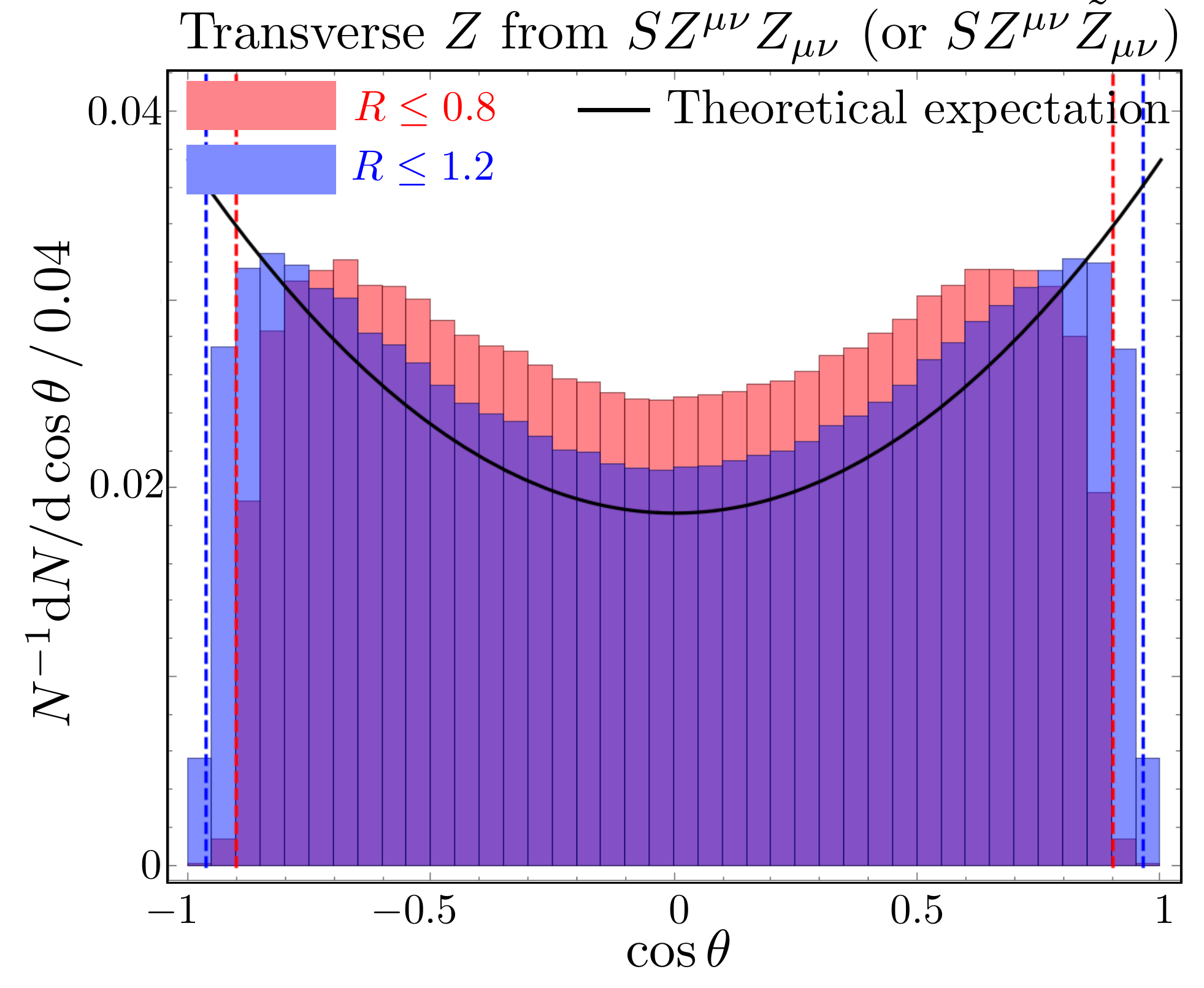}
\caption{Parton-level Monte Carlo simulation results for $S \rightarrow HH/ZZ \rightarrow 4b$ with $M_S=1$ TeV. The upper-left panel shows unit-normalized $\Delta R_{bb}$ distributions for the Higgs boson (orange histogram) and the $Z$ gauge boson (blue histogram) cases. The red and blue dashed lines mark the positions corresponding to $R=0.8$ and $R=1.2$, respectively.
The other three panels (upper-right for $H$, lower-left for $Z_L$, and lower-right for $Z_T$) show unit-normalized $\cos\theta$ distributions with cone sizes for MJs $R\leq 0.8$ (red histogram) and $R\leq 1.2$ (blue histograms) and compare them with corresponding theory predictions (solid black lines).
Dashed vertical lines represent the upper bounds on $|\cos\theta|$ for a given $R_\MJ$ according to eq.~\eqref{eq:cosUpper}.
\label{fig:parton} }
\end{figure*}

To visualize this observation, we exhibit $\cos\theta$ distributions of quarks (say, $b$) from Higgs or $Z$ gauge boson decays. 
To minimize any effects on the angular distributions from their production, we assume that a pair of $H$ or $Z$ bosons are produced via the decay of a heavy scalar $S$, for example, $gg \rightarrow S \rightarrow HH/ZZ$.
Trivially, $\cos\theta$ for the Higgs boson case has a flat distribution.
On the other hand, a $Z$ boson has transverse and longitudinal polarization components, and thus its coupling to particle $S$ is described in a somewhat complicated manner. 
Denoting $M_Z$ and $\Lambda$ as the $Z$ gauge boson mass and a scale parameter, we define the interaction Lagrangian between $S$ and $Z$ as 
\beq
\mathcal{L}_{\textrm{int}} = \kappa_1 \frac{M_Z^2}{\Lambda}\,S\, Z^\mu\, Z_\mu + \frac{\kappa_2}{\Lambda} S \,Z^{\mu\nu} Z_{\mu\nu} + \frac{\kappa_3}{\Lambda} S\,Z^{\mu\nu}\, \tilde Z_{\mu\nu}\,,
\eeq
where $Z^{\mu\nu}$ and $\tilde{Z}^{\mu\nu}$ are the field strength tensor and the dual field strength tensor for the $Z$ boson, respectively. 
In $M_S \gg M_Z$ limit, the first term takes care of the interaction of the longitudinal polarization component while the other two describe that of the transverse polarization components~\cite{Choi:2002jk, Gao:2010qx}, and the resulting differential cross section in $\cos\theta$ is given by
\beq
\frac{\ud \sigma}{\ud\cos\theta}  \sim  2 \kappa_1^2 (1- \cos^2\theta) + \left(\kappa_2^2 + \kappa_3^2\right) (1+ \cos^2\theta) +\mathcal{O}\left(\frac{M_Z^4}{M_S^4}\right)\, .
\eeq

Figure~\ref{fig:parton} displays our numerical results with parton-level Monte Carlo simulation for which the input mass of the heavy resonance $S$ is 1 TeV for illustration. 
As mentioned above, we take the decay process of $H$ or $Z$ into a bottom quark pair. 
The upper-left panel shows the unit-normalized distributions of $\Delta R_{bb}$ for the Higgs boson (orange histogram) and the $Z$ gauge boson (blue histogram). 
The red and the blue dashed lines mark the locations corresponding to $R=0.8$ and $R=1.2$, respectively, allowing us to develop our intuition on what fraction of events are tagged.
The other three panels (upper-right for the Higgs boson, lower-left for the longitudinal $Z$, and lower-right for the transverse $Z$) demonstrate the unit-normalized $\cos\theta$ distributions with $R\leq 0.8$ (red histogram) and $R\leq 1.2$ (blue histogram) and compare them with the corresponding theory expectations represented by solid black lines. 
We clearly observe that a fixed cone size for MJs distorts the shape of differential distributions. 
Hence, when investigating physics governing experimental signatures with kinematic distributions including angular observables, one should conduct a careful examination on how much of partonic information would be missing by the introduction of a fixed cone size for MJs in reconstructing final state objects. 

\section{Angular correlations among final state particles \label{sec:angular}}

As in the case of the SM Higgs boson whose first signature appeared in the final states with $\gamma\gamma$ and $ZZ$, 
if a heavy new particle $X$ respects the SM electroweak gauge symmetry, it may appear as a resonance in the final states with $ZZ$, $WW$, $Z\gamma$, and $\gamma\gamma$. 
We divide them into three categories according to the number of angular degrees of freedom measured in the rest frame of particle $X$.
\begin{itemize}
\item[(a)] $X\to \gamma\gamma$: Two angular degrees of freedom as $(\theta^*, \Phi^*)$
\item[(b)] $X\to Z\gamma$: Four angular degrees of freedom as $(\theta^*, \Phi^*, \theta_1, \phi_1)$
\item[(c)] $X\to ZZ/W^+ W^-$: Six angular degrees of freedom as $(\theta^*, \Phi^*, \theta_1, \phi_1, \theta_2, \phi_2)$ 
\footnote{
These angles are not suitable for the spin and parity analysis in $X \rightarrow W^+W^- \rightarrow \ell^+\nu\ell^- \bar \nu$ channel because the two neutrinos are not detected. Instead, we can use the azimuthal angle between two leptons, $\Delta \phi_{\ell\ell}$, the dilepton invariant mass, $m_{\ell\ell}$, and the transverse mass of the dilepton system, $m_T$, to distinguish spin and parity hypotheses~\cite{Aad:2013xqa,CMS-PAS-HIG-13-003}.
}
\end{itemize}
%
 \begin{figure*}[t!]
 \centering
 \includegraphics[width=0.95\textwidth]{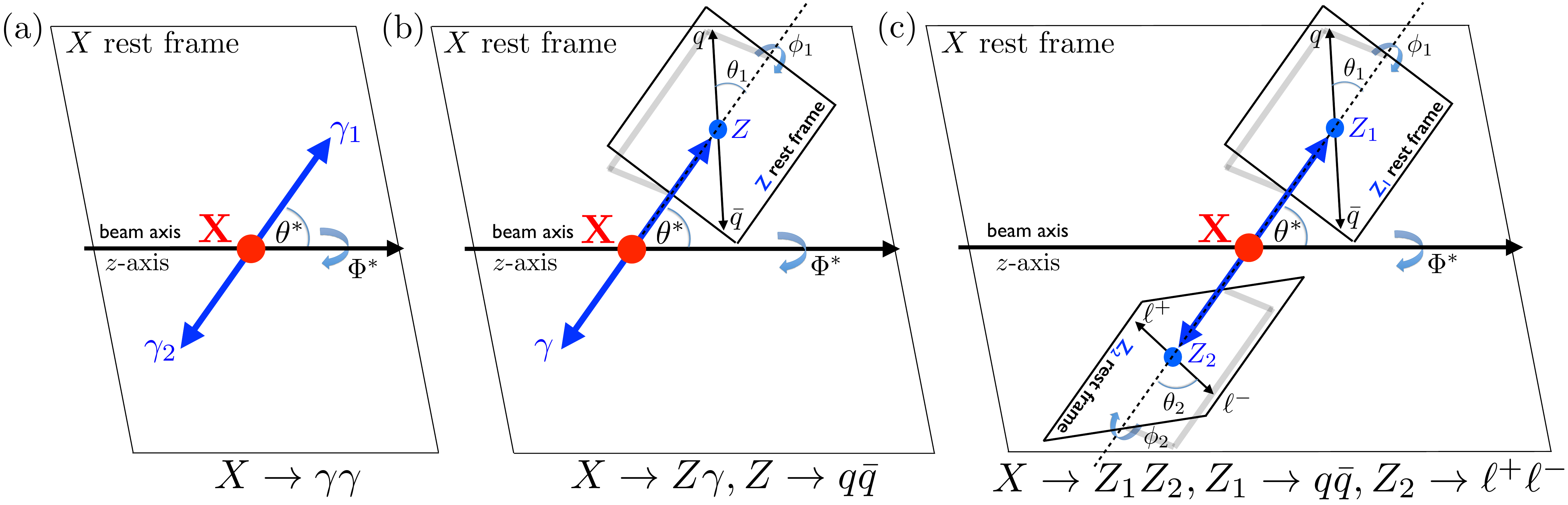}
  \caption{Angular degrees of freedom in $\gamma\gamma$ (the (a) panel), $Z\gamma$ (the (b) panel), and $ZZ/W^+W^-$ (the (c) panel) processes. 
For a sufficiently heavy $X$ (i.e.,  $m_X \gg 2 M_Z$), we can neglect the possibility of off-shellness of internal gauge boson propagators. 
Then the processes in (a), (b), and (c) panels have two, four, and six degrees of freedom, respectively, at the $X$ rest frame. }
\label{fig:DOF}
\end{figure*}
%
We schematically show angular configurations for three cases in Figure~\ref{fig:DOF}, matching the item numbers with the panel ones.
The decay of $X$ into two gauge bosons $V_1$ and $V_2$ involves two degrees of freedom, polar angle $\theta^*$ and azimuthal angle $\Phi^*$ of $V_1$ (or equivalently $V_2$) about the beam axis. 
In a similar manner, each of the two gauge bosons (except the photon) involves two degrees of freedom, polar angle $\theta_i$ and azimuthal angle $\phi_i$ of one of the decay products  relative to the $V_i$ boost direction in the $V_i$ rest frame. 
Another degree of freedom comes with the rapidity of the whole decay system which encodes the information of initial state partons through the parton distribution functions. 
However, imposing a rapidity cut on the reconstructed heavy resonance, we anticipate that any of its associated impact upon kinematic observables becomes mild~\cite{Avery:2012um}.

We begin with the observables related to the decay process of $X$ itself, which are the two angles $\theta^*$ and $\Phi^*$. 
They can be evaluated as follows:
\bea
\cos\theta^*&=&\hat{p}_{V_1}\cdot \hat{z}|_X\,, \\
\cos\Phi^* &=& \hat{x}\cdot(\hat{z}\times \hat{p}_{V_1})|_X\,,
\eea
where $|_X$ implies that all relevant physical quantities are measured in the rest frame of particle $X$.
Here $\hat{z}$ lies on the beam direction as usual, while $\hat{x}$ is chosen to be an azimuth reference direction on the plane perpendicular to $\hat{z}$. 
The determination of the helicity/spin of $X$ by variable $\Phi^*$ or $\theta^*$ is closely connected to the production mechanism for it. 
The azimuthal angle $\Phi^*$ carries the helicity information of $X$, which becomes available if there is interference among different helicity states~\cite{Buckley:2008pp}. 
If $X$ is produced in association with another particle, its helicity state is obtained by a linear superposition of various helicity states with corresponding amplitudes given in terms of relevant Clebsch-Gordan coefficients.
Under a spatial rotation around the $X$ momentum axis by say, $\Phi$, each helicity state obtains a phase factor $e^{i\lambda \Phi}$ where $\lambda$ denotes the helicity value of the state.   
Therefore, the sum over various helicity states give rise to non-trivial interference among the corresponding quantum amplitudes in the resulting cross section, which will be imprinted in the $\Phi^*$ distribution. 
On the other hand, if $X$ is singly produced, its helicity state is uniquely fixed by initial partons, rendering the helicity sum incoherent. 
Thus we do not expect to observe distinctive features in the $\Phi^*$ distribution. 
When it comes to polar angle $\theta^*$, the spin state of $X$ can be inferred from the distribution in $\theta^*$~\cite{Artoisenet:2013puc}.
At the tree level, the matrix element contains a projection of the $X$ helicity onto the beam direction. 
In more detail, the Wigner $d$-function, which depends on the net spin between the initial and the final states, describes the amplitude of this projection whose angle is $\theta^*$. 
Therefore, the $\theta^*$ distribution can be a good observable for identifying the production mechanism and the spin of $X$. 

We next consider angular variables related to the decay of $V_i$. 
As we demonstrated explicitly in Section~\ref{sec:mergedJet}, the impact of a fixed $R_{\MJ}$ upon   $\cos\theta_i$ differs in polarization states (see also the bottom panels in Figure~\ref{fig:parton}). 
This implies that we can infer the $V_i$ polarization from its decaying angles $\theta_i$, which are crucial in understanding the coupling of $X$-$V_1$-$V_2$, and they are defined as follows:
\bea
\cos\theta_1 &=& \hat p_{\bar q} \cdot \hat p_{V_2} |_{V_1}\, , \\
 \cos\theta_2 &=& \hat p_{\ell^+} \cdot \hat p_{V_1} |_{V_2}\, ,
\eea
where the decay products of $V_1$ and $V_2$ are distinguished merely to avoid any potential notational confusion (see also Figure\,\ref{fig:DOF}(c) for relevant decay products).

It turns out that the remaining angles $\phi_i$ pertain to the CP state, which is
one of the highly non-trivial properties to be identified in collider analyses. 

Indeed, the difference between two azimuthal angles of the $V_1$ and $V_2$ decaying planes, i.e., $\phi \equiv \phi_1-\phi_2$, provides the strongest discriminating power between different CP states\,\cite{Choi:2002jk, Gao:2010qx, Bolognesi:2012mm},\footnote{In Ref.\,\cite{DeRujula:2010ys}, the authors considered $J_{PC}=0^{++}$ with the SM-like Higgs boson case where a scalar interacts mostly with the longitudinal polarization vector of gauge bosons through an interaction of $H Z^\mu Z_\mu$.} 
and this quantity is evaluated by
\bea
\cos \phi \equiv \cos( \phi_1 -\phi_2) = \frac{\left(\hat p_q \times \hat p_{\bar q}\right) \cdot \left(\hat p_{\ell^-} \times \hat p_{\ell^+}\right)}
{|\hat p_q \times \hat p_{\bar q}| \, |\hat p_{\ell^-} \times \hat p_{\ell^+}|}\,\Bigg|_X\, .
\eea

In the rest of this paper, we focus on the determination of the CP state of $X$ assuming that $X$ is a scalar $S$, as other properties such as the spin of $X$ or the interaction to a longitudinal or transverse component of $V_i$ can be measured by other angular variables explained above. 
We remark that if there are interactions between CP-even scalar and the longitudinal polarization of $V_i$ through either a tree level coupling $SV_i^{\mu}V_{i\mu}$ 
or a higher dimensional operator $S\,D^\mu H^\dagger D_\mu H$, we can easily distinguish them from the corresponding interactions with CP-odd scalar because the latter mostly interacts with the transverse polarization vector of $V_i$.
We therefore consider only higher dimensional operators of dimension 5, for which identifying the CP state is more challenging.
Before the breakdown of the SM electroweak gauge symmetry $SU(2)_L \times U(1)_Y$, relevant Lagrangians for CP-even and CP-odd state scalars are
\bea
&&\mathcal{L}_{0^{++}} \ni  
\frac{c_Y}{\Lambda} S \, B_{\mu \nu} B^{\mu \nu}+\frac{c_W}{\Lambda} S \,W^a_{\mu \nu} W^{a\mu \nu}
 \, , \\
&&\mathcal{L}_{0^{-+}} \ni  
\frac{\tilde c_Y}{\Lambda} S \, B_{\mu \nu} \tilde B^{\mu \nu}+\frac{\tilde c_W}{\Lambda} S \,W^a_{\mu \nu} \tilde W^{a\mu \nu}
\, ,
\eea
where $W^a_{\mu \nu}$ and $B_{\mu \nu}$ are field strength tensors of $SU(2)_L$ and $U(1)_Y$, respectively, while $\tilde{W}^a_{\mu \nu}$ and $\tilde{B}_{\mu \nu}$ are their corresponding dual field strength tensors. 
After electroweak symmetry breaking, the couplings between $S$ and mass eigenstate vector bosons can be described as
\bea
&&\mathcal{L}_{0^{++}} \ni
  \frac{c_{WW}}{\Lambda} S\, W^+_{\mu \nu} W^{-\mu \nu}+  \frac{c_{ZZ}}{\Lambda} S\, Z_{\mu \nu} Z^{\mu \nu}+  \frac{c_{\gamma \gamma}}{\Lambda} S\, A_{\mu \nu} A^{\mu \nu} 
  +   \frac{c_{Z \gamma}}{\Lambda} S Z_{\mu \nu} A^{\mu \nu}  \, , \\
  &&\mathcal{L}_{0^{-+}} \ni
  \frac{\tilde c_{WW}}{\Lambda} S\, W^+_{\mu \nu} \tilde W^{-\mu \nu}+  \frac{\tilde c_{ZZ}}{\Lambda} S\, Z_{\mu \nu} \tilde Z^{\mu \nu}+  \frac{\tilde c_{\gamma \gamma}}{\Lambda} S\, A_{\mu \nu} \tilde A^{\mu \nu} 
  +   \frac{\tilde c_{Z \gamma}}{\Lambda} S Z_{\mu \nu}\tilde A^{\mu \nu}  \, ,
\eea
where new coupling constants $c_{WW}$, $c_{ZZ}$, $c_{\gamma\gamma}$, and $c_{Z\gamma}$ are related to $c_Y$, $c_W$, and the Weinberg angle $\theta_w$ as follows:
\bea
 c_{WW} &=&2\,c_W , \label{eq:cww}\\ 
 c_{ZZ} &=& c_W\cos^2\theta_w+c_Y \sin^2\theta_w , \\
 c_{\gamma \gamma}&=& c_Y\cos^2\theta_w+c_W \sin^2\theta_w , \\
 c_{Z\gamma}&=& (c_W-c_Y)\,\sin2\theta_w. \label{eq:czgamma}
\eea

Similarly, we have $\tilde c_{WW}$, $\tilde c_{ZZ}$, $\tilde c_{\gamma\gamma}$, and $\tilde c_{Z\gamma}$ in terms of $\tilde c_Y$ and $\tilde c_W$ as in eqs.~\eqref{eq:cww} through~\eqref{eq:czgamma}.

As two coupling constants $c_Y$ and $c_W$ determine four decay modes of $S$, at least two decay channels should be non-vanishing.
For example, if the $S\to\gamma\gamma$ channel is observed, one can expect to observe at least either $S\to ZZ$ or $S\to Z\gamma$ channel as well. 
However, $S\to W^+W^-$ may not be available, as it depends only on $c_W$
which could vanish if $S$ were $SU(2)_L$-singlet.

As briefly discussed before, $\phi$ plays an important role in determining the CP state of resonance $S$. 
In this sense, $Z\gamma$ and $\gamma\gamma$ final states are irrelevant because they do not involve two decaying planes.
In our numerical study, we focus on $S\to ZZ$ which subsequently decay semileptonically, i.e., $q\bar{q}\ell^+\ell^-$.
One reason for this choice is that the $q\bar{q}\ell^+\ell^-$ final state is expected to offer a better handle in inferring the underlying decay mode than the fully hadronic decay channel in which there exists non-negligible chance to misidentify observed events as $S\rightarrow W^+W^-$ due to the issue of jet mass resolution\,\cite{Allanach:2015hba,Buschmann:2016kwr}.\footnote{One could study the $S\to W^+W^- \to q\bar q\ell \nu$ channel by reconstructing the four vector of a neutrino $(\nu)$ using the energy-momentum conservation.}
Compared to the fully leptonic channel,
the semileptonic channel certainly enjoys higher statistics due to the larger branching fraction of $Z$ into quark pairs, allowing us to have better signal sensitivity.
However, in a more realistic situation, this naive expectation is not straightforwardly applied.  
Once we take SM backgrounds into consideration, we are forced to impose severe cuts to suppress huge backgrounds including $Z$+jets so that we may end up with a similar order of sensitivity compared to the $4\ell$ channel. 
More specifically, it turns out that for $m_S \gtrsim 700$ GeV,
the signal sensitivity expected from the semileptonic channel becomes comparable to that from the fully leptonic channel~\cite{Khachatryan:2015cwa, Gouzevitch:2013qca}.
Remarkably, the jet substructure techniques come into play in this high-mass regime.
Note again that a merged jet from major backgrounds contains a single quark together with additional QCD activities from radiation, whereas a signal merged jet consists of two partons. Therefore, jet substructure techniques enable us to reduce SM backgrounds more efficiently, hence get them under control. 

On top of background rejection, we pro-actively utilize jet substructure methods 
to extract partonic information from a merged jet initiated by $V_i\to q\bar q$.
As explicitly demonstrated in Section~\ref{sec:mergedJet}, the procedures in the methods {\it effectively} restrict relevant phase space of final states, and in particular, the accessible region in $\theta_i$ angles may be significantly affected.
The coefficients for the differential distributions in $\phi$ are related to $\theta_i$ in the narrow width approximation (NWA) as follows~\cite{Gao:2010qx}:
\bea
\frac{\ud^3 \sigma_{0^{++}}}{\ud \cos\theta_1 \ud \cos\theta_2 \ud \phi} \propto 
&&2 \sin^2\theta_1  \sin^2\theta_2 +\cosh^2\left(2\eta\right) \left(1+\cos^2\theta_1\right)\left(1+\cos^2\theta_2\right) \nonumber \\
&&-\cosh\left(2\eta\right) \sin\left(2\theta_1\right) \sin\left(2\theta_2\right)  \cos\phi  \nonumber\\
&&+\cosh^2\left(2\eta\right) \sin^2\theta_1 \sin^2\theta_2 \cos\left(2\phi\right) \, , 
\label{eq:three1}\\
\frac{\ud^3 \sigma_{0^{-+}}}{\ud \cos\theta_1 \ud \cos\theta_2 \ud \phi} \propto 
&&\left(1+\cos^2\theta_1\right)\left(1+\cos^2\theta_2\right) -\sin^2\theta_1 \sin^2\theta_2 \cos\left(2\phi\right) \, ,
\label{eq:three2}
\eea
where we average contributions from different quark and anti-quark flavors as we cannot discern them. 
Here $\eta$ defines a Lorentz boost factor as $\cosh\eta=M_S / (2 M_Z)$.
Certainly, the above expressions imply that jet clustering procedures alter $\phi$ distributions by limiting $\theta_i$ angles. 
If there were no restrictions on $\theta_i$, integrating $\theta_i$ over the full ranges of $(0, \pi)$ would give rise to differential distributions in $\phi$ as
\bea
\frac{\ud \sigma_{0^{++}}}{ \ud \phi} \propto && 2+\cosh^2\left(2\eta\right) \left[4+\cos\left(2\phi\right)\right] \xrightarrow[M_S \gg M_Z ]{} 4+\cos\left(2\phi\right) \, ,
\label{eq:cpEVEN} \\
\frac{\ud \sigma_{0^{-+}}}{ \ud \phi} \propto && 4-\cos\left(2\phi\right)\, .
\label{eq:cpODD}
\eea
However, as we pointed out in the previous section, fixing the angular separation between relevant subjets results in shrinking accessible phase space with respect to $\theta_i$ (see also eq.\,\eqref{eq:cosUpper}), and therefore, to appropriately interpret outputs from any data analyses for discriminating the CP state of $S$, we should be armed with a solid understanding of relevant effects. 

We shall closely look at this observation in the next section, taking a couple of benchmark points (BPs) with different jet size parameters in Cambridge/Aachen (C/A) algorithm~\cite{Dokshitzer:1997in,Wobisch:1998wt}. The following BPs are chosen to cover different kinematical regions: one for the moderately boosted $Z$ and the other for a highly boosted kinematics of $Z$.
\begin{itemize}
\item BP1\,: $M_S = 750\GeV$ with a large jet size of $R_{\MJ}=1.2$,
\item BP2\,: $M_S=1500\GeV$ with a decent jet size of $R_{\MJ}=0.6$.
\end{itemize}
For the mass choice in BP1, we expect moderately boosted phase space in which the associated merged jet analysis becomes comparable to analyses based on a normal jet size because the efficiency for tagging a single merged jet with C/A of $R=1.3$ becomes similar to that for tagging two ordinary jets with the anti-$k_t$ algorithm of $R=0.5$~\cite{Gouzevitch:2013qca, ATLAS-CONF-2016-082, CMS-PAS-B2G-16-010}.
We find that typical Lorentz boost factors in the two BPs are large enough (e.g., $\cosh\eta\simeq 4.17$ for BP1 and $\cosh\eta\simeq 8.33$ for BP2) to simplify eq.~\eqref{eq:three1} as
\beq
\frac{\ud^3 \sigma_{0^{++}}}{\ud \cos\theta_1 \ud \cos\theta_2 \ud \phi} \propto  \left(1+\cos^2\theta_1\right)\left(1+\cos^2\theta_2\right) +
 \sin^2\theta_1 \sin^2\theta_2 \cos\left(2\phi\right)\, .
 \label{eq:three1A}
\eeq
Note that the second term in this expression differs from that of eq.~\eqref{eq:three2} by the sign.
Denoting two relevant coefficients by $C_1$ and $C_2$, we have
\bea
C_1  &\equiv &\left[
\int_{(\cos \theta_1)_{\min}}^{(\cos \theta_1)_{\max}}
\ud \cos \theta_1 
\int_{(\cos \theta_2)_{\min}}^{(\cos \theta_2)_{\max}}
\ud \cos \theta_2
\right] \left(1+\cos^2\theta_1\right)\left(1+\cos^2\theta_2\right)\, ,\label{eq:C1exp} \\ 
C_2  &\equiv &
\left[
\int_{(\cos \theta_1)_{\min}}^{(\cos \theta_1)_{\max}}
\ud \cos \theta_1 
\int_{(\cos \theta_2)_{\min}}^{(\cos \theta_2)_{\max}}
\ud \cos \theta_2
\right]  \sin^2\theta_1 \sin^2\theta_2 \, , \label{eq:C2exp}
\eea
from which we find
\bea
\frac{\ud\sigma_{0^{\pm +}}}{\ud\phi}\propto 1 \pm \frac{C_2}{C_1} \cos(2\phi)=1\pm R_{\phi} \cos(2\phi)\,, \label{eq:RphiDef}
\eea
where we define the ratio of $C_2$ to $C_1$ as $R_{\phi}$. 
Hence, a better discriminating power is expected with a larger $R_{\phi}$.  
The expressions in eqs.\,\eqref{eq:cpEVEN} and~\eqref{eq:cpODD} suggest that this ratio at the parton level without any restriction on the phase space should converge to a quarter.
\beq
R_\phi \xrightarrow[\textrm{Full phase space}]{}  \frac{1}{4}
\label{eq:Ratio}
\eeq 

\begin{figure}[t]
\centering
\includegraphics[width=8cm]{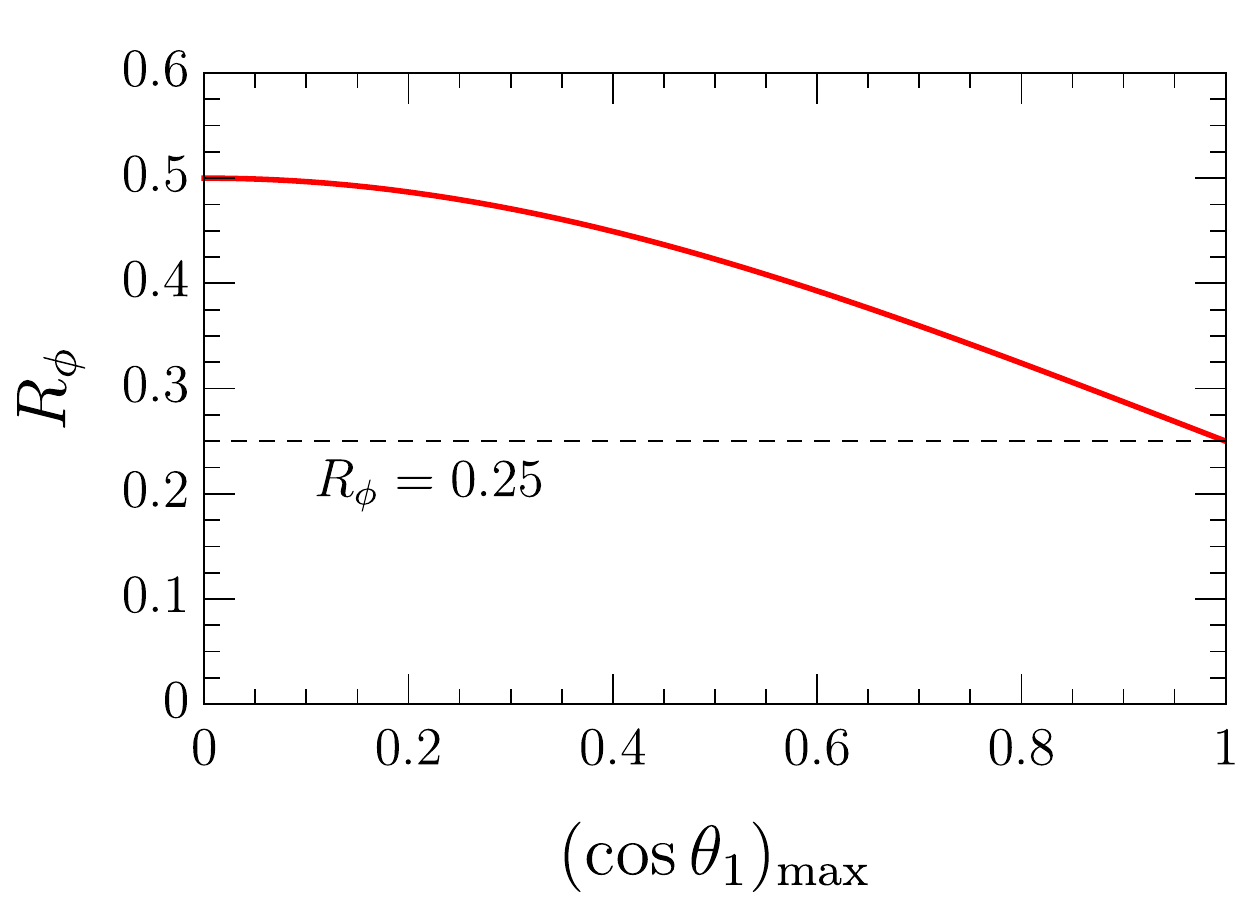}
\caption{\label{fig:Rphi} The functional behavior of $R_{\phi}$ according to $(\cos\theta_1)_{\max}$. The black dashed, horizontal line represents the $R_{\phi}$ value in the full phase space limit.}
\end{figure}

As discussed in the previous section, finding hadronic decaying $Z$ boson by a merged jet causes a phase space reduction toward the plane orthogonal to the $Z$ boson propagation direction as illustrated in eq.~\eqref{eq:cosUpper}. 
In this context, it is interesting to look into the behavior of $R_{\phi}$ as we restrict the phase space. 
Assigning $\theta_1$ and $\theta_2$ to hadronic and leptonic branches, respectively, we restrict $\theta_1$ under the assumption that $(\cos \theta_1)_{\max} = -(\cos \theta_1)_{\min}$ and $\theta_2$ is unrestricted for simplicity.\footnote{In more realistic situations, there arises some mild phase space reduction even on the leptonic side. However, we here isolate the effect induced in jet substructure techniques for developing the relevant insight.}
Noting that the integrands in eqs.~\eqref{eq:C1exp} and~\eqref{eq:C2exp} are even in $\cos\theta_1$, we find that $R_{\phi}$ can be expressed as
\bea
R_{\phi}=\frac{1}{2}\cdot\frac{(\cos\theta_1)_{\max}-\frac{(\cos\theta_1)_{\max}^3}{3}}{(\cos\theta_1)_{\max}+\frac{(\cos\theta_1)_{\max}^3}{3}}\,, 
\eea
from which we see that $R_{\phi}$ becomes $1/4$ with $(\cos \theta_1)_{\max}$ approaching to 1 (i.e., full phase space). 
Figure~\ref{fig:Rphi} shows the functional behavior of $R_{\phi}$ over $(\cos\theta_1)_{\max}$, wherein $R_{\phi}$ monotonically increases as $(\cos\theta_1)_{\max}$ decreases. 
This implies that phase space reduction by a fixed cone size renders $R_{\phi}$ greater than 1/4 (dashed black horizontal line in Figure~\ref{fig:Rphi}), {\it remarkably} achieving better identification on the CP state. 
In the next section we shall confirm this observation with Monte Carlo simulation at both parton and detector levels. 

\section{Results with jet substructure techniques \label{sec:results}}

In this section, we present our results with jet substructure techniques, using Monte Carlo simulation. For a more realistic study, we consider various effects such as parton shower, hadronization/fragmentation, and detector responses.
To this end, we take a chain of simulation programs.
We first create our model files using FeynRules\,\cite{Alloul:2013bka} and plug them into a Monte Carlo event generator \textsc{MadGraph5}\,\cite{Alwall:2014hca} with parton distribution functions parameterized by \textsc{NN23LO1}\,\cite{Ball:2012cx}.
The generated events are further pipelined to \textsc{Pythia 6.4}\,\cite{Sjostrand:2006za} for taking care of showering and hadronization/fragmentation, and to \textsc{Delphes-3.3.2}\,\cite{deFavereau:2013fsa} with a CMS detector model for taking care of detector responses.
In order to form jets from the final state particles, we employ the particle-flow algorithm in \textsc{Delphes-3.3.2} and feed resultant particle-flow objects to \textsc{FastJet}\,\cite{Cacciari:2011ma,Cacciari:2005hq}.

\subsection{Event reconstruction}

\begin{figure}[t!]
\begin{center}
\begin{tabular}{cc}
\includegraphics[width=7.2cm]{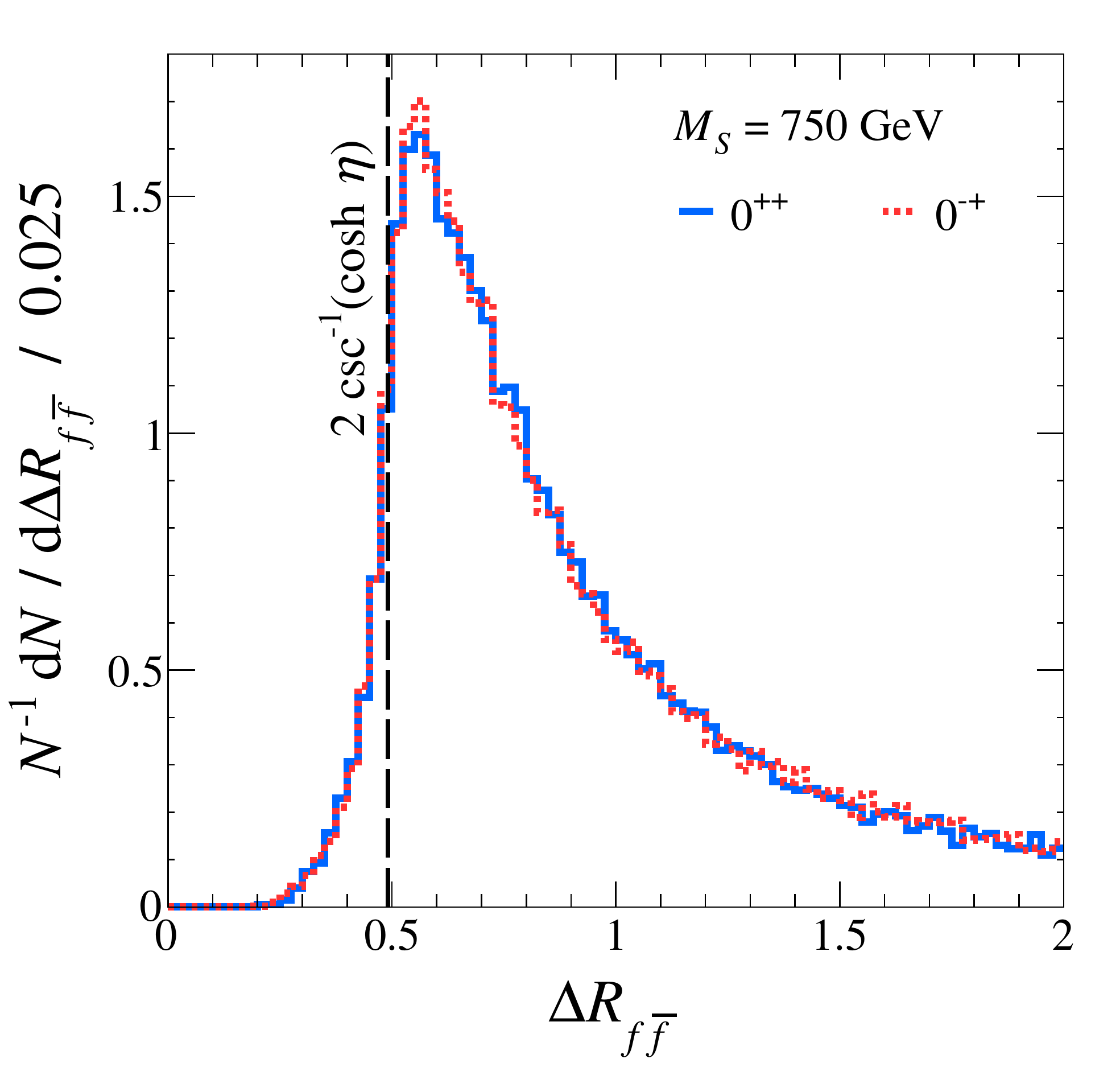} \quad &
\includegraphics[width=7.2cm]{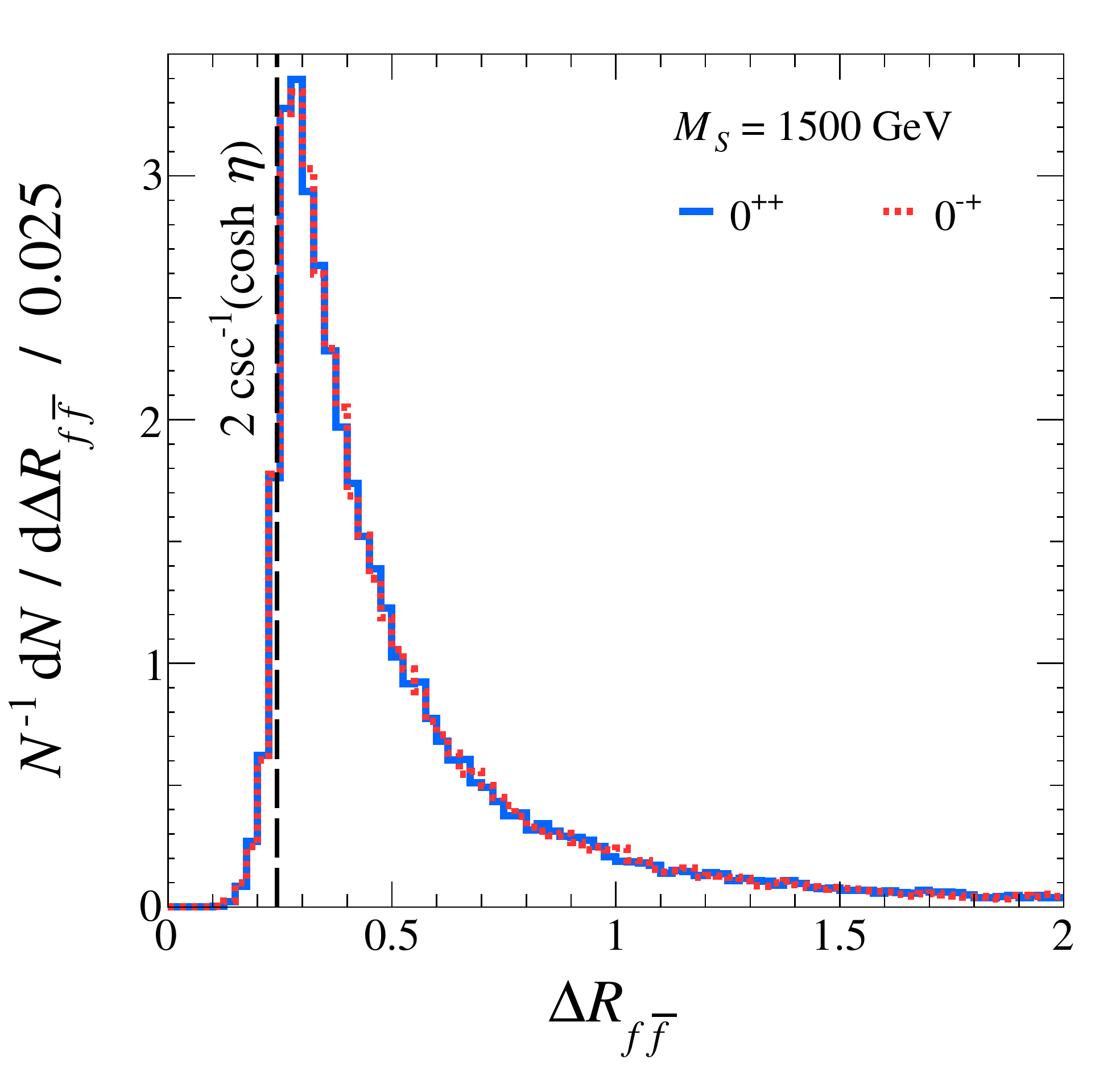} \\
\end{tabular}
\end{center}
\caption{\label{fig:separation_parton} Parton level distributions of $\Delta {R_{f\bar{f}}}$ for $M_S = 750$ GeV (left panel) and $M_S=1500\GeV$ (right panel) for which $f$ denotes any fermionic decay product of the $Z$ gauge boson.
We select events with a mass window of $|M_{Z} - m_{f\bar{f}}| < 15 \,\Gamma_Z$. 
Dashed black lines mark the expected lower bounds of $\Delta {R_{f\bar{f}}}$ as in eq.\,\eqref{eq:Rsize} for which $P_T$ of a $Z$ boson is localized at the Jacobean peak of $M_Z \sinh\eta$ with $\cosh\eta$ being a boost factor. 
}
\end{figure}

As we discussed earlier, for our benchmark points belonging to a high mass regime, $Z$ decay products are likely to be highly collimated. 
Denoting the angular distance between the two (fermionic) decay products as $\Delta R_{f\bar f}$, its distribution develops a peak as shown in Figure~\ref{fig:separation_parton}, which is inherited from a Jacobian peak in the $Z$-boson transverse momentum distribution. 
The last statement can be understood by restricting eq.\,\eqref{eq:Rsize} into the transverse plane, i.e.,
\bea
\Delta R_{f\bar f} \approx 2\csc^{-1} \left(\frac{\sqrt{P_{T(f\bar f)}^2+ m_{f\bar f}^2}}{m_{f\bar f}}\right)\,,
\eea
where $m_{f\bar{f}}$ is the invariant mass of two decay products. 
The minimum opening distance is obtained by setting the numerator to be half the mass of $S$
\bea
\Delta R_{f\bar f} \ge 2\csc^{-1}\left(\frac{M_S}{2 m_{f\bar f}}\right) \simeq 2\csc^{-1}\left(\frac{M_S}{2 M_Z}\right)\,.
\label{eq:parton_dis}
\eea
Note that $m_{f\bar{f}}$ follows the usual Breit-Wigner distribution around $M_Z$, and therefore, some small fraction of events can populate even below the expected minimum value $2\csc^{-1}(M_S/(2 M_Z))$ in the $\Delta R_{f\bar{f}}$ distributions exhibited in Figure\,\ref{fig:separation_parton}.\footnote{Note that $\csc^{-1}(x)$ is a monotonically decreasing function in terms of $x$.}
Predicated upon this parton level assessment, we determine an isolation criteria for leptonic decay products of $Z$ bosons in Section~\ref{sec:LepIsol} and a jet size for clustering merged jets to capture hadronic decaying $Z$ bosons into a single jet in Section~\ref{sec:TagMJ}. 

\subsubsection{Lepton isolation criteria \label{sec:LepIsol}}

To reconstruct individual $Z$ boson-induced leptons without any confusion with heavy flavor quark-induced leptons, we require the following isolation criteria:
\beq
I = \frac{1}{p_{T,\ell}} \sum_{i\neq \ell} p_{T,i} <  I_\textrm{iso} \, ,
\eeq
where $\ell$ is a candidate for an isolated lepton, and $i$'s are any particles in the vicinity of the lepton candidate $\ell$ which satisfies
\beq
 \Delta R_{i\ell} < R_\textrm{iso} \textrm{ and } p_{T,i} \ge p_T^{\min}\, .
\eeq
Isolation parameters for each benchmark point are tabulated in Table~\ref{tab:lepton_reco}, for which the values are conventional~\cite{Khachatryan:2015hwa,Chatrchyan:2012xi}
except that for $R_{\textrm{iso}}$.
We choose $R_{\textrm{iso}}$ so as to have an isolated lepton according to the observation made in Figure \ref{fig:separation_parton}.
\begin{table*}[t!]
\centering
\begin{tabular}{|c|c|c|}
\hline
$M_S$ & 750 GeV & 1500 GeV\\
\hline
$R_{\textrm{iso}}$                   & 0.3     & 0.2 \\
$p_T^{\min}$          & 0.5 GeV & 0.5 GeV\\
$I_{\textrm{iso}}$            & 0.12  & 0.12 \\
\hline
\end{tabular}
\caption{\label{tab:lepton_reco} Isolation parameters for each benchmark point to reconstruct an individual lepton from a $Z$ boson decay.  }
\end{table*}

\subsubsection{Tagging a merged jet \label{sec:TagMJ}}

We begin with applying C/A algorithm to cluster particles from a hadronically decaying $Z$ boson. 
As this is a sequential recombination algorithm based on the angular separation between two objects, it is useful for us to access sub-clusters by the angular order, in particular, to evaluate the $\phi$ angular variable.
The algorithm combines two objects, which have the smallest angular distance, by adding up their momenta. 
This combining process continues until every clustered object is isolated from the others by an angular distance $R_\MJ$.
Here $R_\MJ$ defines the jet size for a merged jet in the C/A algorithm. 
In language of the $k_T$ algorithm~\cite{Ellis:1993tq}, the C/A algorithm is equivalent to the sequential clustering with a metric between objects,  
\beq
\label{eqn:CA_metric}
d_{ij} = \frac{\Delta R_{ij}^2 }{R_\MJ^2}, \quad d_{iB} = 1 \,,
\eeq
where $B$ denotes the beam line.\footnote{Here $B$ is a legacy notation of $k_T$ algorithm, as $d_{iB}$ for C/A algorithm does nothing with the beam line and it is just related to the threshold angular scale $R_{\mathrm{MJ}}$.}
For each iteration, two objects which have the smallest $d_{ij}$ are combined. 
If $d_{iB}$ is smallest, the object $i$ is promoted to a C/A jet and escapes from clustering.
The iteration terminates if all objects are identified as jets. 
After completion of clustering, we can obtain an angular hierarchy of sub-clusters by simply rewinding the clustering procedure.
We then match sub-clusters to partons from a $Z$ gauge boson, imposing relevant cuts to reduce the possibility of mistagging a QCD jet as a $Z$-induced one.
To achieve this goal, we employ the Mass Drop Tagger (MDT)~\cite{Butterworth:2008iy} whose procedure is briefly summarized below. 
The MDT essentially traces back the clustering sequences of a C/A jet and attempts to find subjets satisfying the symmetric conditions.
\begin{table}[t]
\begin{center}
\begin{tabular}{|c|c|c|}
\hline
$M_S$ & 750 GeV & 1500 GeV\\
\hline \hline
$R_\MJ$   & $1.2$                  & $0.6$ \\
\hline
$(\mu_*\, , y_*)$  & \multicolumn{2}{c|}{$(0.67\, , 0.09)$} \\
\hline
$R_*$     &  0.3 &  0.2 \\
\hline \hline
Jet size & \multicolumn{2}{c|}{Efficiency } \\
\hline
$R_\MJ - 0.2$     &  10.5\%  & 12.9\%  \\
$R_\MJ $          &  10.4\%  & 13.4\%  \\
$R_\MJ + 0.2$     &  10.4\%  & 13.3\%  \\
$R_\MJ + 0.4$     &     -     & 13.0\%  \\
\hline
\end{tabular}
\end{center}
\caption{\label{table:MJ_reco} Parameters of the MDT procedures for the two benchmark points. Here we adopt the same $\mu_*$ and $y_*$ as in the original BDRS Higgs boson tagger~\cite{Butterworth:2008iy} for $M_Z \sim M_H \sim  \mathcal{O}(100)\GeV$. $Z$-tagging efficiencies in the table are evaluated after selection cuts in Table~\ref{tab:cut_flow} for a given merged jet size. A jet size $R_\MJ$ is chosen such that tagging efficiencies with different $R_\MJ$'s remain unchanged.  }
\end{table}
\begin{enumerate}
\item[(1)]
Clustering: We cluster energy deposits in calorimeters using the C/A algorithm of a jet radius $R=R_\MJ$. 
\item[(2)]
Splitting to look into a substructure: We rewind the last clustering sequence of a jet $j$, labelling two subjets as $j_1$ and $j_2$ with $m_{j_1} > m_{j_2}$.
\item[(3)]
Checking symmetric conditions:  Major backgrounds in our case would be $Z(\to \ell^+\ell^-)+j$s where a quark-initiated jet appear as a merged one.
In this case, most of the energy deposits are inclined to be localized along the momentum direction of the initial quark, so that there is a high chance of unbalanced energy sharing between two subjets including mass and transverse momentum. 
In contrast, a signal MJ consists of two prongs (i.e., two quarks) that ensure democratic energy sharing in two subjets. 
To quantify this difference, the MDT demands an upper bound $\mu_*$ and a lower bound $y_*$ on MDT parameters $\mu$ and $y$, respectively:
\beq	
\mu\equiv \frac{m_{j_1}}{m_j} < \mu_* \,, \quad y\equiv\frac{ \min \left( P_{T(j_1)}^2, P_{T(j_2)}^2 \right) }{m_j^2} \Delta R^2_{j_1 j_2} > y_*\, .
\label{eq:MDTcuts}
\eeq
This procedure is useful to discriminate prongs in subjects from soft showering, on top of reducing backgrounds.
If subjets do not satisfy above criteria, the MDT procedure redefines $j_1$ as $j$ and repeats the rewinding procedure in (2). 
\end{enumerate}
Once the MDT tags a signal MJ and locates two prongs in the MJ, it decontaminates QCD corruptions in subjets by reclustering energy deposits in the MJ again with the C/A algorithm of a small radius jet size $R_\textrm{filt}$, 
\begin{enumerate}
\item[(4)]
Filtering: We reculster constituents of an MJ with the C/A algorithm of radius,
\beq
R_{\textrm{filt}}=\min\left(R_*\,,\frac{\Delta R_{j_1j_2}}{2}\right)
\eeq 
to find $n$ new subjets $\{s_1, s_2,\cdots, s_n\}$  ordered in descending $P_T$. 
Here $R_*$ is the maximum allowed size for subjets in order to minimize the QCD contamination. The MDT takes into account an $\mathcal{O}(\alpha_s)$ correction from hard radiation, by allowing up to three subjets in redefining an MJ as 
\beq
p_\MJ^\mu = \sum_{i=1}^{\min(n,3) } p_{s_i}^\mu\, .
\eeq
\item[(5)]
Assigning subjets to prongs from a $Z$: If we have only two subjets $\{s_1, s_2\}$, we take these two subjets as two particles from a $Z$ boson. 
In the case where we have three subjets  $\{s_1, s_2, s_3\}$, we merge $s_3$ with
other subjet $s_i$ which has the smaller angular distance from $s_3$. By this merging process, we identify subjets $\{j_{1}, j_{2}\}$ in an MJ as
\beq
\{p_{j_{1}}^\mu, p_{j_{2}}^\mu\} = 
\begin{cases}
\{p_{s_1}^\mu, p_{s_2}^\mu\} \text{ for }n=2\, , \\
P_T\text{-ordered }\{p_{s_i}^\mu+p_{s_3}^\mu, p_{s_j}^\mu\} \text{ with }\Delta R_{s_i\,s_3} < \Delta R_{s_j\, s_3}\text{ for } n = 3\, .
\end{cases}
\label{step:MJ}
\eeq
\end{enumerate}
We summarize parameters of the MDT procedures for two benchmark points in Table~\ref{table:MJ_reco}. 
As an MDT procedure has a cut $y_*$ on the phase space, we expect certain effects on the angular distributions in return as the cone size of a merged jet 
restricts the polar angles of decaying particles from $Z$ bosons.
Since a jet clustering procedure with the MDT is a key process to recover the parton-level information from the corresponding reconstruction-level information, we investigate effects from the MDT to understand phase-space distortion in reconstruction-level analyses.

\subsection{Phase-space distortion from a jet substructure}
\label{sec:JCUT}

As discussed earlier, constructing a merged jet to capture partons from the decay of a heavy (boosted) particle often accompanies cuts to suppress the rate to misidentify an ordinary QCD jet as an MJ. 
In the MDT procedure, symmetric cuts $\mu_*$ and $y_*$ are utilized to reduce single-prong jets from QCD backgrounds. 
While the $\mu_*$ cut does not give any strong
restriction on signal MJs, the $y_*$ cut may result in a limit on the phase space of the subjets from a $Z$ boson. 
Suppose that the softer subjet $j_2$ carries away $z$ fraction of the total momentum, i.e., $zP_{T(\MJ)}$. 
We then find that symmetric cut $y$ in eq.~\eqref{eq:MDTcuts} can be expressed as
\beq
\frac{ \min ( P_{T(j_1)}^2, P_{T(j_2)}^2 ) }{m_\MJ^2} \Delta R^2_{j_1 j_2} \simeq \left(\frac{z^2\, P_{T(\MJ)}^2}{m_\MJ^2}\right)\left(\frac{1}{\sqrt{z (1-z)}} \frac{m_\MJ}{P_{T(\MJ)}}\right)^2 > y_* \, ,\label{eq:zinequi}
\eeq
where in the first step we make use of eq.~\eqref{eq:R} in the limit of $P_{T(\MJ)} \gg m_\MJ$. 
A similar expression is readily available for the harder subjet $j_1$ which takes away the momentum of $(z-1)P_{T(\MJ)}$.
Solving the two inequalities for $z$ (one from eq.~\eqref{eq:zinequi} and the other from the corresponding one for $j_1$), we find that for a given $y_*$, the momentum sharing $z$ should be confined to a region defined by
\beq
\frac{y_*}{1+y_*} < z < \frac{1}{1+y_*}\, ,
\eeq
which, in turn, restricts the angular distance between the two subjets in terms of $y_*$, 
\beq
R_\textrm{est}^{(y_*)} \le \frac{1+y_*}{\sqrt{y_*}} \frac{m_\MJ}{p_{T(\MJ)}}\, .
\label{eq:MDT_R}
\eeq

\begin{figure}[t]
\begin{center}
 \includegraphics[width=0.97\textwidth]{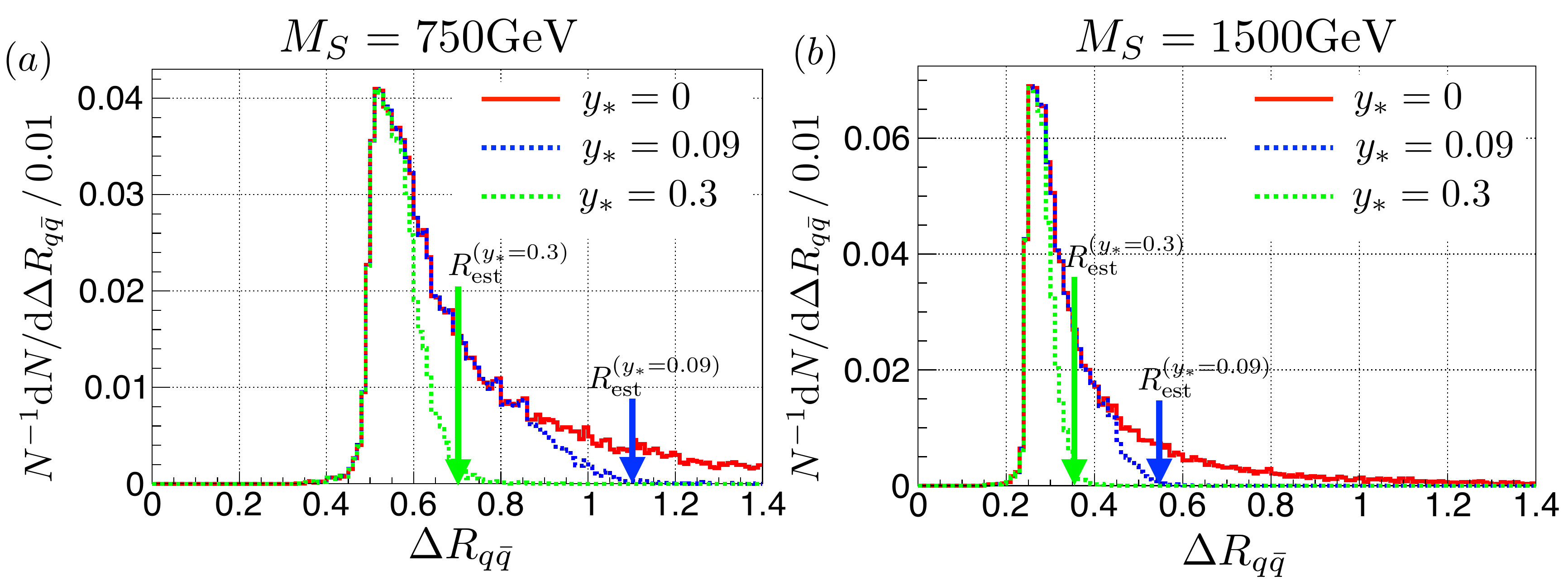}
 \end{center}
   \caption{$\Delta R_{q\bar q}$ distributions between the two quarks from a $Z$ boson decay for (a) $M_S =750\GeV$ and (b) $M_S=1500\GeV$ with a basic cut of  $P_{T(\MJ)}> 0.40\,M_S$.  
$R_{\textrm{est}}^{(y_*)}$ is an estimated distance between subjets evaluated from eq.~\eqref{eq:MDT_R} with a symmetric cut $y_*$ in the BDRS tagger.}
\label{fig:ycut}
\end{figure}

To develop our intuition on the effects from this restriction, we apply symmetric conditions of the MDT to parton-level simulation data for $S\rightarrow ZZ \rightarrow q\bar{q}\ell^+\ell^-$.
 In a parton-level simulation, only the $y_*$ cut in eq.~\eqref{eq:MDTcuts} remains effective. Thus we impose a symmetric cut $y_*$ to the two quarks from a $Z$ boson.
We then plot distributions of $\Delta R_{q\bar{q}}$ with the events whose $y$ values are greater than a certain $y_*$. 
Figure~\ref{fig:ycut} shows those distributions for three different $y_*$ values (red solid histogram for $y_*=0$, blue dotted histogram for $y_*=0.09$, and green dotted histogram for $y_*=0.3$) with $M_S = 750$ GeV (left panel) and $M_S=1500$ GeV (right panel). 
We impose a basic cut of $P_{T(\MJ)}> P_{T(\MJ)}^*= 0.40\,M_S$ as well, which is well-motivated in the sense of reducing backgrounds by focusing on the central region.
\begin{figure}[t]
\begin{center}
\begin{tabular}{cc}
\includegraphics[width=7.2cm]{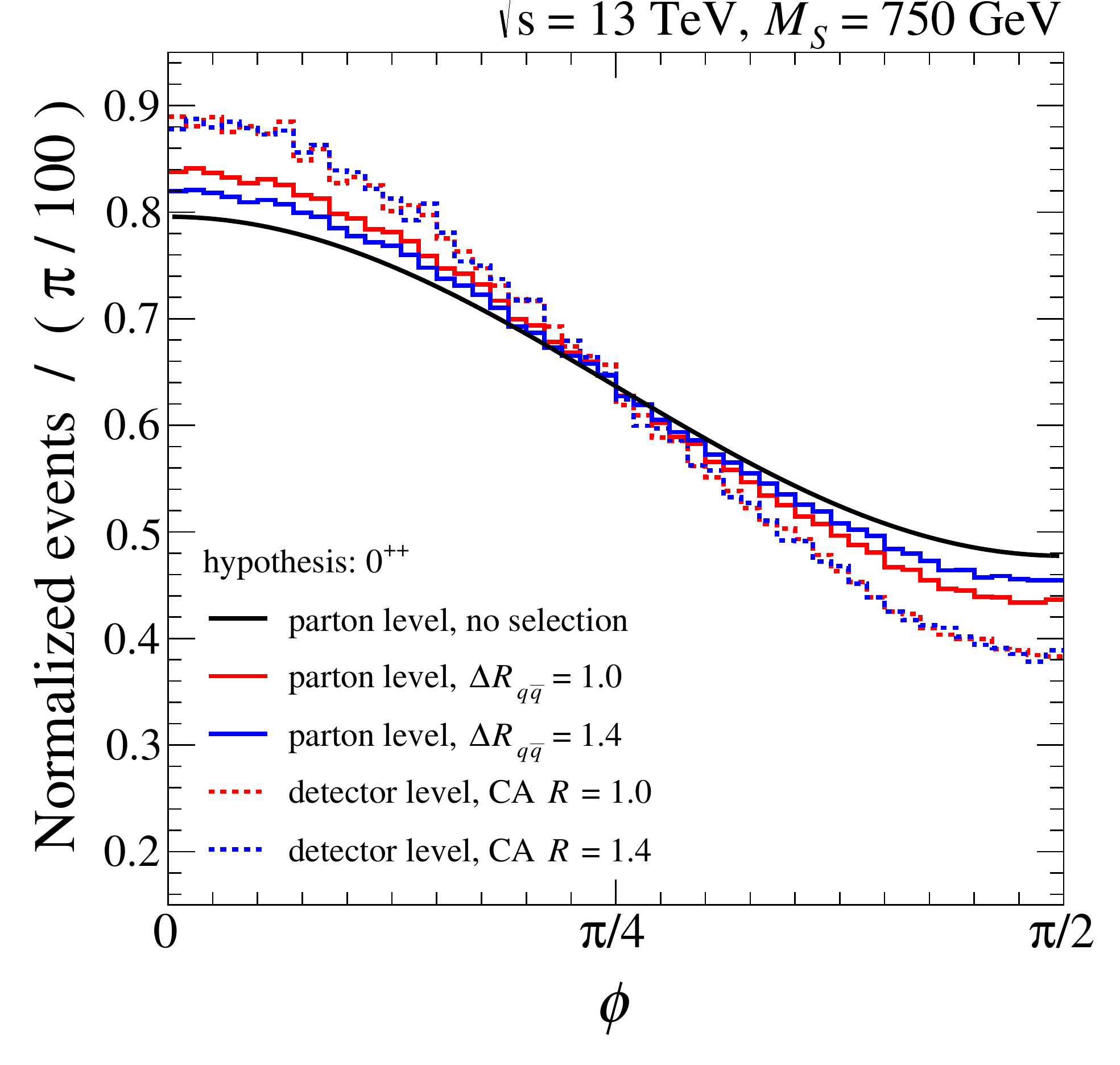} &
\includegraphics[width=7.2cm]{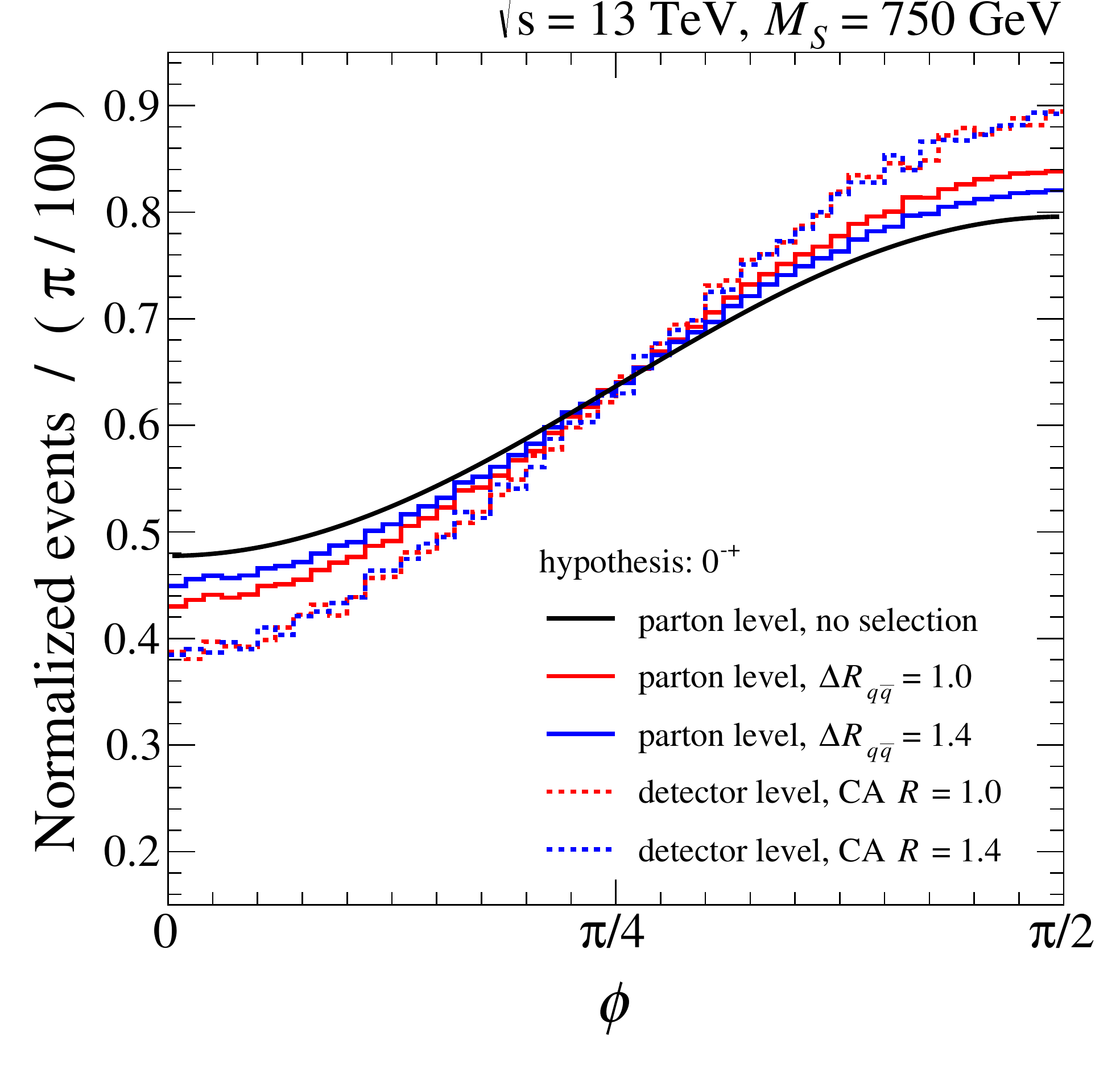} \\ 
\includegraphics[width=7.2cm]{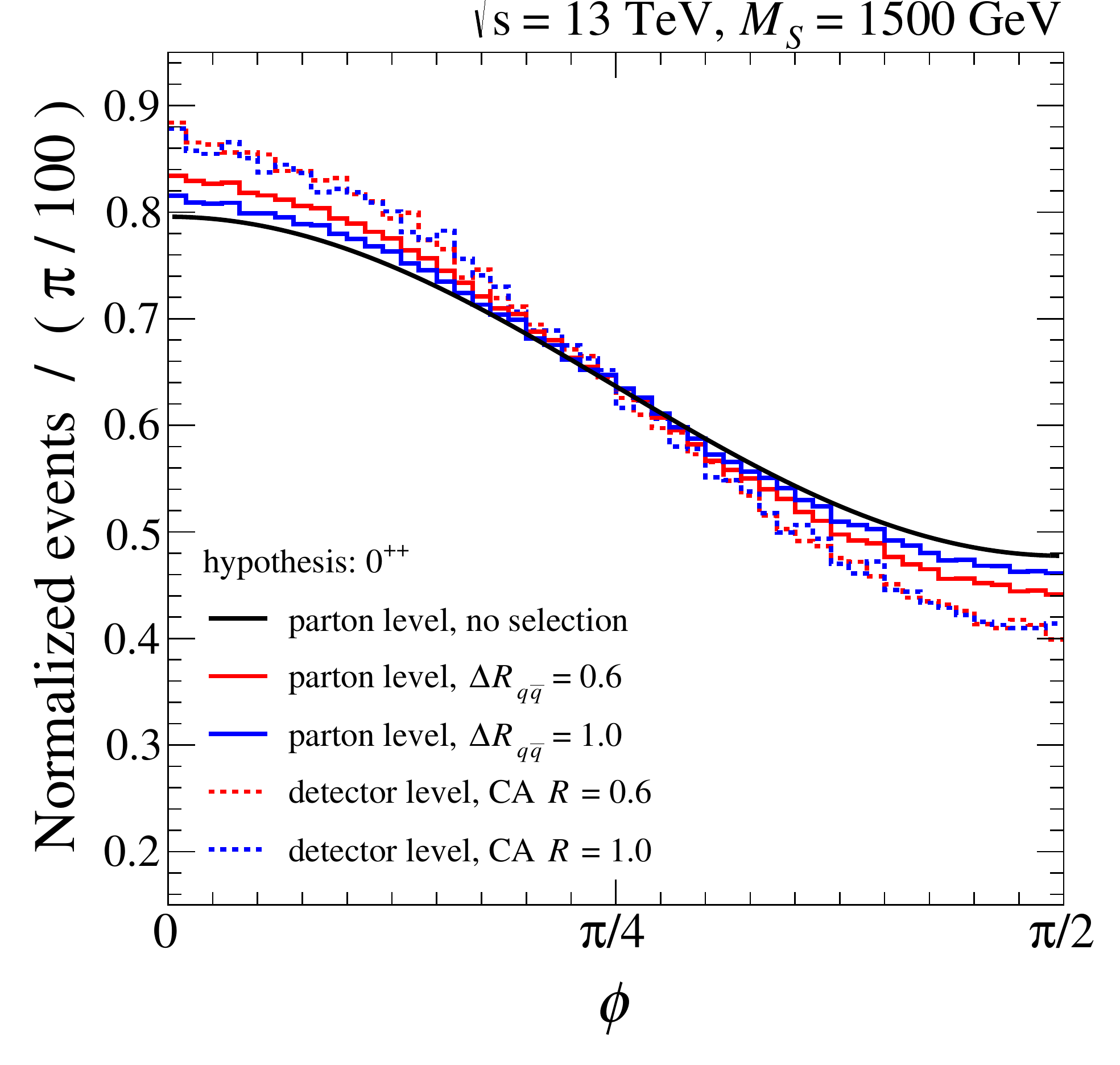} &
\includegraphics[width=7.2cm]{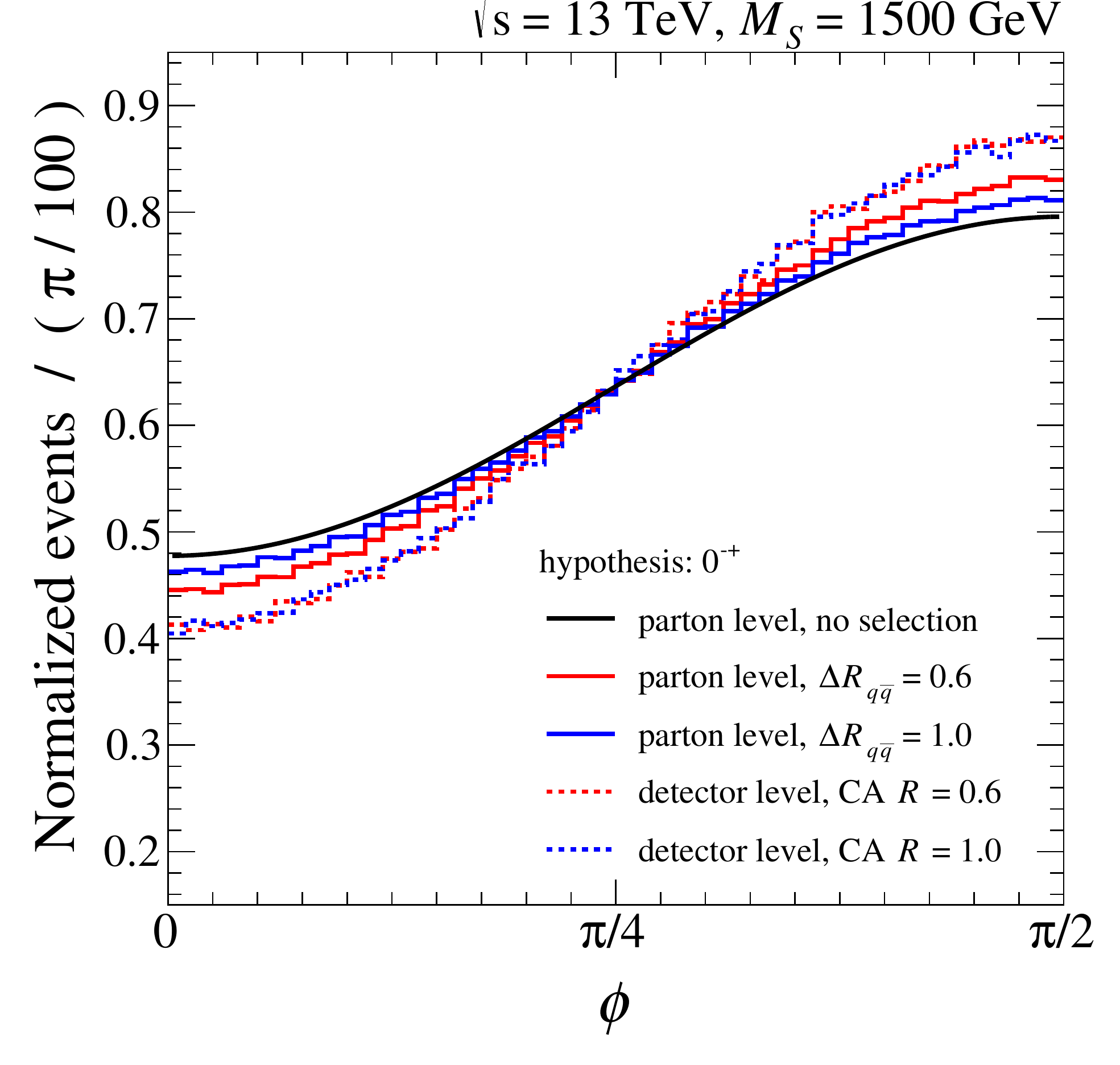}  
\end{tabular}
\end{center}
\caption{\label{fig:histogram_phi} $\phi$ distributions under the hypotheses $0^{++}$ (left column) and $0^{-+}$ (right column) at the parton level (solid lines) and the detector level (dotted lines). Black solid lines in the four panels are theoretical expectations without any restriction on the angular distance between two quarks as in eqs.\,\eqref{eq:cpEVEN} and\,\eqref{eq:cpODD}. 
}
\end{figure}

We clearly observe that as we increase the $y_*$ cut, more phase space with large $\Delta R_{q\bar{q}}$ is removed. 
This implies that even though we begin with a sufficiently large cone size $R_\MJ$ to retain most of phase space as in Table~\ref{table:MJ_reco}, the $y_*$ cut in the MDT procedure effectively restricts the available region of $\Delta R_{q\bar{q}}$ below $R_\textrm{est}^{(y_*)}$.
Moreover, we find that $\Delta R_{q\bar{q}}$ is smaller than typical choices of $R_{\MJ}$, for example, 
\bea
&&\Delta R_{q\bar q} \le R_\text{est}^{(y_*=0.09)} \simeq 1.1 < R_\MJ\,(=1.2) \text{ for }M_S=750\GeV \, ,
\label{eq:ycutBP1}\\
&&\Delta R_{q\bar q} \le R_\text{est}^{(y_*=0.09)} \simeq 0.55 < R_\MJ\,(=0.6) \text{ for }M_S=1.5\TeV\, ,
\label{eq:ycutBP2}
\eea
which are marked by blue arrows in Figure~\ref{fig:ycut}. The resulting restriction on $\cos\theta$ (i.e., the $Z$ rest-frame polar angle of the harder quark relative to the $Z$ boost direction) can be derived from eqs.~\eqref{eq:cosUpper} and~\eqref{eq:MDT_R}:
\beq
|\cos\theta| \le \sqrt{1- \left(\frac{m_\MJ}{P_{T(\MJ)}^*}\right)^2\cot^2\left(\frac{R_\MJ}{2}\right)}  \ 
\simeq \sqrt{1-\frac{4y_*}{(1+y_*)^2}+\frac{2}{3}\left(\frac{m_\MJ}{P^*_{T(\MJ)}}\right)^2 } \, , 
\label{eq:jet_dist}
\eeq
where the approximation is valid up to the second order in $\left(R_\MJ/2\right)$.
Thus as far as $R_\MJ$ is larger than $R_\text{est}^{(y_*)}$, the cone size $R_\MJ$ does not invoke any direct deformation on the phase space, compared to cuts in the MDT procedure, which are introduced to reduce background QCD jets. 

Our Monte Carlo study indeed confirms this observation. 
In Figure\,\ref{fig:histogram_phi}, we contrast the $\phi$ distributions at the parton level with those at the detector level.
For parton-level distributions, we restrict the angular distance between two quarks from a $Z$ boson decay by two different upper bounds of $\Delta R_{q\bar q}$ for each benchmark point.
In the case of $M_S=750\GeV$ (upper panels), the two upper bounds are chosen to be $1.0$ (red lines) and $1.4$ (blue lines) to have $R_\MJ=1.2$ between them. 
Similarly, in the case of $M_S=1.5$ TeV (lower panels), they are chosen to be $0.6$ (red lines) and $1.0$ (blue lines).
We clearly see that $\phi$ distributions depart further from the theory expectation (solid black curves) with the smaller cone size $\Delta R_{q\bar q}$, whether the resonance is CP-even (left panels) or CP-odd (right panels). 
When it comes to detector-level analyses, however, once we introduce a fairly hard $y_*$ cut resulting in $R_\text{est}^{(y_*)}<R_\MJ$, the above-discussed parton-level effect simply disappears. 
Corresponding dotted lines in Figure\,\ref{fig:histogram_phi} clearly support our expectation that final $\phi$ distributions are not much different even with different $R_\MJ$ values.\footnote{Additional cuts including a detector geometry cut and object selection cuts (especially $P_T$) give further restrictions on the phase space. Thus angular distributions are distorted further, compared to parton-level distributions.}
We also understand this point from ``constant'' MJ-tagging efficiencies even with different C/A jet sizes in Table~\ref{table:MJ_reco}. 
Although we vary the size of MJs with different $R_\MJ$ values, the overall cut on the angular distance of a quark pair is determined by $R_\text{est}^{(y_*)}$, allowing us to have a ``stable'' MJ-tagging rate. 

Another important message that one may realize from this series of exercises is that the impact of analysis cuts upon detector-level reconstructed objects is in {\it more favor} of our goal of discriminating CP states, unlike typical expectations in detector-level data analyses. 
More specifically, the difference of $\phi$ distributions between CP-even and CP-odd cases appears enhanced even after incomplete integrations over angular variables such as polar angles $\theta_i$ in eqs.~\eqref{eq:three1} and~\eqref{eq:three2} (see also Figure~\ref{fig:Rphi}). 
This enhancement overcomes the adverse effects of detector resolution which often degrade subsequent data analyses.

\section{Analysis with matrix element methods \label{sec:MEManal}}

In this section, we discuss further analyses with matrix element methods using four-momentum information of subjets obtained by the jet substructure technique delineated in the previous section. 
We begin with a general overview for CP state discrimination with various measures, followed by matrix element methods and our main results with them.

\subsection{Determining the CP property}

To deal with experimental systematics properly and maximize distinctive asymmetric features between the $\phi$ differential distributions for CP-even and CP-odd resonances, a simple measure $A_{\phi}$ has been introduced for the $S\to ZZ \to 4\ell$ channel~\cite{Chala:2016mdz}:
\beq
A_{\phi} = \frac{ N \left(\phi > \frac{\pi}{4} \right) - N \left(\phi < \frac{\pi}{4} \right)}{N \left(\phi > \frac{\pi}{4} \right) + N \left(\phi < \frac{\pi}{4} \right)}\, ,
\label{eq:asymmetric}
\eeq
where $N$ simply denotes the number of events. 
One may make use of the below-defined cumulative probability over $A_{\phi}$ as a measure to determine the unusualness for any observation $A_{\phi}^{\mathrm{obs}}$ under a given hypothesis,
\bea
 p_{0^{++}} (A_\phi^{\mathrm{obs}};A_\phi) & = & P(A_\phi \ge A_\phi^{\mathrm{obs}} | 0^{++} )\, ,
 \label{eq:Asympp} \\
 p_{0^{-+}} (A_\phi^{\mathrm{obs}};A_\phi) & = & P(A_\phi \le A_\phi^{\mathrm{obs}} | 0^{-+} ) \, ,
 \label{eq:Asymnp}
\eea
where $P$ implies the associated probability. 

An alternative method to obtain a probability density function ({\it pdf\,}) is the kernel density estimation (KDE). One may estimate a {\it pdf\,} from simulated data, and use the estimated function $f_{\mathrm{KDE}}$ in performing a log-likelihood ratio test as
\beq
q_{\phi} = \sum_{i=1}^{N_{\textrm{evt}}} \ln 
\left(\frac{f_{\mathrm{KDE}}(\phi_i | 0^{++}) }{f_{\mathrm{KDE}}(\phi_i | 0^{-+})} \right) \, ,
\label{eq:LLphi}
\eeq
to obtain the most powerful test between two simple hypotheses at a given significance level $\alpha$ according to the Neyman-Pearson lemma~\cite{Neyman289,Agashe:2014kda}. 
The {\it pdf\,} $P(q_\phi | 0^\textrm{PC})$ for a test statistic $q_\phi$ with a given hypothesis $0^\textrm{PC}$ and the given number of events $N_\textrm{evt}$ is calculated from a huge number of pseudo-experiments which are generated with hypothesis $0^\textrm{PC}$.
The corresponding cumulative probabilities based on $q_{\phi}$ are  
\bea
 p_{0^{++}} (q_\phi^{\mathrm{obs}};q_\phi) & = & P(q_\phi \le q_\phi^{\mathrm{obs}} | 0^{++} )\, ,
 \label{eq:LLphipp} \\
 p_{0^{-+}} (q_\phi^{\mathrm{obs}};q_\phi) & = & P(q_\phi \ge q_\phi^{\mathrm{obs}} | 0^{-+} ) \, .
 \label{eq:LLphinp}
\eea

However, the above approaches, which are based on $\phi$ distributions, rely on the projection of our observed momenta of visible particles, $\{{\bf p}^{\textrm{reco}} \} = \{p_{j_{1}}, p_{j_{2}}, p_{\ell^-} , p_{\ell^+} \}$, into a single angular variable $\phi$. Although in our study, phase-space reduction by cuts in jet substructure methods can enhance the difference between two CP hypotheses as we have observed in the previous section, it does not guarantee whether this projection attains the best sensitivity in cases where there exist at least three correlated angular variables as in eqs.\,\eqref{eq:three1} and\,\eqref{eq:three2}. 
In the next section, we instead directly convert the observed momenta into a probability under a given model hypothesis. 
We then utilize this probability as a likelihood ratio test between different hypotheses on the CP state of a scalar resonance $S$ in our study.

\subsection{Matrix Element Method}

As briefly mentioned in eq.~\eqref{eq:MEM}, the probability based on the matrix element in a given hypothetical process $\alpha$ is given by
\beq
\label{eq:MEM2}
\mathcal{P}\left(\{{\bf p}^{\textrm{reco}}\} | \alpha \right) = \frac{1}{\sigma_\alpha}
\int\ud x_1 \ud x_2  \frac{f_{p_1}(x_1) f_{p_2}(x_2)}{2s\, x_1 \, x_2} \int\ud \Pi_{q_i} \, \mathcal{W}\left( {\bf q_i} , \{\bf{p}^{\textrm{reco}}\}\right) \, 
\Big|\mathcal{M} \left({\bf q}_j \,; \alpha\right) \Big|^2  \, ,
\eeq
where $f_{p_i}(x_i)$ is a parton distribution function  of parton $p_i$ inside the beam with a fractional energy of $x_i$. 
$\Pi_{q_i}$ describes the phase space of parton-level particles $q_i$ which are related to observed momenta $\{{\bf p}^{\textrm{reco}}\}$ of corresponding particles.
If detectors were perfect, such a relation would be trivial.  
However, as instrumental effects including detector smearing and responses become important factors in precise measurements, transfer functions 
$\mathcal{W}\left( {\bf q_i} , \{\bf{p}^{\textrm{reco}}\}\right)$ are introduced to map the information from reconstructed particles to the parton-level input for the MEM by modelling energy smearing, in particular, effects in jet reconstruction stemming from showering, hadronization/fragmentation, and jet energy scales with gaussian functions that were obtained in the course of understanding top-quark properties in the Tevatron experiments~\cite{EstradaVigil:2001eq, Canelli:2003id, Abazov:2004cs, Abazov:2004ym, Abazov:2014dpa,D0:2016ull}. 
To reduce the dependence on the transfer function in~\eqref{eq:MEM2}, one may use a deeper substructure of merged jets, e.g., finer subjet analyses as in the shower deconstruction method~\cite{Soper:2011cr,Soper:2012pb,Soper:2014rya}. 
Fine structure analyses often benefit the studies based on parton-showering-sensitive features, e.g., distinguishing merged jets from ordinary QCD jets. 

We, however, emphasize that the deeper pattern of parton showering is less relevant to identifying the CP state of resonance $S$ with merged jets. 
In our study, we instead take a simplified but conservative approach for which we set $ \mathcal{W}\left( {\bf q_i} , \{\bf{p}^{\textrm{reco}}\}\right)$ to be a delta function of momenta of quarks from the decay of $Z$ boson at the point of momenta of the two prong subjets from the mass drop tagger as we are not aware of precise information on detector responses. 
Ignorance of details of parton showering and the detector response significantly simplifies the probability in~\eqref{eq:MEM2} at the cost of maximal sensitivity suggested by the Neyman-Pearson lemma.
Indeed, such details are less relevant as long as reconstructed subjets do depict quarks from the $Z$ boson decay reasonably well.
We can further minimize potential impact from ignorance of higher-order parton showering by selecting a merged jet with its mass around $m_Z$.
To regain sensitivity from above projections, we model a {\it pdf\,} based on the reconstruction-level distributions as we describe below, instead of modeling the transfer function.

We remark that for the case at hand, all kinematic information can be restored with measured four-momenta of visible particles, meaning that the $x_i$'s in parton distribution functions become fixed. 
Hence, the probability evaluated from a matrix element can be simplified as follows: 
\beq
\mathcal{P}\left(\{{\bf p}^{\textrm{reco}}\} | \alpha \right) \simeq \frac{1}{\sigma_\alpha}
\frac{f_{p_1}(x_1) f_{p_2}(x_2)}{2s\, x_1 \, x_2} \, 
\Big|\mathcal{M} \left(\{{\bf p}^{\textrm{reco}}\}  \,; \alpha\right) \Big|^2  \, .
\eeq
We then decompose the matrix element into the production part $p_1 p_2\to S$ and the decay part $S\to j_{1},\, j_{2}\,, \ell^+\,,\ell^-$  through a narrow width approximation (NWA) which is valid as long as the decay width $\Gamma_S$ of resonance $S$ is negligible compared to its mass $M_S$. 
We also note that $S$ is a scalar particle so that any helicity connections with partons in production part are disconnected unlike higher spin cases~\cite{Artoisenet:2013puc}.
Therefore, we have the ratio of probabilities with different CP hypotheses $0^{\textrm{PC}}$ as 
\beq
\frac{\mathcal{P}\left(\{{\bf p}^{\textrm{reco}} \} | 0^{++} \right)}{\mathcal{P}\left(\{{\bf p}^{\textrm{reco}}\} | 0^{-+} \right)} 
\simeq 
\frac{\big|\mathcal{M}_{\left(S \to q\bar q \ell^+\ell^-\right)} \left(\{{\bf p}^{\textrm{reco}}\} \,; 0^{++}\right) \big|^2 }
{\big|\mathcal{M}_{\left(S \to q\bar q \ell^+\ell^-\right)} \left(\{{\bf p}^{\textrm{reco}}\}  \,; 0^{-+}\right) \big|^2 } \, ,
\label{eq:prob}
\eeq
where we dropped the common parts involving a production mode. Here we use the fact that cross sections $\sigma_{0^{\textrm{PC}}}$ are fixed by the observed value. 
There is a subtlety in calculating a matrix element $\mathcal{M}$ as the current jet algorithms cannot specify the charge or flavor for light quarks. 
In order to deal with this issue, we symmetrize a matching between subjet $(j_{1},\, j_{2})$ and $(q,\, \bar q)$ as 
\bea
|\mathcal{M}_{\left(S \to q\bar q \ell^+\ell^-\right)} \left(\{{\bf p}^{\textrm{reco}}\}  \,; 0^{\textrm{PC}}\right) \big|^2_{\textrm{sym}} 
\equiv
&& \sum_{q\in \{u,d\}} \Bigg(\big|\mathcal{M}_{\left(S \to q\bar q \ell^+\ell^-\right)} \left(\{ j_{1}, j_{2},\ell^+,\ell^-\}  \,; 0^{\textrm{PC}}\right) \big|^2 \nonumber \\
+&&\big|\mathcal{M}_{\left(S \to q\bar q \ell^+\ell^-\right)} \left(\{ j_{2}, j_{1},\ell^+,\ell^-\}  \,; 0^{\textrm{PC}}\right) \big|^2\Bigg)\,,
\eea
which alters the above-given probability ratio to the symmetrized ratio called a kinematic discriminant (KD) 
\beq
\textrm{KD} \equiv \frac{\mathcal{P}\left(\{{\bf p}^{\textrm{reco}} \}| 0^{++} \right)}{\mathcal{P}\left(\{{\bf p}^{\textrm{reco}} \} | 0^{-+} \right)} 
\simeq 
\frac{\big|\mathcal{M}_{\left(S \to q\bar q \ell^+\ell^-\right)} \left(\{{\bf p}^{\textrm{reco}} \} \,; 0^{++}\right) \big|^2_{\textrm{sym}} }
{\big|\mathcal{M}_{\left(S \to q\bar q \ell^+\ell^-\right)} \left(\{{\bf p}^{\textrm{reco}} \}  \,; 0^{-+}\right) \big|^2_{\textrm{sym}} } \, .
\label{eq:prob_sym}
\eeq
Predicated upon the KD, we construct two {\it pdf\,}'s $P_{0^\textrm{PC}}(\mathrm{KD}) = P(\ln\textrm{KD} | 0^\textrm{PC})$ with a large number of reconstructed events for each hypothesis that are prepared with Monte Carlo simulation at the detector level, ensuring the consideration of various experimental effects. 
Assuming that each pseudo-experiment is independent and identically distributed, we set two likelihoods $\mathcal{L}(0^\textrm{PC})$ with the fixed number of events $N_\textrm{evt}$
\beq
\mathcal{L}(0^\textrm{PC}) \equiv \prod_{i=1}^{N_\textrm{evt}}P_{0^\textrm{PC}}(\textrm{KD}_i).
\eeq
Corresponding test statistic $q_\mathcal{M}$ is defined as the log-likelihood ratio,
\beq
q_{\mathcal{M}} \equiv \ln \frac{\mathcal{L}(0^{++})}{\mathcal{L}(0^{-+})} = \sum_{i=1}^{N_\textrm{evt}} \ln 
\left(\frac{P_{0^{++}}(\textrm{KD}_i)}{P_{0^{-+}}(\textrm{KD}_i)} \right)\, .
\label{eq:LL_MEM}
\eeq
A {\it pdf} of  $P(q_{\mathcal{M}} | 0^\textrm{PC})$ for a test statistic $q_{\mathcal{M}}$ with a given hypothesis $0^\textrm{PC}$ and a given number of events $N_\textrm{evt}$ is calculated from a huge number of pseudo-experiments. Cumulative probabilities based on $q_{\mathcal{M}}$  are given by 
\bea
 p_{0^{++}} (q_{\mathcal{M}}^{\mathrm{obs}};q_{\mathcal{M}}) & = & P(q_{\mathcal{M}} \le q_{\mathcal{M}}^{\mathrm{obs}} | 0^{++} )\, ,
 \label{eq:LLMpp} \\
 p_{0^{-+}} (q_{\mathcal{M}}^{\mathrm{obs}};q_{\mathcal{M}}) & = & P(q_{\mathcal{M}} \ge q_{\mathcal{M}}^{\mathrm{obs}} | 0^{-+} ) \, .
 \label{eq:LLMnp}
\eea

\subsection{Results}

\begin{table}
\begin{center}
\begin{tabular}{|c|c|r|r|}
\hline
Cut flow & selection &$750\GeV$ &$1500\GeV$ \\

\hline
parton level              &                                             & 100.0 \%    & 100.0 \% \\
\hline
object tagging            & one merged jet, two $\ell$                  &  61.0 \%    &  63.4 \% \\ 
lepton $P_T$              & $P_T > 25\GeV$                              &  52.0 \%    &  58.8 \% \\
$m_{(\ell^+,\,\ell^-)}$   & $[83,99]$ GeV                               &  47.4 \%    &  53.5 \% \\
$m_{\MJ}$                 & $[75,105]$ GeV                              &  20.6 \%    &  25.5 \% \\
$y_{ZZ} $                 & $|y_{ZZ}| < 0.15$                           &  16.3 \%    &  21.3 \% \\
$P_{T(\MJ)} $       & $P_{T(\MJ)}> 0.4 \, m_{(\MJ,\,\ell^+,\,\ell^-)} $ &  11.5 \%    &  14.7 \% \\
\hline                                                                      
\multirow{2}{*}{$m_{(\MJ,\,\ell^+,\,\ell^-)}$}
  & within $M_S\pm 50 \GeV $     & 10.4 \%  &  - \\
  & within $M_S\pm 100 \GeV $     &-   & 13.4 \% \\
\hline
\end{tabular}
\end{center}
\caption{\label{tab:cut_flow} 
Event selection criteria  and corresponding efficiencies for each benchmark point. 
As rapidities of final particles do not depend on the CP states of a spin 0 particle, we expect that the efficiencies in different CP states of $S$ are the same.   
}
\end{table}

We finally present our main results on distinguishing CP-even and CP-odd states in this section, comparing three methods, two with angular variables $A_{\phi}$ and $\phi$ , and the other with an MEM-based variable. 
To maximize relevant performances, we first construct {\it pdf}\,s for test statistics in both methods based on the log-likelihood ratio. 
In our analyses we do not consider backgrounds since (1) we compare the performance of each method in the best case, and (2) background subtraction can be performed with $_{\textsc{s}}$\textsc{Plot}~\cite{Pivk:2004ty}. 
We rather focus on studying effects from cuts to reduce backgrounds. 
Detailed information on potential backgrounds and recipes to take them into consideration in KD-based analyses shall be provided in Appendix~\ref{sec:BKG}, and we simply continue our discussion here, having the dominant reducible backgrounds in our mind. 

\begin{figure}[t!]
\begin{center}
\begin{tabular}{ccc}
\includegraphics[width=4.7cm]{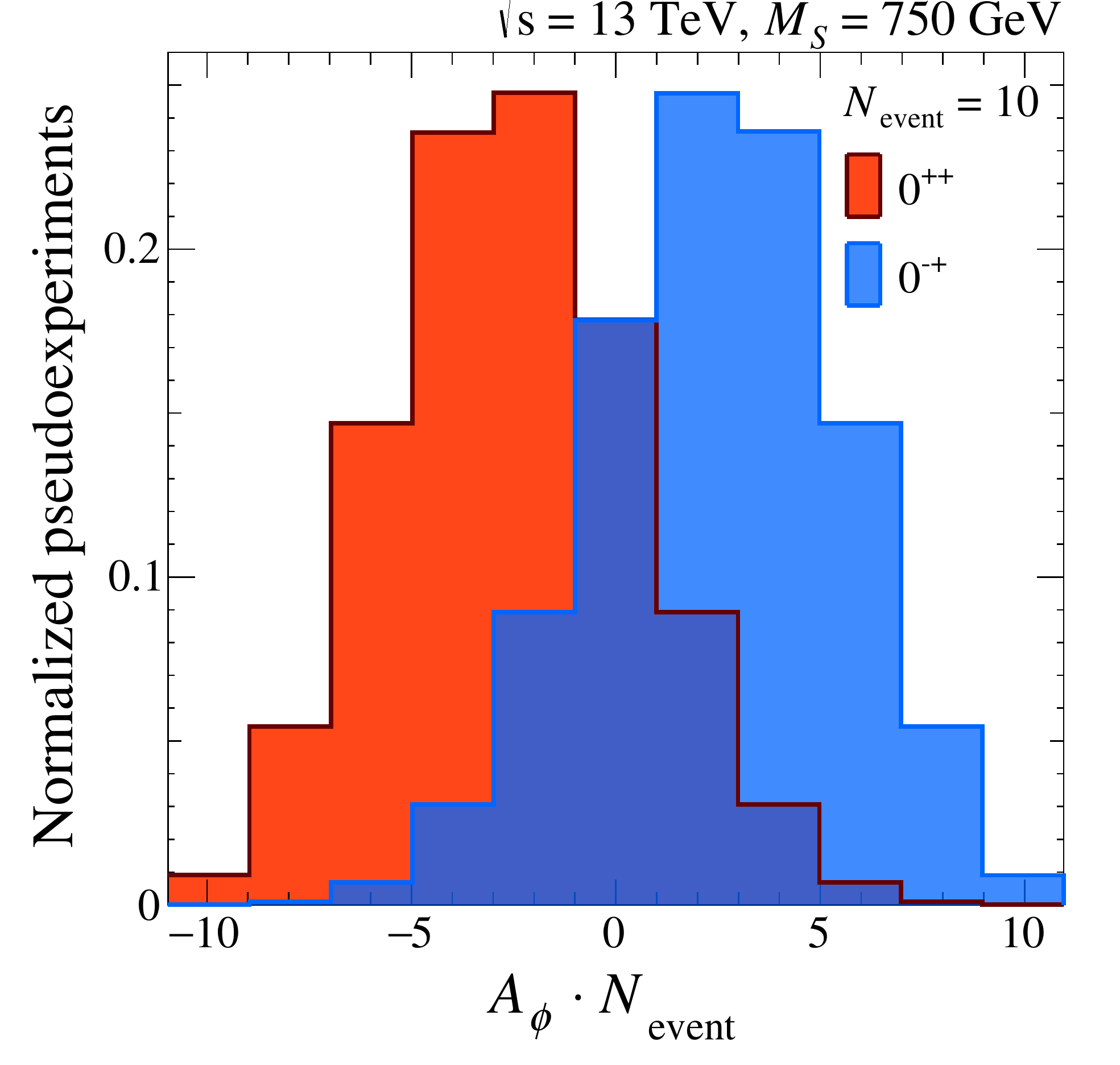} &
\includegraphics[width=4.7cm]{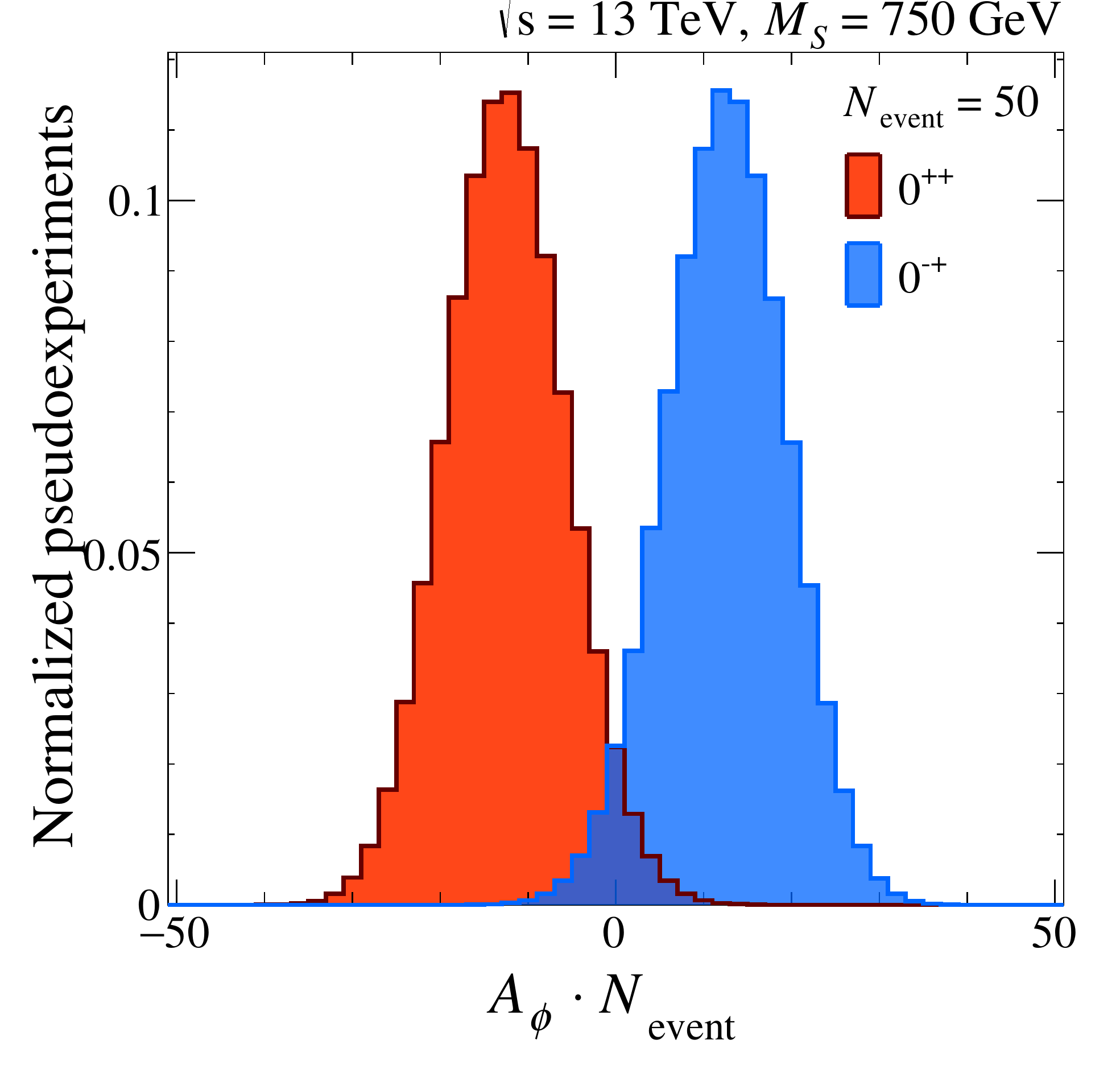} &
\includegraphics[width=4.7cm]{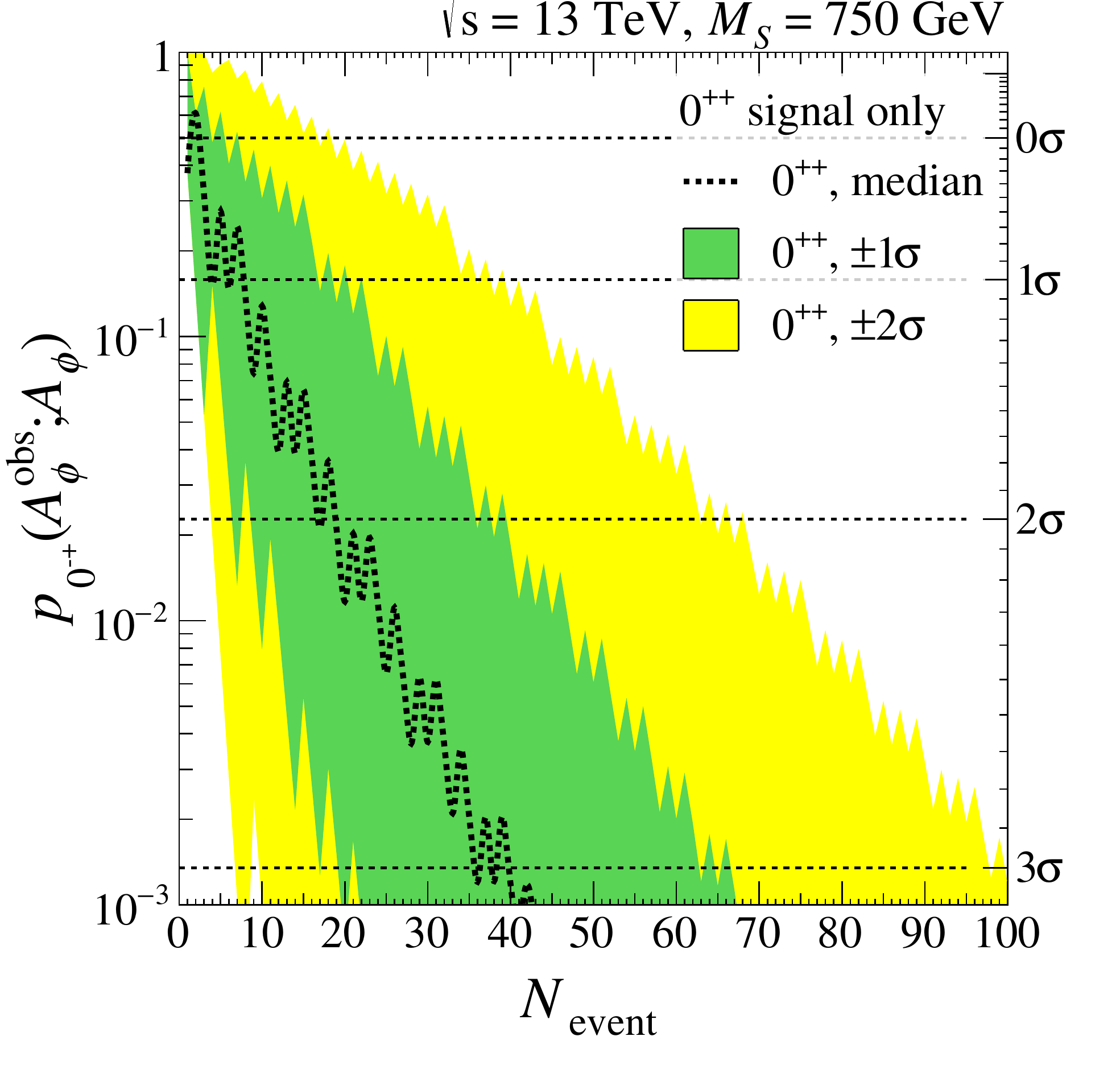} \\
\includegraphics[width=4.7cm]{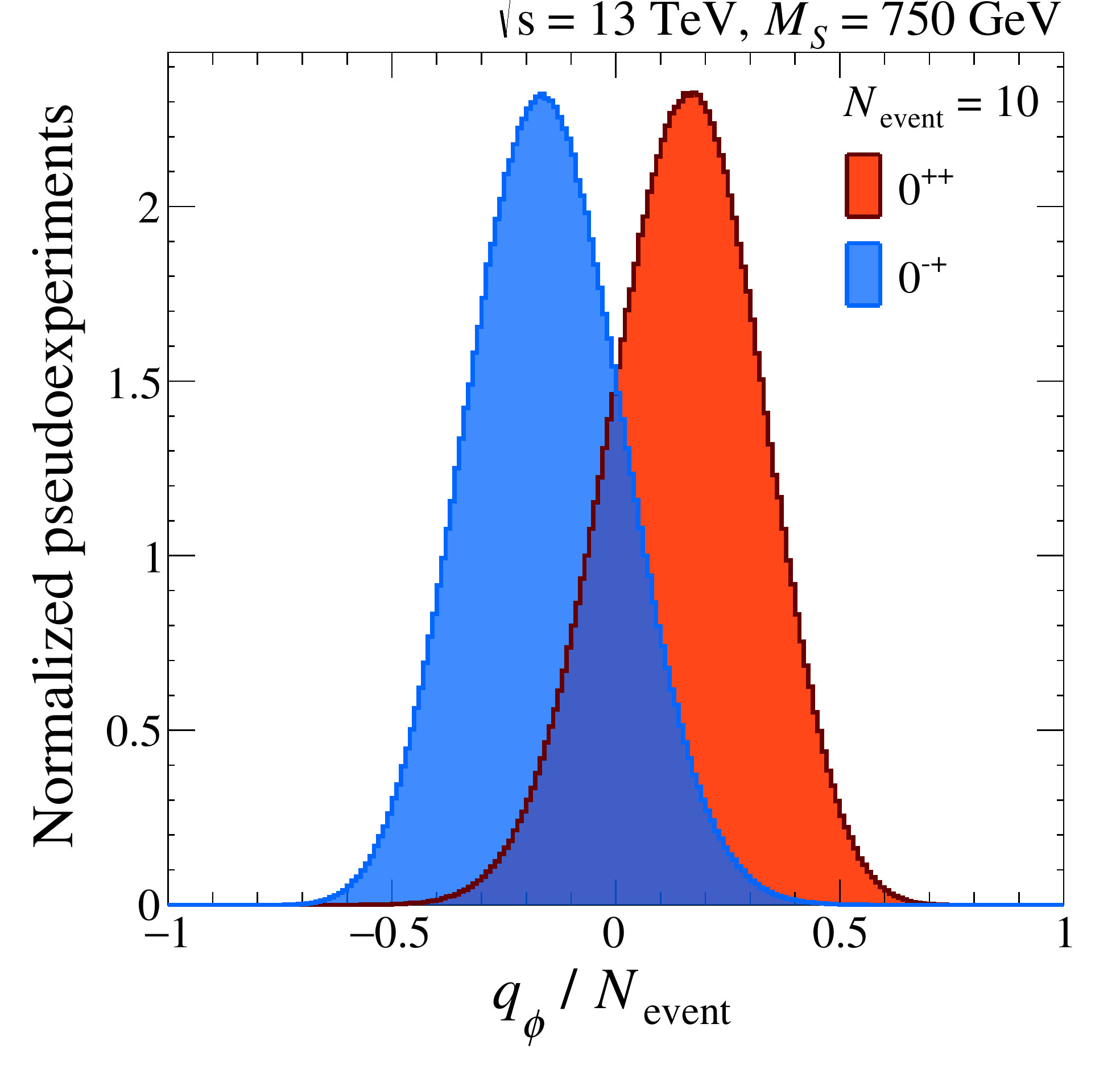} &
\includegraphics[width=4.7cm]{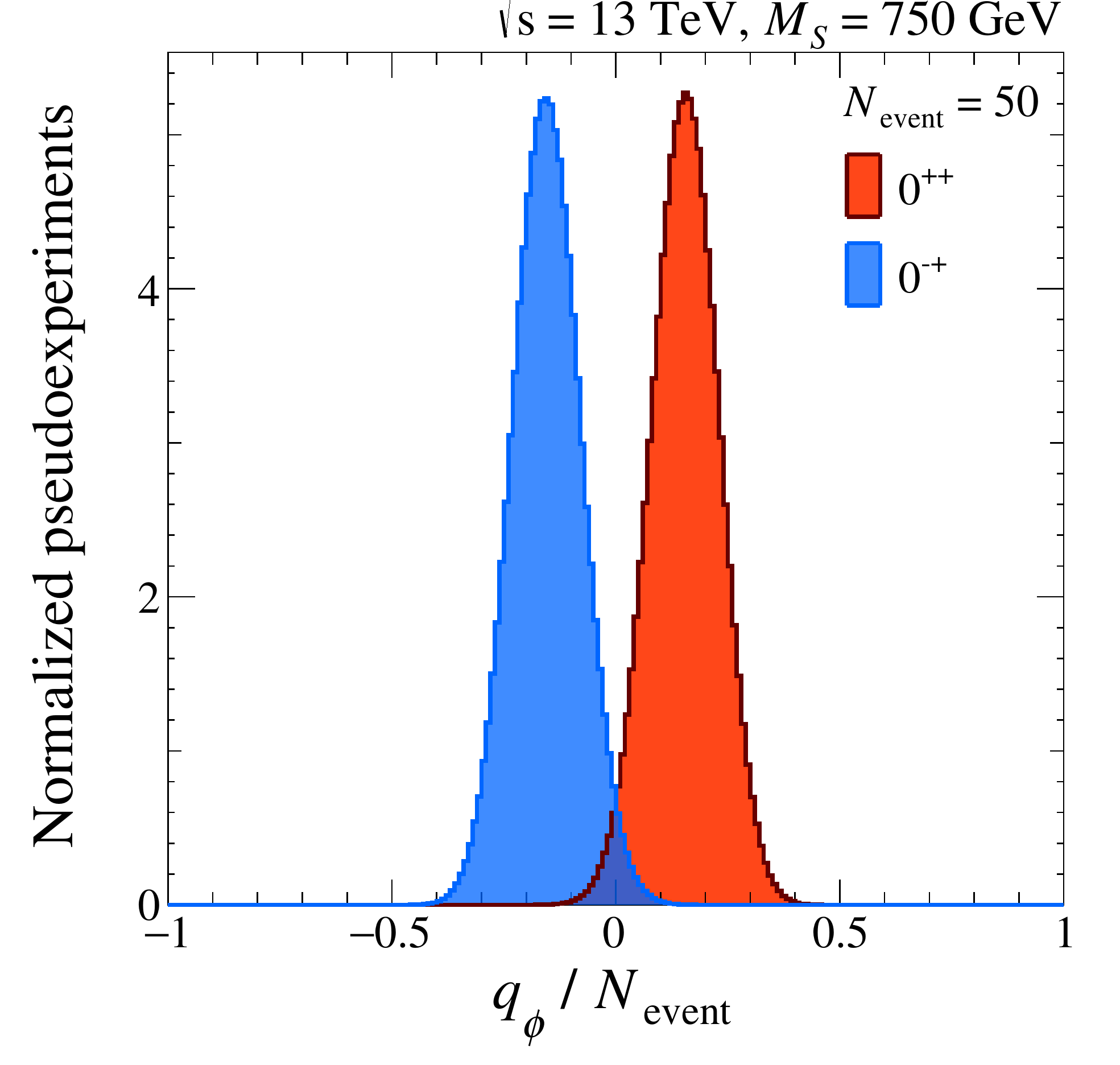} &
\includegraphics[width=4.7cm]{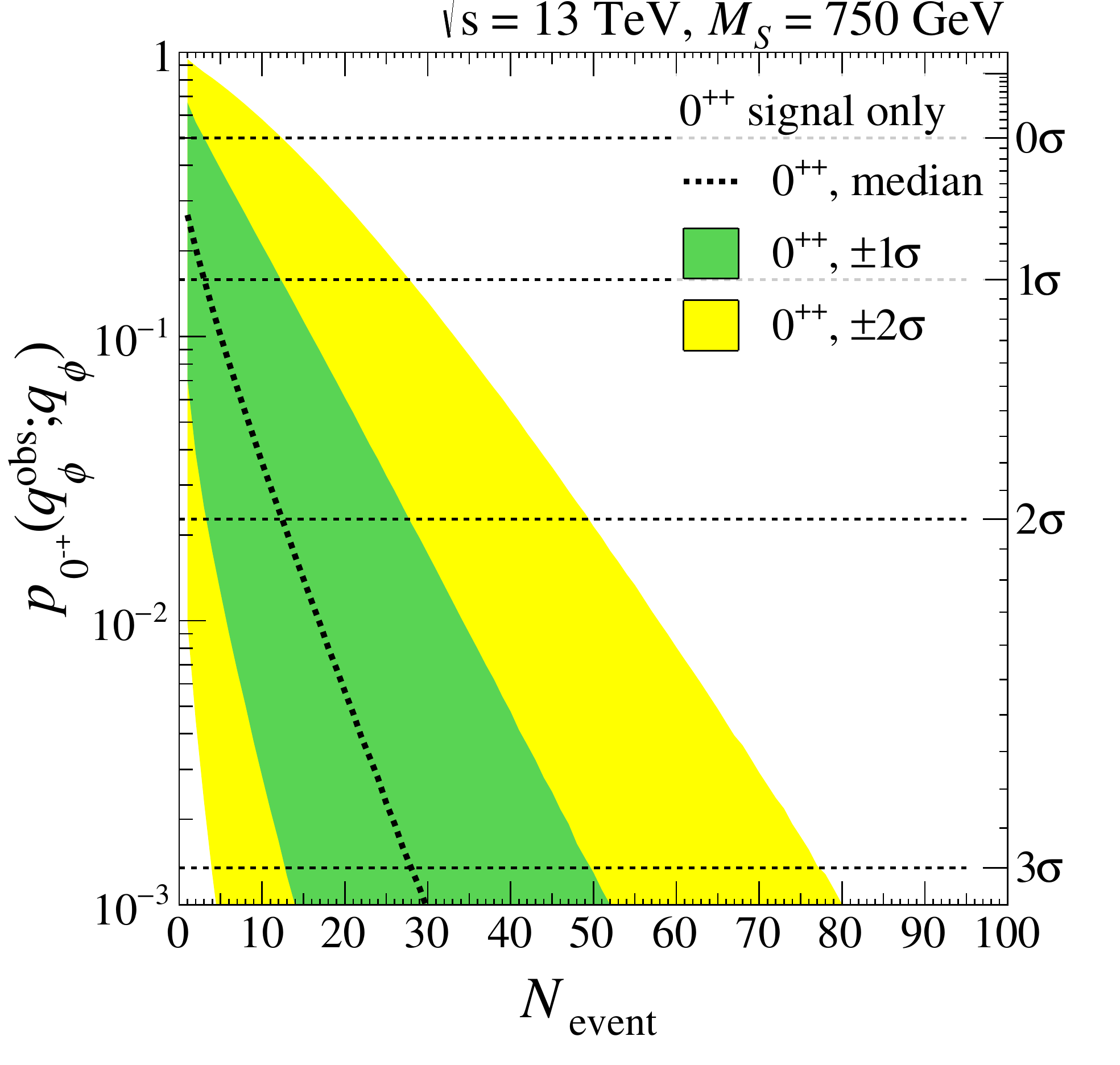} \\
\includegraphics[width=4.7cm]{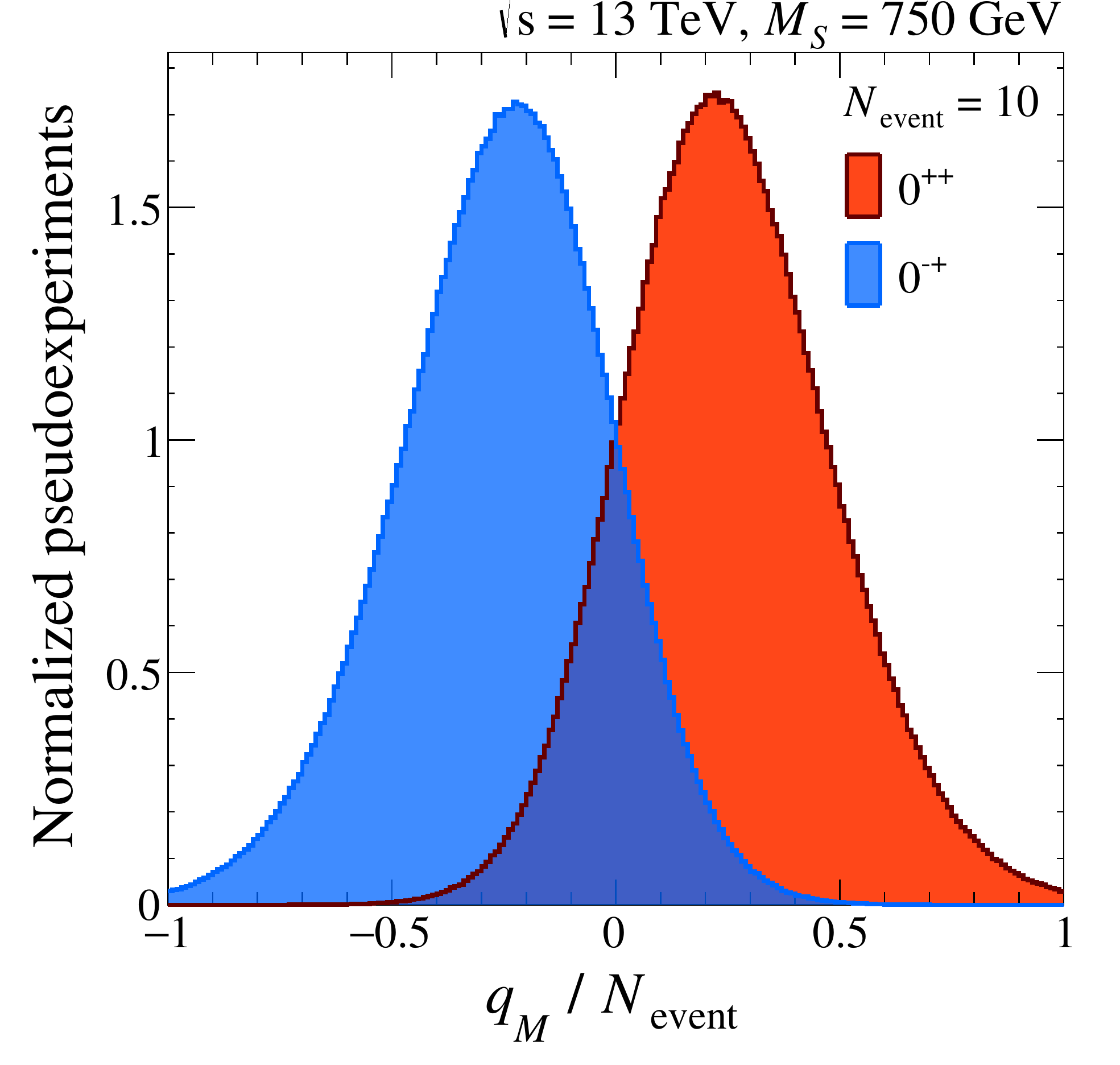} &
\includegraphics[width=4.7cm]{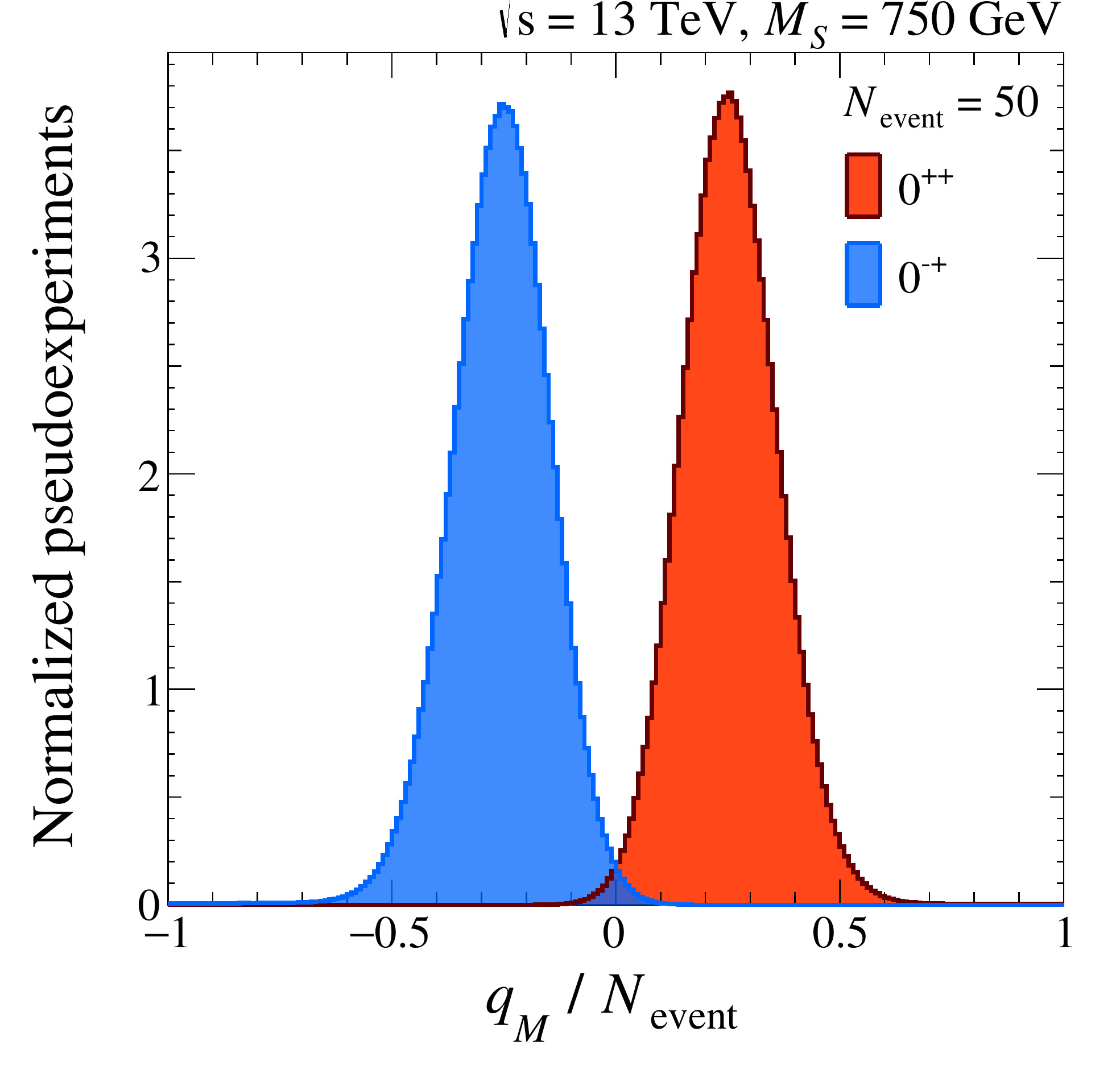} &
\includegraphics[width=4.7cm]{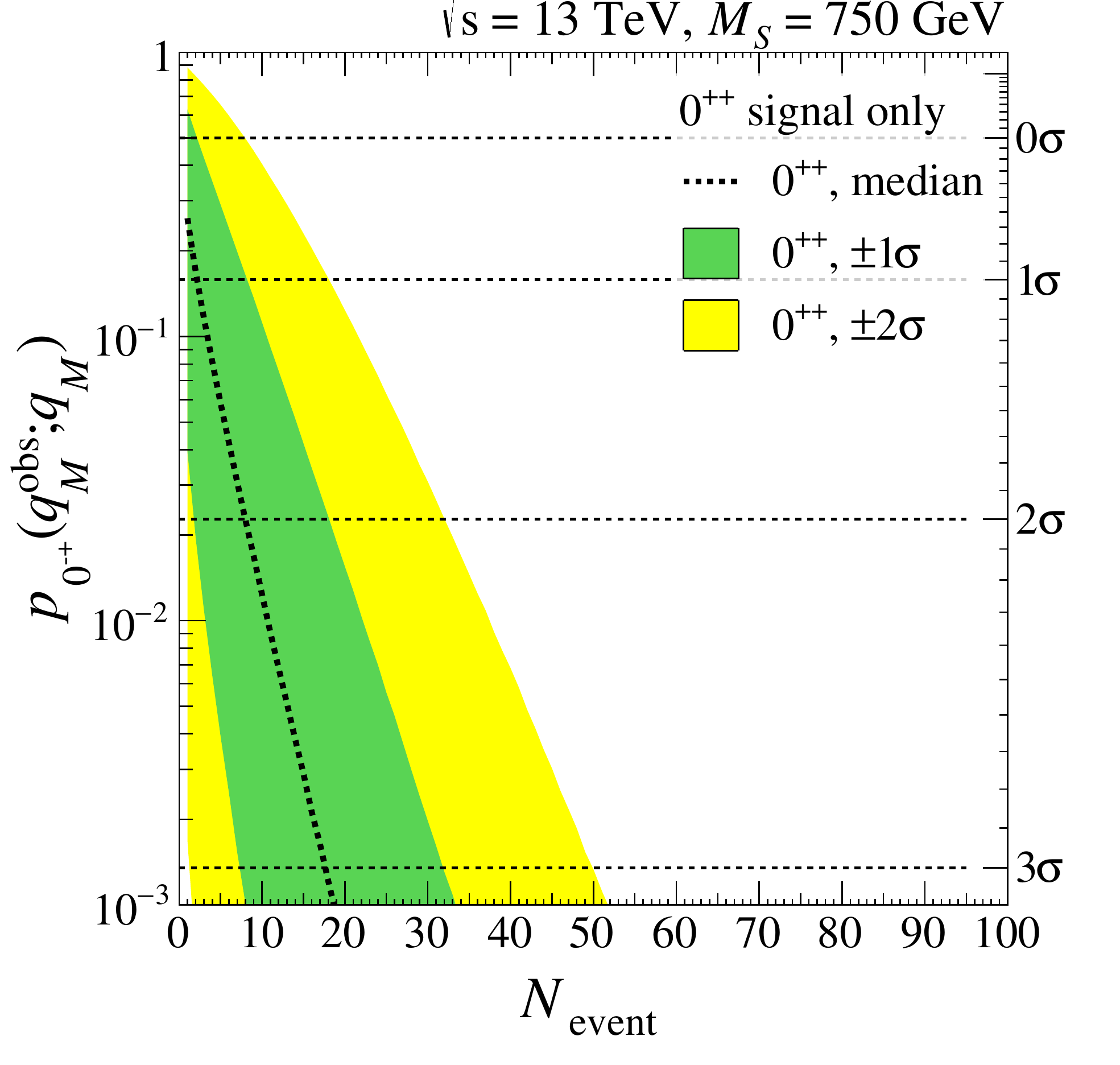}
\end{tabular}
\end{center}
\caption{\label{fig:results_750} 
Performance comparisons among various methods for the BP of $M_S=750\GeV$ with $R_\MJ=1.2$: the method based on $A_\phi$ variable of eq.~\eqref{eq:asymmetric} in the first row, a log-likelihood ratio test based on the $\phi$ angular distributions from eq.~\eqref{eq:LLphi} in the second row, and a log-likelihood ratio test based on the MEM of eq.~\eqref{eq:LL_MEM} in the third row. 
}
\end{figure}

\begin{figure}[ht!]
\begin{center}
\begin{tabular}{ccc}
\includegraphics[width=4.7cm]{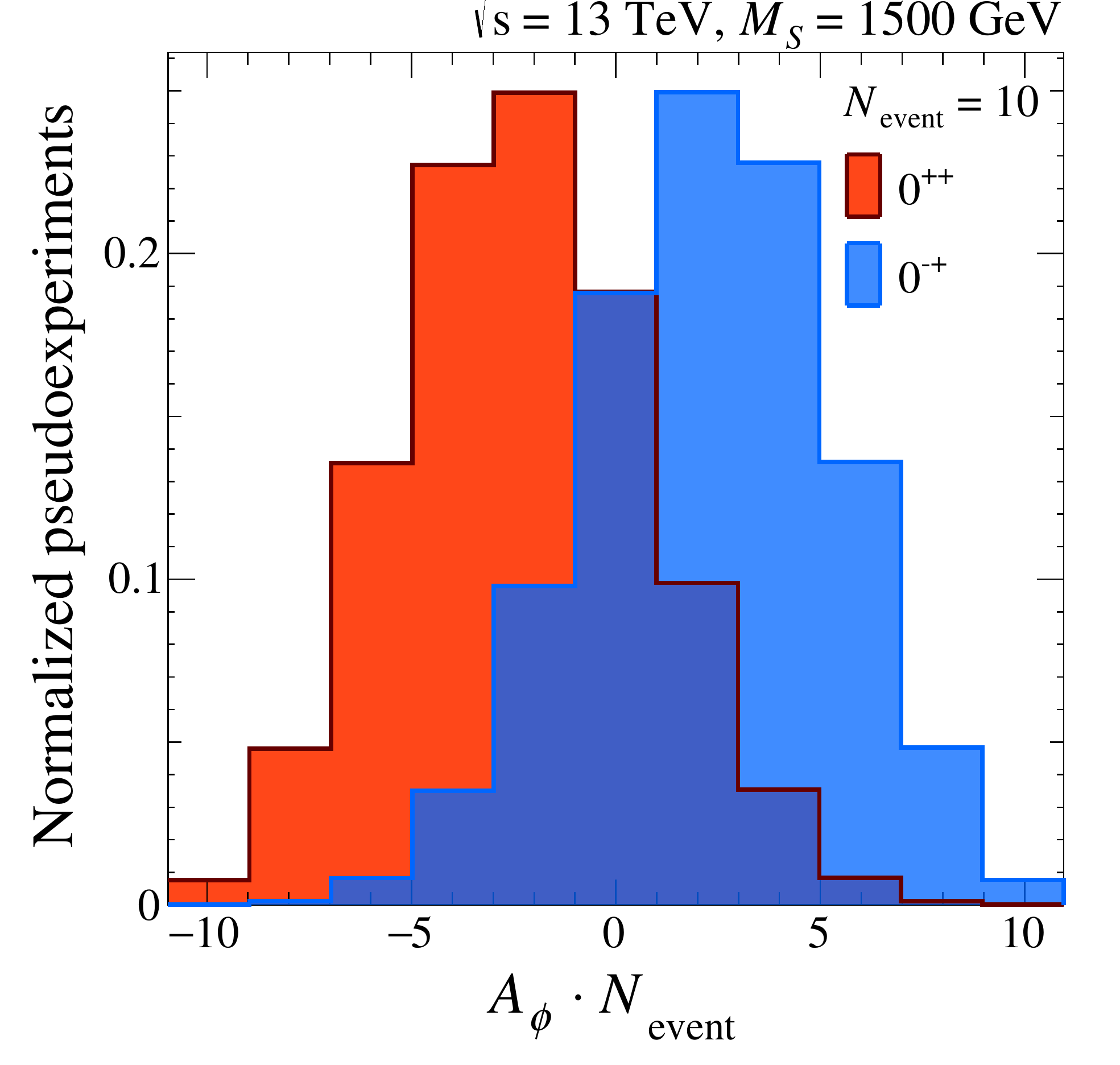} &
\includegraphics[width=4.7cm]{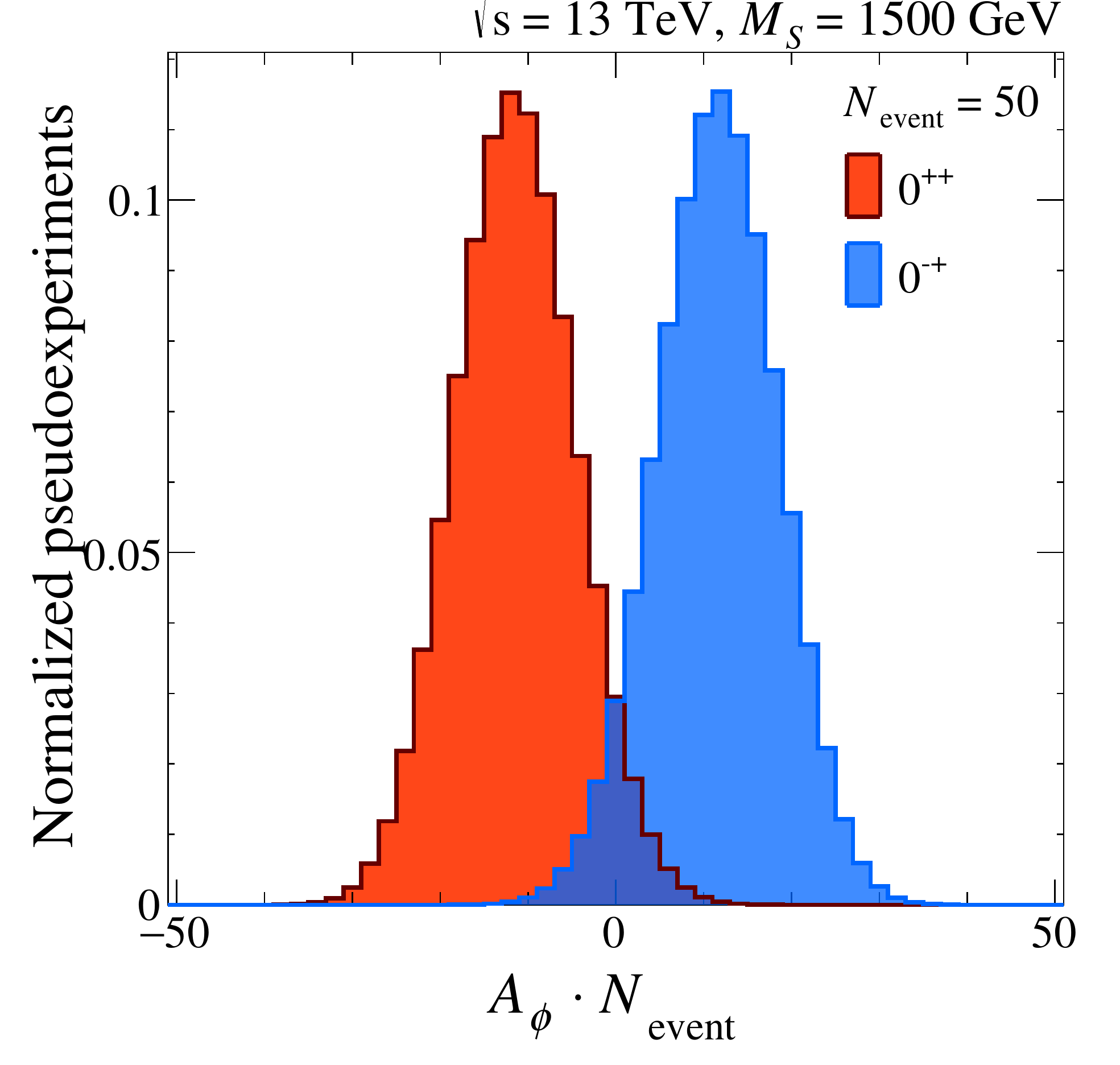} &
\includegraphics[width=4.7cm]{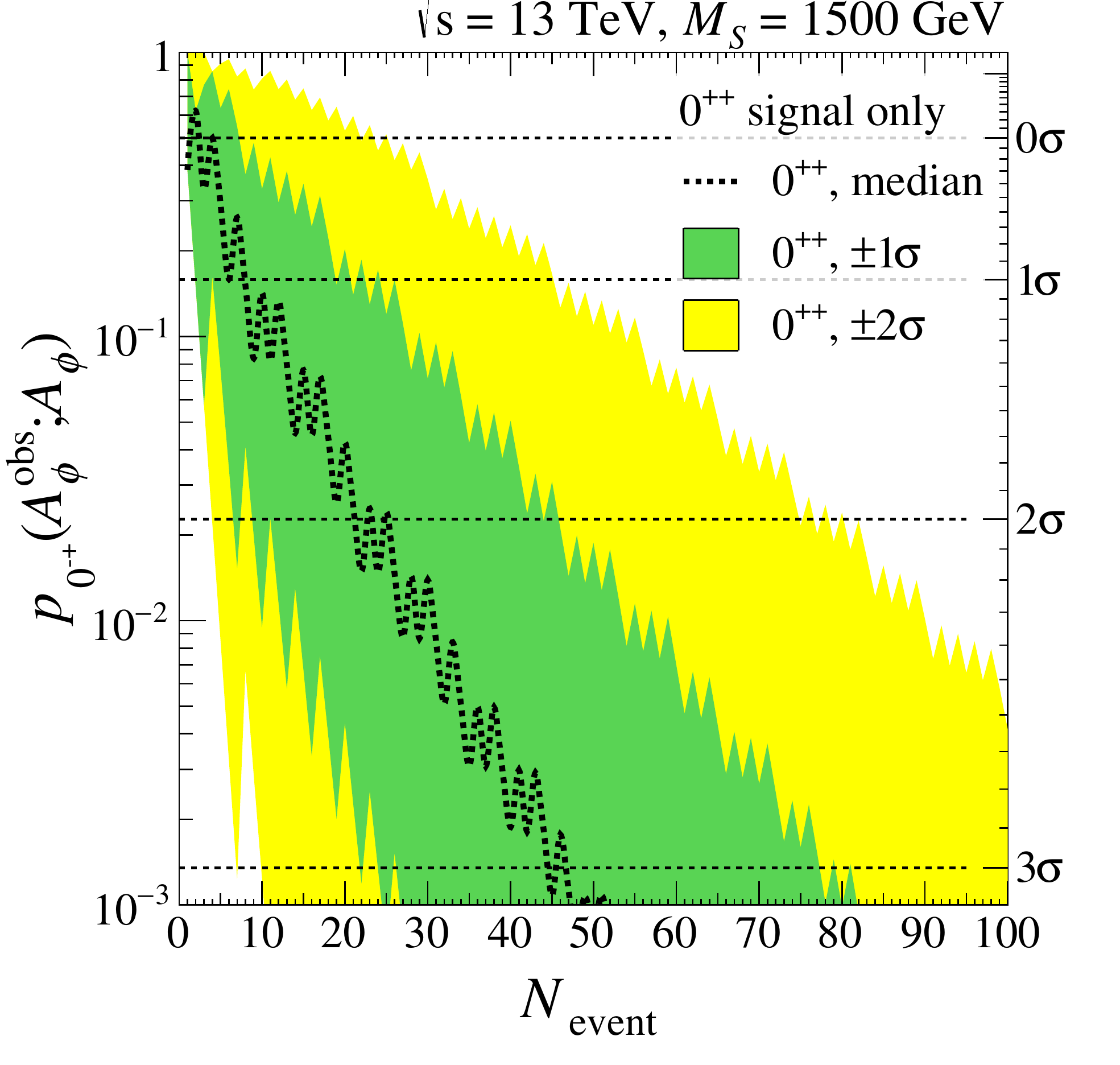} \\
\includegraphics[width=4.7cm]{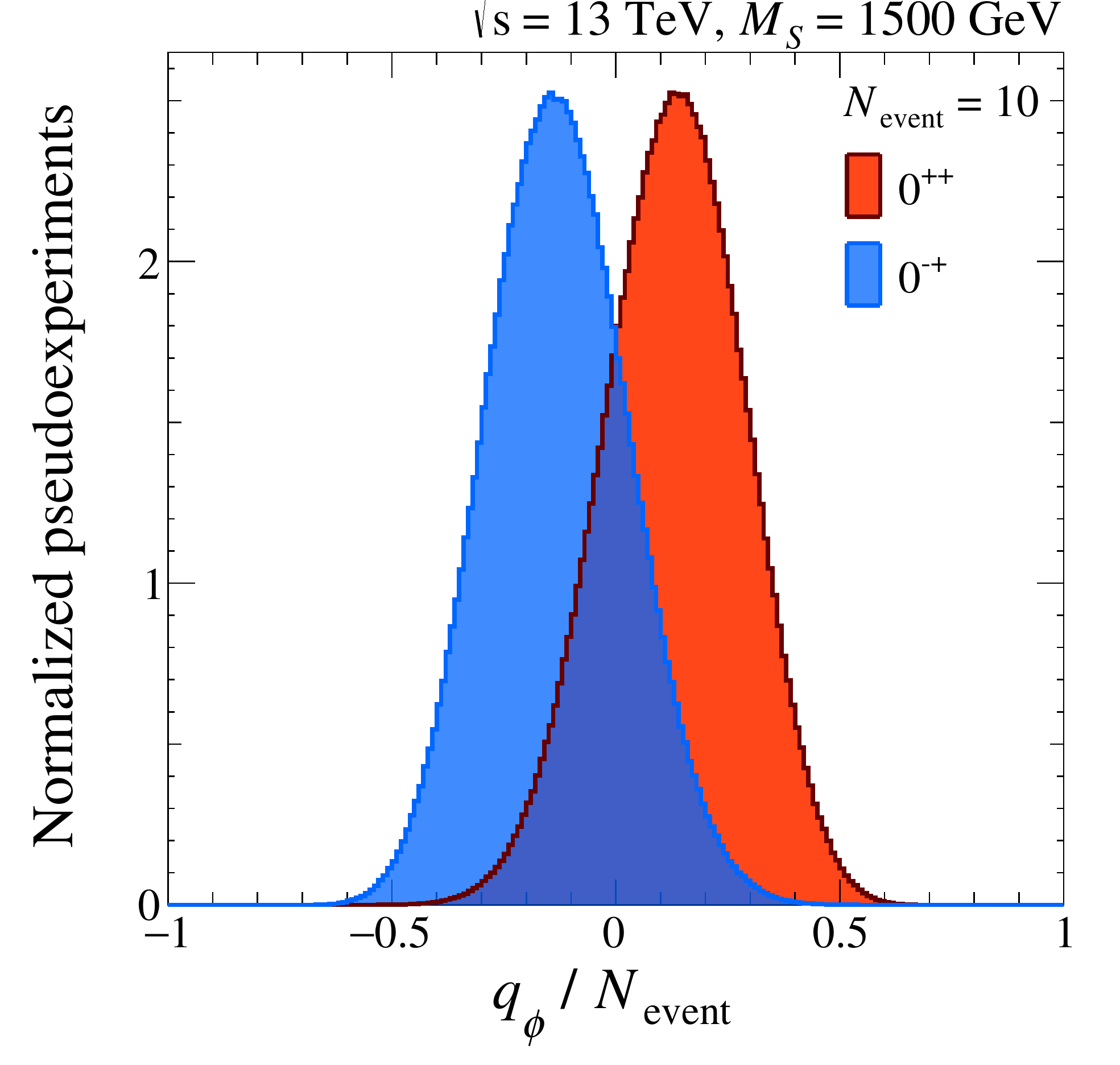} &
\includegraphics[width=4.7cm]{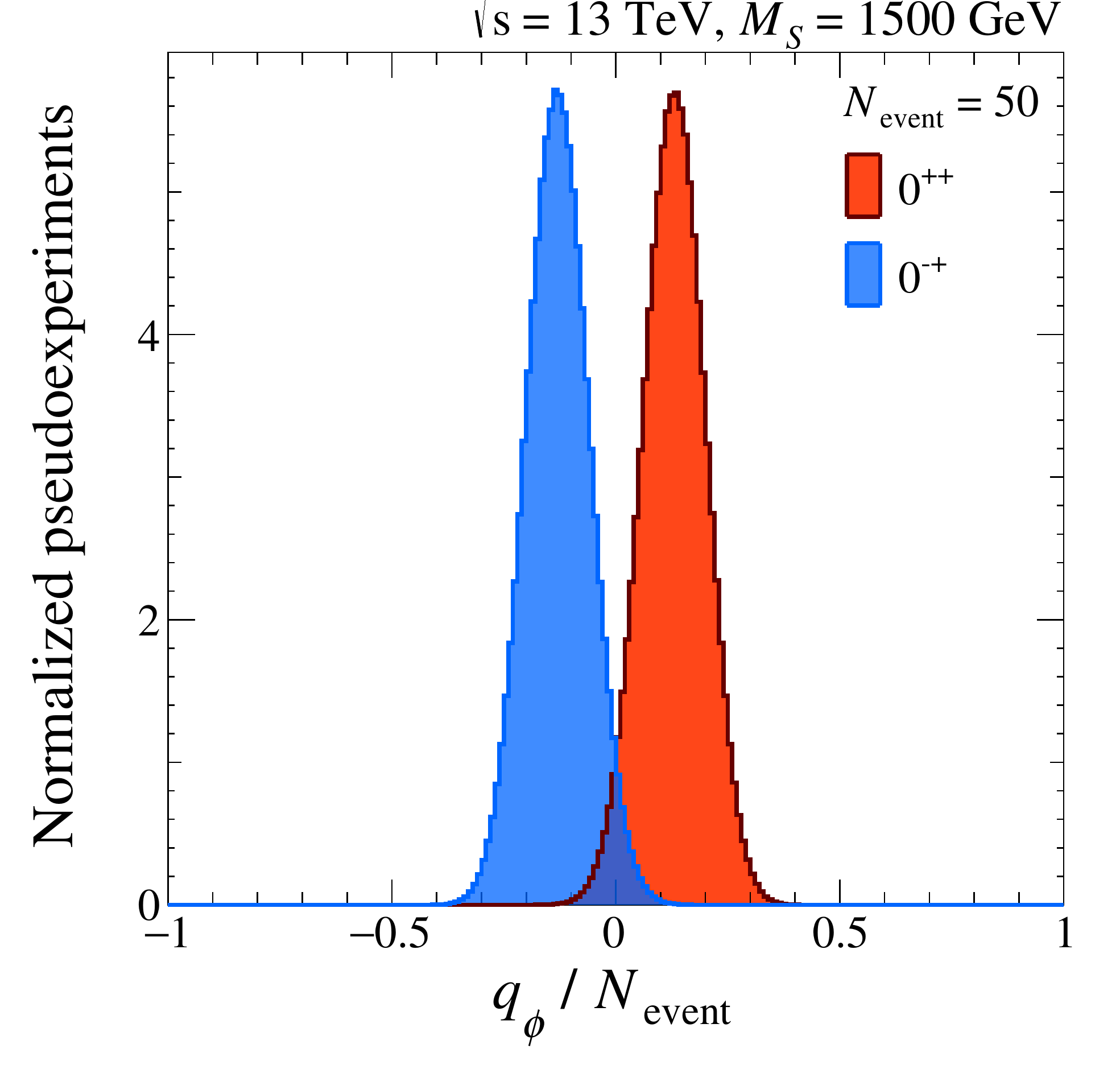} &
\includegraphics[width=4.7cm]{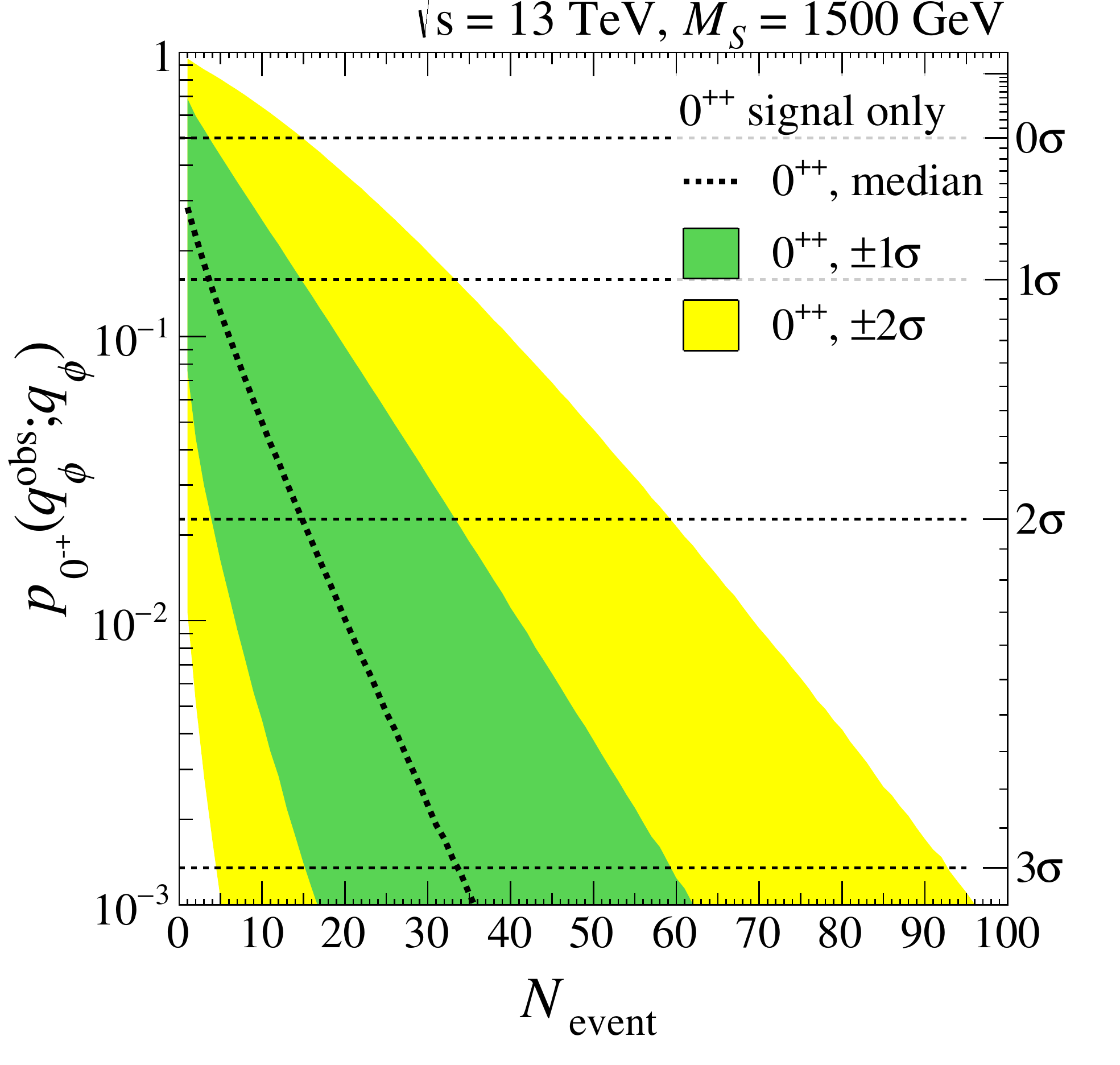} \\
\includegraphics[width=4.7cm]{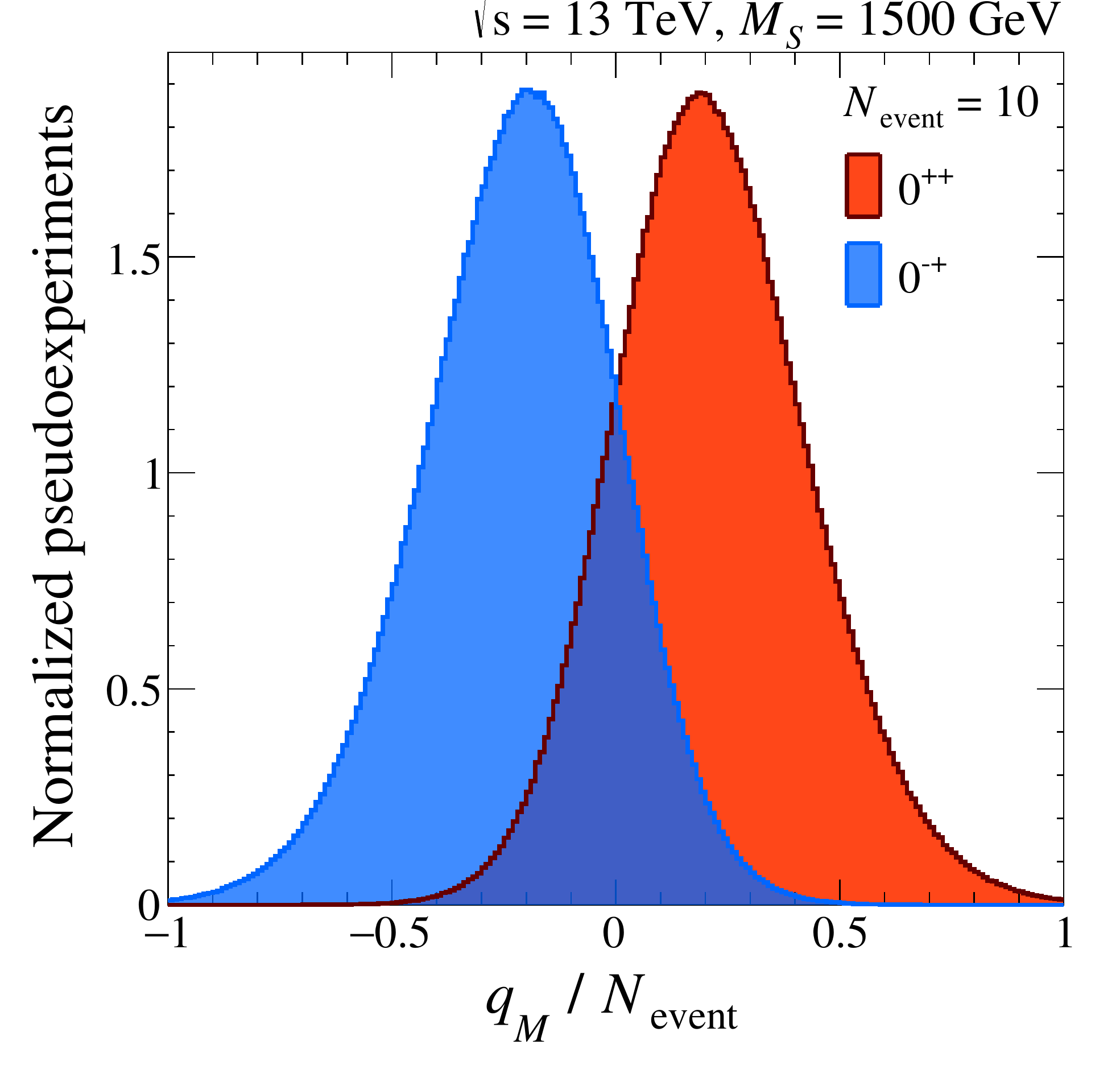} &
\includegraphics[width=4.7cm]{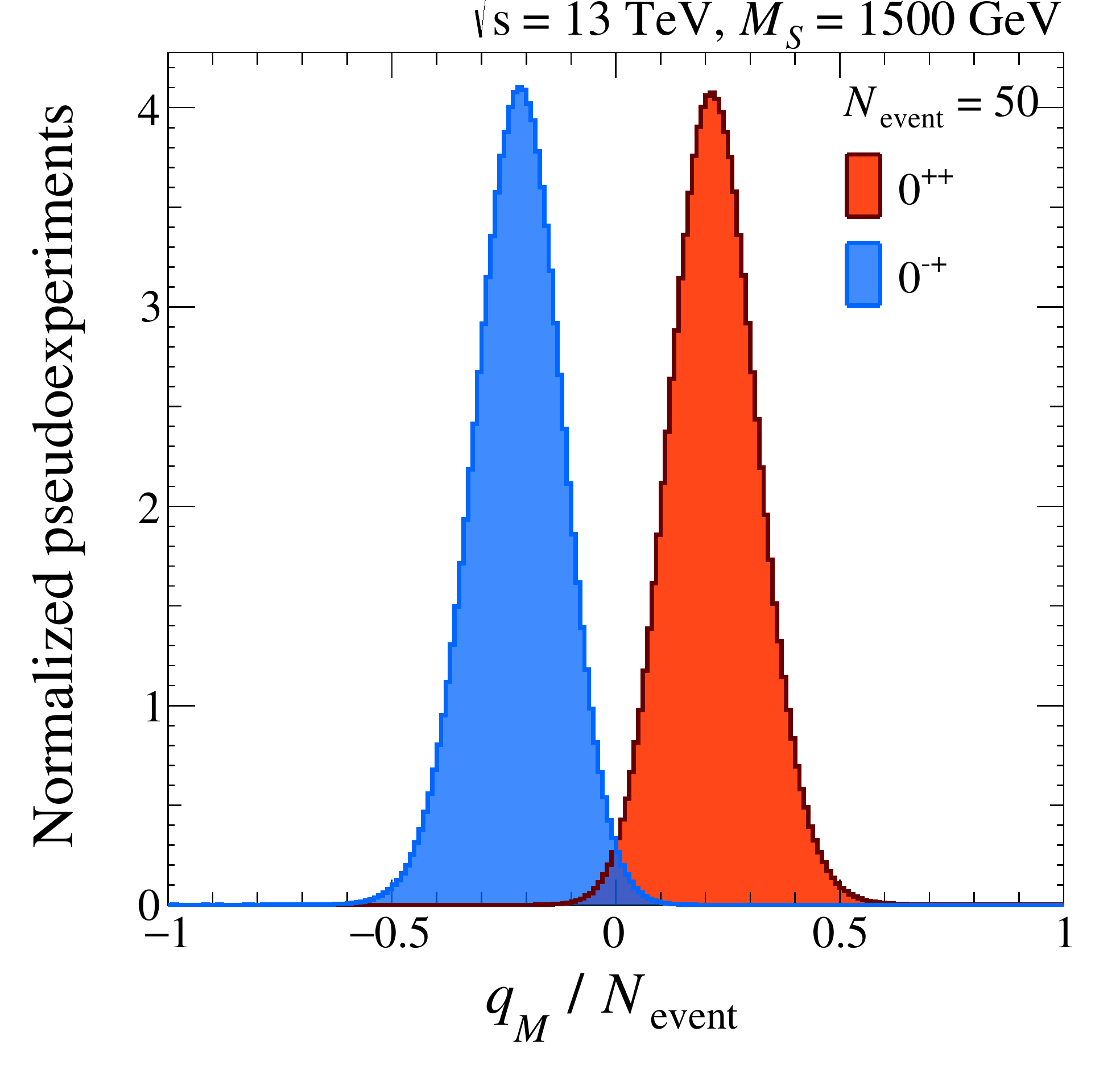} &
\includegraphics[width=4.7cm]{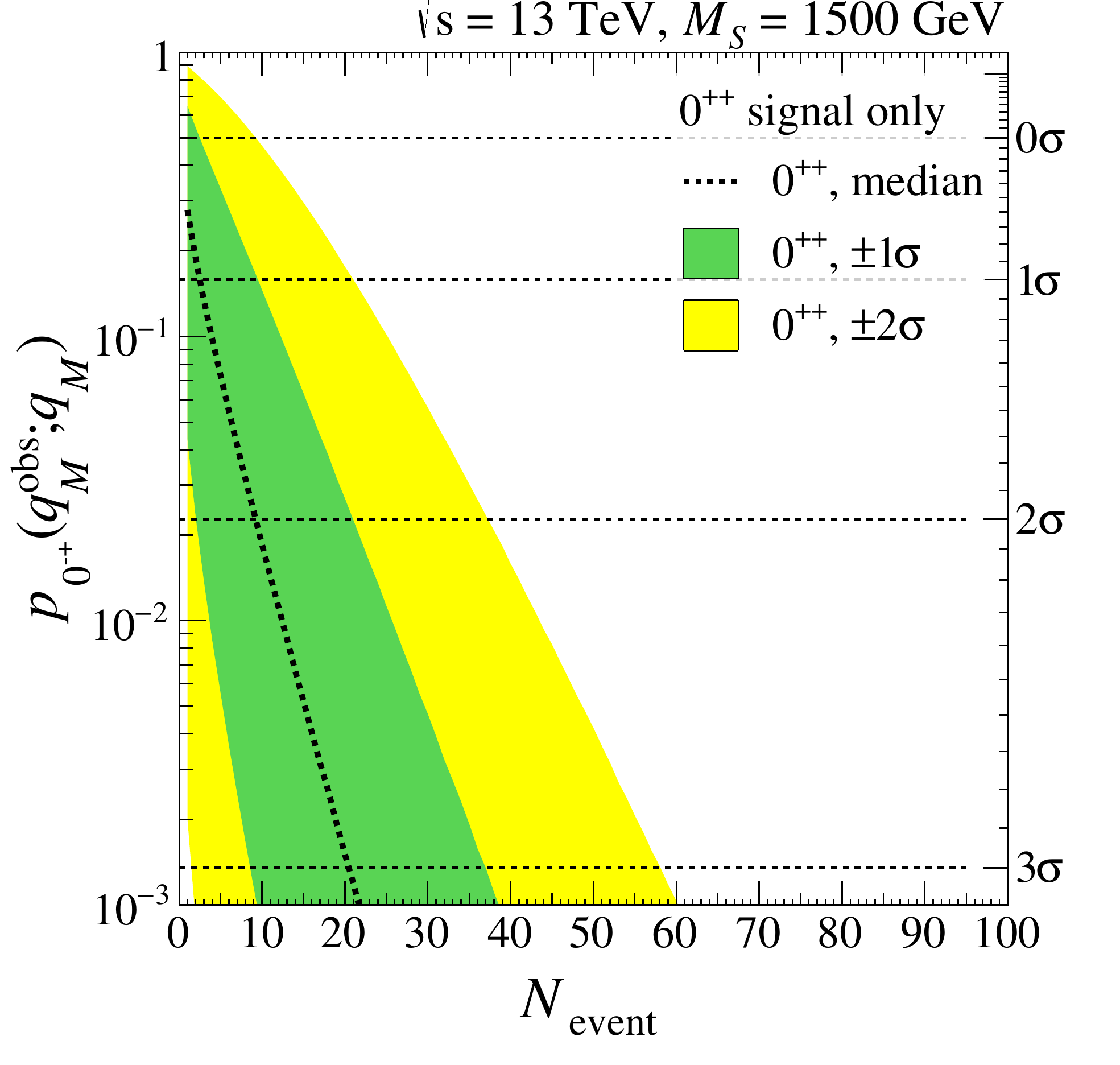}
\end{tabular}
\end{center}
\caption{\label{fig:results_1500}  Similar results to Figure~\ref{fig:results_750} for the BP of $M_S=1500\GeV$ with $R_\MJ=0.6$.
}
\end{figure}

As discussed earlier, the main SM backgrounds to two leptons plus a single MJ are $Z(\to \ell^+\ell^-)+j$s where a QCD jet can mimic an MJ by dressing up a mass due to QCD contaminations~\cite{Aad:2014xka, Aad:2015owa, Aaboud:2016okv, Khachatryan:2015cwa, CMS-PAS-B2G-16-010}.
Obviously, the resonance mass window cut is useful to suppress backgrounds, i.e., the invariant mass formed by an MJ and a lepton pair should fall into the range around the mass of $S$. 
We set a different mass range in each benchmark point to consider effects of smearing.
To reduce backgrounds further, it is noteworthy that for a signal event, a merged $Z$-jet and a di-leptonic $Z$ are typically symmetric since they originate from a single resonance, whereas for a background event, the corresponding objects are {\it a}symmetric because a quark-initiated jet and $Z$ are expected to have a sizable mass gap between them. 
This observation motivates us to introduce a $P_T$-asymmetric variable $y_{ZZ}$~\cite{Aad:2015owa} that is expected to be a reasonable choice to reduce $Z+j$s  backgrounds:   
\beq
y_{ZZ} = \frac{ P_{T(\MJ)} - P_{T(\ell^+\ell^-)} }{ P_{T(\MJ)} + P_{T(\ell^+\ell^-)}}\, .
\eeq
Cut-efficiency flows for our benchmark points are summarized in Table~\ref{tab:cut_flow}.
Note that the efficiencies here are the same for both CP-even and CP-odd states. 
Basically, the reconstruction efficiency through detector geometry (i.e., rapidity coverage) and $P_T$ selection depends on the rapidity of visible objects. 
However, their rapidity depends on $\theta^*$ and $\theta_i$, the former of which has nothing to do with the CP state (as discussed in Section~\ref{sec:angular}) and the latter of which is sensitive only to the $Z$ decay structure. 
Imposing those cuts on Monte Carlo event samples,
we conduct posterior analyses to determine the CP state of $S$ using the {\it pdf\,}s from three methods listed below.
\begin{itemize}
\item[1.] $A_\phi$ variable in eq.\,\eqref{eq:asymmetric}
\item[2.] Log-likelihood ratio $q_\phi$ on $\phi$ distributions in eq.\,\eqref{eq:LLphi}
\item[3.] Log-likelihood ratio $q_\mathcal{M}$ based on the MEM in eq.\,\eqref{eq:LL_MEM} 
\end{itemize}
We present our results for $M_S=750\GeV$ in Figure~\ref{fig:results_750} and $M_S=1500\GeV$ in Figure~\ref{fig:results_1500}: $A_{\phi}$ in the first row, $q_{\phi}$ in the second row, and $q_{\mathcal{M}}$ in the third row.
To ensure enough statistics, we prepare 5 million pseudo-experiments for both BPs at the center-of-mass energy of 13 TeV.
The first two columns show their distributions under the $0^{++}$ hypothesis (red histogram) and the $0^{-+}$ hypothesis (blue histogram) with different numbers of events $N_{\textrm{event}}=10$ (first column) and $N_{\textrm{event}}=50$ (second column). 

In discriminating different hypotheses $0^{++}$ and $0^{-+}$ with a test static $\chi$, 
we calculate $p_{0^{-+}}(\chi^{\textrm{obs}} ; \chi)$ to reject the $0^{-+}$ hypothesis in favor of $0^{++}$ (Type I error $\alpha$)~\cite{DeRujula:2010ys}, exhibiting ``Brazilian'' plots in the third column, i.e., 1$\sigma$ (green) and 2$\sigma$ (yellow) bands around the peak in $p_{0^{++}}(\chi^{\textrm{obs}} ; \chi)$ according to the number of required events to separate the two hypotheses. 
We clearly observe that the most CP-sensitive method is to utilize a test static with the MEM-based log-likelihood ratio. 
In our study, we find that the separating power of the MEM-based method {\it remarkably} does not change much with resonance mass $M_S$, providing its robustness over a wide range of mass space.
In BP1 of $M_S=750\GeV$ representing the moderate boost region, the required number of events to have Type I error in the level of $3\sigma$ is $N_{\textrm{event}}^{(3\sigma)}=18^{+14}_{-10}$ within $1\sigma$ deviation for the $0^{++}$ hypothesis. 
For BP2 of $M_S=1.5$ TeV representing highly boosted region, the corresponding required number of events is $N_{\textrm{event}}^{(3\sigma)}=21^{+17}_{-12}$\,.
This can be understood in terms of the distortion in the differential distribution in $\phi$ as it is the most crucial angular variable encapsulated in the MEM to determine the CP state.  
If we neglect restriction on the phase space of leptons as lepton isolation $R_\textrm{iso}$ is larger than the minimum angular distance $\Delta R_{f\bar f}$ in eq.~\eqref{eq:parton_dis}, the most relevant one is from the phase-space reduction in the jet clustering procedure as in eq.\,\eqref{eq:jet_dist}. 
The corresponding coefficient ratios $R_\phi$, which are defined in eq.\,\eqref{eq:RphiDef}, for the two BPs are
\bea
R_\phi \simeq 0.299 \textrm{ for BP1} \, , \\
R_\phi \simeq 0.309 \textrm{ for BP2} \, .
\eea
As the shapes in $\phi$ distributions between the two benchmark points become similar after MDT procedures,  
we anticipate that $N_{\textrm{event}}^{(3\sigma)}$  for both BPs are of same order, accordingly. 

\section{Conclusion \label{sec:conclusion}}
As the second phase of the LHC accumulates more data, resonance searches in the TeV mass range are readily available in many channels. 
A wide range of new physics models predict such massive particles which often decay into heavy SM states such as top quark, Higgs, and $W/Z$ gauge bosons. 
While their leptonic decay modes enjoy the clean nature of the associated final state, the hadronic ones, which typically come with larger branching fractions, are anticipated to play an important role in discovery potential as well as property measurement at {\it earlier} stages. 
Nevertheless, relevant analyses are often challenging because (hadronic) decay products are inclined to be highly collimated due to the substantial mass gap between the heavy resonance and the SM heavy states involved in the process of interest.

In this paper, we tackled this challenge with the aid of jet substructure techniques, and showed that it is possible to measure physical properties of new particles using subjet information. 
More specifically, we illustrated that the Matrix Element Method can be a powerful method for identifying properties of a new particle in the final state with two jets and two leptons via a pair of SM gauge bosons.
As a concrete example, we focused on discriminating the CP property of a spin-0 resonance which decays into a pair of SM gauge bosons. 
For a systematic approach, we adopted the prescription of the ``merged'' jet to capture two quarks from the decay of an SM gauge boson, which also helps to reduce combinatorial issues. 
We then studied effects from jet clusterings and associated jet substructure methods on the phase space for visible particles. 

A certain extent of prejudice in performing data analyses with detector-level reconstructed particles is typically expected in comparison with relevant theoretical expectations. 
However, our study based on both analytical calculations and Monte Carlo simulation demonstrated that restrictions on the phase space invoked by jet clustering procedures could enhance the difference in angular distributions for new particles with different CP states, {\it unlike} the naive expectation stated above. 
We also showed that the performance of our data analyses does not significantly depend on the size of MJs, as internal cuts in jet grooming procedures affect the phase space for visible particles stronger. 
We believe that our finding here benefits the determination of a reasonable size of MJs to make a balance between analysis performance and enhancement of the ratio of signal-over-background.

In our analyses with the MEM, we refrained from integrating partonic phase space through transfer functions which map reconstructed objects to the partonic phase space. 
We rather modeled a probability density function ({\it pdf}\,), generating many pseudo-experiments with Monte Carlo simulation for a given signal hypothesis. This procedure can take into account various effects from jet clustering procedures together with detector effects as well as offer computational advantages compared to the situation where $2^{N_{\textrm{vis}}}$-dimensional integration is required in dealing with transfer functions.

Two benchmark points were selected to cover various phase space regions from the moderately boosted one to the highly boosted one. According to our numerical studies, discriminating different CP states requires $\mathcal{O}(20)$ signal events at the level of $3\sigma$ significance over the mass range of a new particle from over $700\GeV$ that the current LHC has the equal level of sensitivity in a merged jet with dileptons compared to a full leptonic channel \cite{Khachatryan:2015cwa}.

\section*{Acknowledgements}
M.P. appreciates useful discussions with Veronica Sanz about the efficiency in tagging MJs in terms of the  resonance mass.
M.P. also appreciates the kind hospitality of Kavli-IPMU while part of this work had been performed. 
M.P., K.K., and D.K. thank the organizers of the CERN TH institute funded by CERN-KOREA program where 
a part of this work was completed.
K.K thanks the hospitality and support from PITT PACC. 
This work is supported by IBS under the project code, IBS-R018-D1. 
K.K. and D.K. are supported by DOE (DE-FG02-12ER41809, DE-SC0007863). D.K. is supported in part by the Korean Research Foundation (KRF) through the CERN-Korea fellowship program.
C.H. is supported by World Premier International Research Center Initiative (WPI Initiative), MEXT, Japan.

\appendix

\section{Background consideration in MEM analyses}
\label{sec:BKG}

\begin{table}
\begin{center}
\begin{tabular}{|l|l|r|r|r|}
\hline
 & \multicolumn{4}{c|}{BP1 ($M_S = 750\GeV$)} \\
\hline 
cut flow                          & selection criterion           & $\sigma_{Z+\mathrm{jets}}$ & $\sigma_{ZZ} $ & $\sigma_{ZW}$ \\
\hline
parton level              & $P_T$ of leading jet $\geq 150$ GeV           & 8.65 pb   & 8.19  fb  & 8.96  fb\\
\hline
object tagging            & One merged jet, two $\ell$                    & $44.11\%$ & 55.30\%   & 55.83\%\\
lepton $P_T$              & $P_T > 25\GeV$                                & $33.47\%$ & 44.88\%   & 47.24\%\\
$m_{(\ell^+,\,\ell^-)}$   & $[83,99]$ GeV                                 & $30.54\%$ & 40.91\%   & 42.92\%\\
$m_{\MJ}$                 & $[75,105]$ GeV                                & $1.60\%$  & 12.10\%   & 10.72\%\\
$y_{ZZ} $                 & $|y_{ZZ}| < 0.15$                             & $0.72\%$  & 11.06\%   &  9.83\%\\
$P_{T(\MJ)} $       & $P_{T(\MJ)}> 0.4 \, m_{(\MJ,\,\ell^+,\,\ell^-)} $   & $0.48\%$  &  7.22\%   &  5.29\%\\
$m_{(\MJ,\,\ell^+,\,\ell^-)}$  & within $M_S\pm 50 \GeV $                 & $0.037\%$ &  0.82\%   &  0.68\%\\
\hline
Cross section ($\sigma$)  & -                                                             & 3.16 fb   & 0.0671 fb & 0.0609 fb\\
\hline
\end{tabular}
\end{center}
\caption{\label{tab:cut_flow_bkg}
Cut flows for major backgrounds of BP1. In a signal region defined with a merged jet, $Z+\mathrm{jets}$ becomes the dominant background, compared to irreducible electroweak process $pp\to ZZ$.
}
\end{table}

\begin{figure}[t!]
\begin{center}
 \includegraphics[width=0.5\textwidth]{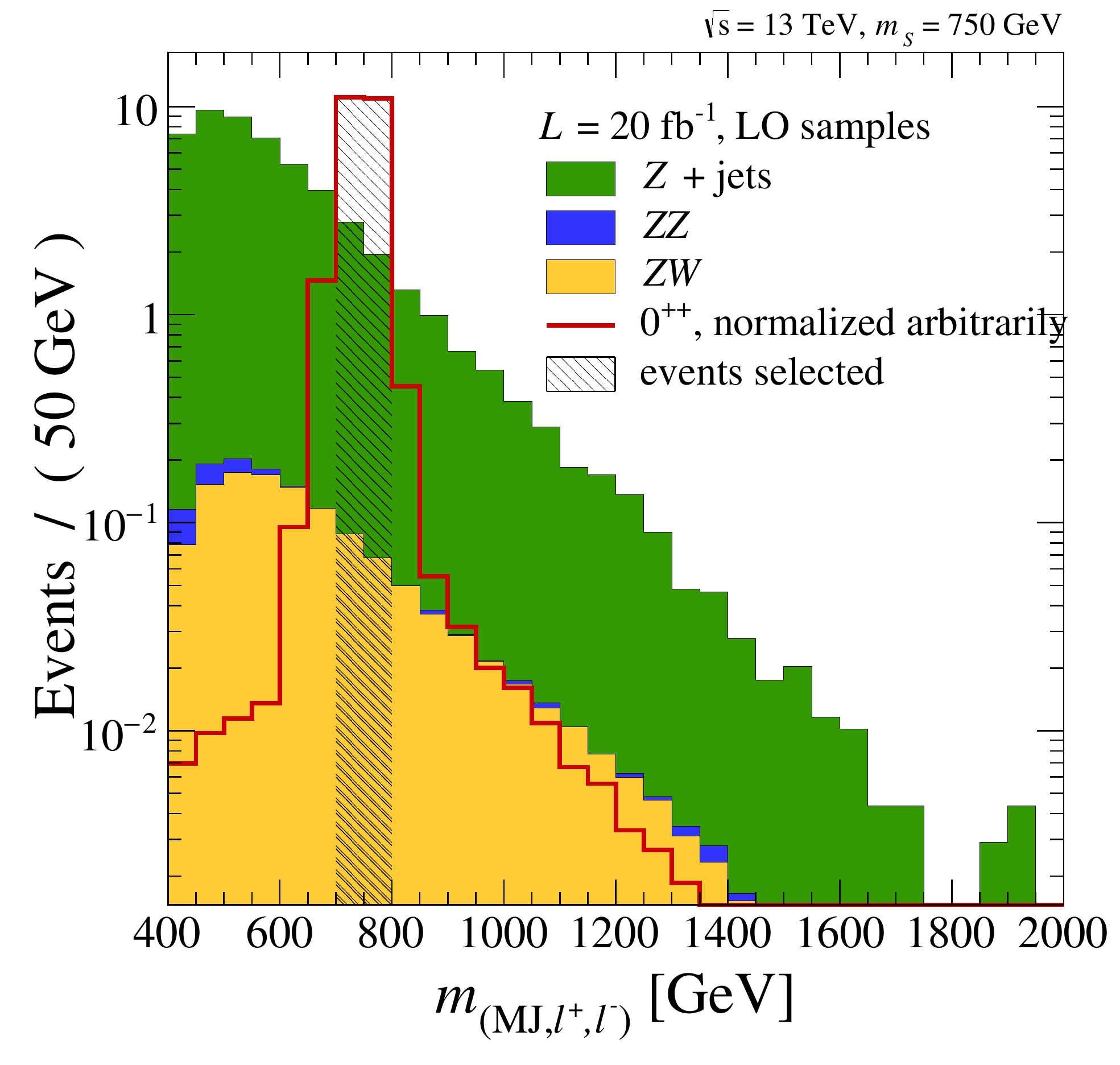}
\end{center}
\vspace{-1.0cm}
   \caption{$m_{(\MJ,\ell^+,\ell^-)}$ distributions of backgrounds and signal after all the cuts except $m_{(\MJ,\ell^+,\ell^-)}$ in Table~\ref{tab:cut_flow_bkg}.} 
\label{fig:bg}
\end{figure}

The major irreducible background in which the final state particles are the same as those in our case (i.e., $jj\ell^+\ell^-$) is $ZV$ pair production. 
Here $V$ includes not only $Z$ but $W^{\pm}$ gauge bosons which decay into two jets because a $W$-induced MJ 
can easily fake a $Z$-induced MJ due to jet mass resolution. On the other hand, the major reducible background is $Z+js$ where a QCD jet $j$ can mimic an $Z$-induced MJ by acquiring a non-vanishing mass due to QCD corruptions.
To estimate contributions from above backgrounds, we perform detector-level Monte Carlo simulation for $ZV$ and $Z+j$s, and summarize the associated cut flows in Table~\ref{tab:cut_flow_bkg}.
We also demonstrate, in Figure~\ref{fig:bg}, $m_{(\mathrm{MJ},\ell^+,\ell^-)}$ distributions for the backgrounds and signal after all the cuts except $m_{(\mathrm{MJ},\ell^+, \ell^-)}$ in Table~\ref{tab:cut_flow_bkg}, suggesting that the $Z+j$s dominates $ZV$ backgrounds by far.

In the presence of backgrounds (denoted as bkg), we express probability density function $\mathcal{P}$ with respect to a set of discriminator variables $\{x\}$ for the signal-plus-background hypothesis as follows:
\bea
\mathcal{P}(\{x\}|0^{\pm +}+\mathrm{bkg}) = r_{S}\mathcal{P}(\{x\}|0^{\pm+})+r_{B}\mathcal{P}(\{x\}|\mathrm{bkg})\,,
\eea
where $r_{S(B)}$ is the ``observed'' fractional signal (background) cross section to the total observed one. 
Here $\mathcal{P}(\{x\}|\alpha)$ is an individual {\it pdf} under hypothesis $\alpha$ (either signal or background), which is built from the associated amplitude.
It turns out that the best discriminator is the set of momenta $\{p\}$ itself. 
The matrix elements for the signal and background processes are needed to convert observed momentum information into a form of probability under a given hypothesis. 
We then define the likelihood ratio $q_{\mathcal{M}}$ as
\beq
q_{\mathcal{M}}  = \sum_i^{N_{\mathrm{eve}}} \ln \frac{\mathcal{P}(\{p\}_i|0^{++} + \mathrm{bkg})}{\mathcal{P}(\{p\}_i|0^{-+} + \mathrm{bkg}) } 
= \sum_i^{N_{\mathrm{eve}}} \ln \left(\frac{  \mathcal{P}(\{p\}_i|0^{++}) + R_\sigma \mathcal{P}(\{p\}_i|\mathrm{bkg}) } {\mathcal{P}(\{p\}_i|0^{-+}) + R_\sigma \mathcal{P}(\{p\}_i|\mathrm{bkg}) } \right)\, ,
\eeq 
where $R_{\sigma}$ is the ratio of the ``observed'' background to the ``observed'' signal cross sections.
If the relevant backgrounds are well under control or sufficiently suppressed, i.e., $R_{\sigma}< 1$, the above expression is approximated to 
\begin{eqnarray}
q_{\mathcal{M}} \nonumber
& = &
\sum_i^{N_{\mathrm{eve}}}  
\ln \frac{\mathcal{P}(\{p\}_i|0^{++})}{ \mathcal{P}(\{p\}_i|0^{-+} ) } 
+ 
\sum_i^{N_{\mathrm{eve}}}  
\ln \left[
\left( 1 + R_\sigma \frac{\mathcal{P}(\{p\}_i| \mathrm{bkg})}{\mathcal{P}(\{p\}_i|0^{++})} \right) \big{/}
\left( 1 + R_\sigma \frac{\mathcal{P}(\{p\}_i| \mathrm{bkg})}{\mathcal{P}(\{p\}_i|0^{+-})} \right)
\right]
\\
& \simeq &
\sum_i^{N_{\mathrm{eve}}}  
\ln \frac{\mathcal{P}(\{p\}_i|0^{++})}{ \mathcal{P}(\{p\}_i|0^{-+} ) }  +  R_\sigma \,\sum_i^{N_{\mathrm{eve}}}  \left(\frac{\mathcal{P}(\{p\}_i| \mathrm{bkg})}{\mathcal{P}(\{p\}_i|0^{++})}-\frac{\mathcal{P}(\{p\}_i| \mathrm{bkg})}{\mathcal{P}(\{p\}_i|0^{-+})}\right) +\mathcal{O}(R_\sigma^2) \, . \label{eq:qMbkg}
\end{eqnarray}
Note that it is not necessary to consider the second term in order to discriminate the CP property of resonance $S$, because the first term already carries relevant information to be used for the hypothesis test as we have seen in Section~\ref{sec:MEManal}.
Certainly, using the second term can improve the discriminating power according to the Neyman-Pearson lemma.
However, the likelihood ratio between signal and background is not easily factorizable into matrix elements, and therefore, we conservatively utilize the first term only for the hypothesis test.

\begin{figure}[t!]
\begin{center}
\begin{tabular}{ccc}
\includegraphics[width=4.7cm]{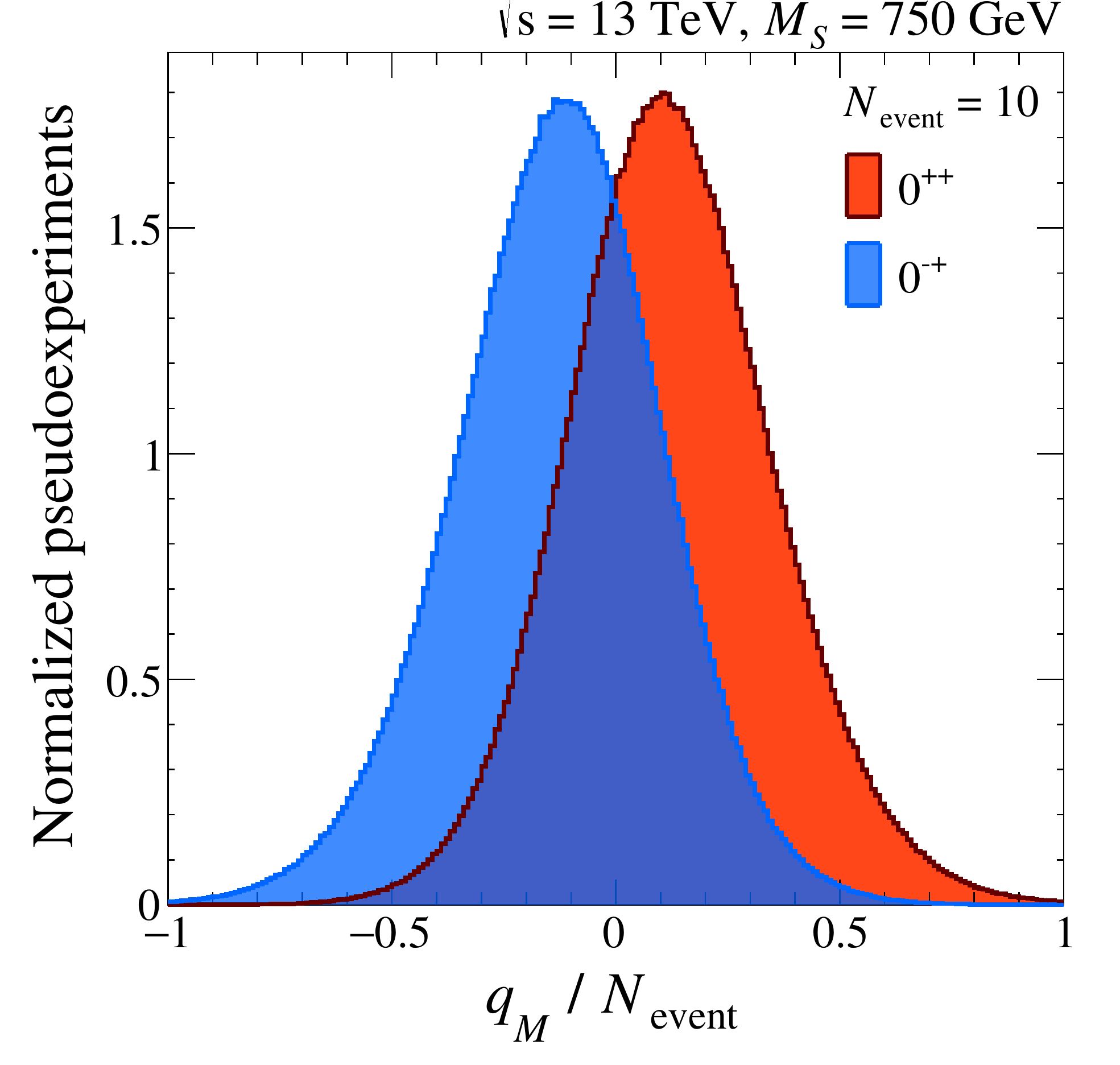} &
\includegraphics[width=4.7cm]{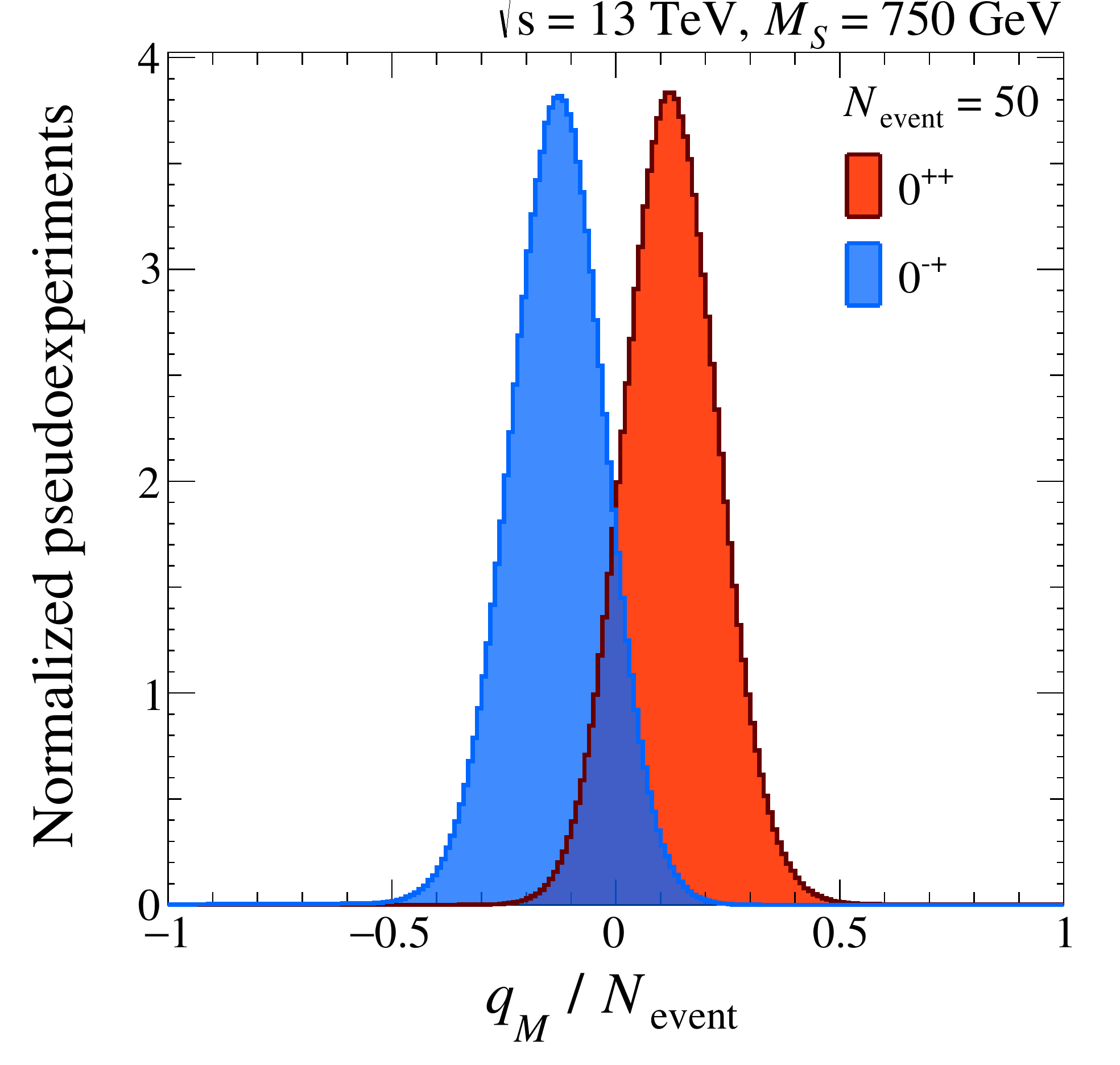} &
\includegraphics[width=4.7cm]{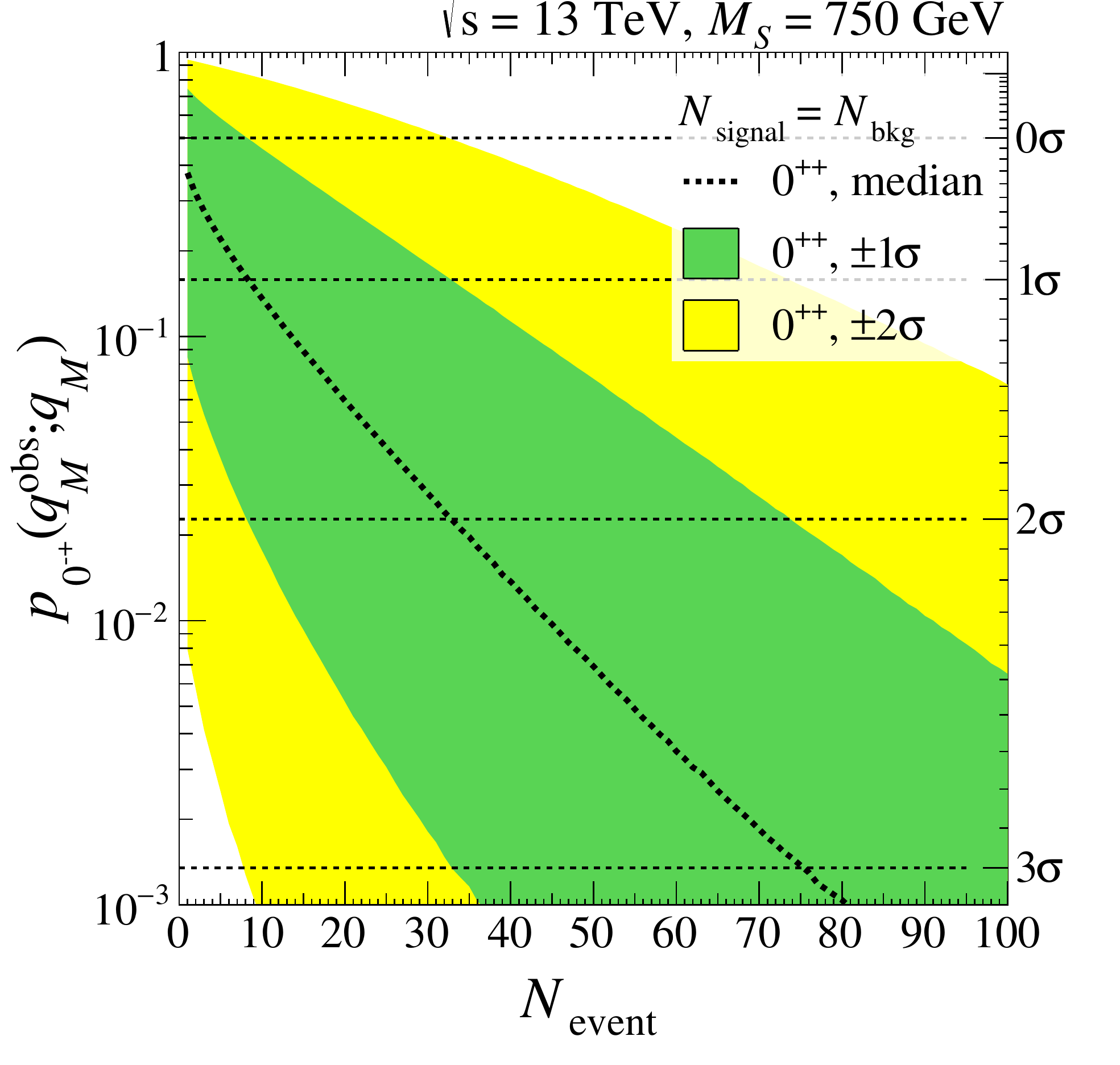}
\end{tabular}
\end{center}
\caption{\label{fig:results_750_sig_bkg}  Similar results to those in Figure~\ref{fig:results_750} for the BP of $M_S=750\GeV$ in the presence of all backgrounds shown in Figure~\ref{fig:bg}.
}
\end{figure}

We conduct similar exercises as in Figure~\ref{fig:results_750} for BP1 including the contribution from all backgrounds shown in Figure~\ref{fig:bg}, and exhibit the resulting discrimination power in Figure~\ref{fig:results_750_sig_bkg}. 
We evaluate $q_{\mathcal{M}}$ using only the dominant term in eq.~\eqref{eq:qMbkg}, setting us free from the {\it pdf} under the background hypothesis. 
For each of the event samples for the CP-even and CP-odd, we take the same numbers of signal and background events.
Comparing them with the corresponding plots in the second row of Figure~\ref{fig:results_750}, we see that (not surprisingly) more events are required to discriminate the CP property of the scalar.  
In this sense, more reduction of background events would help to probe the properties of the new particles.

As the main background is from $Z+j$s in which QCD jets fake MJs, an analytic matrix element of $Z+j$s {\it after} the MDT should be provided in order to implement the background into the MEM more accurately. 
The leading order and next-to-leading order results are shown in Refs.~\cite{Dasgupta:2013ihk,Dasgupta:2013via}. However, it would be challenging to go beyond next-to-leading logarithmic accuracy since the original definition of the MDT carries non-global logarithms~\cite{Dasgupta:2013ihk,Dasgupta:2013via}.  
While a modified Mass Drop Tagger or a soft drop have been proposed~\cite{Dasgupta:2013ihk,Dasgupta:2013via,Larkoski:2014wba} and next-to-next-to-leading logarithmic accuracy has been shown recently~\cite{Frye:2016aiz}, dedicated examinations along the line are certainly beyond the scope of the paper in which a simple MDT is employed. 
We therefore do not provide any further discussion on the MEMs including backgrounds.

\section{Phase space restriction from other jet substructure methods \label{sec:appB}}

Besides the MDT, there are other jet substructure methods which have been widely used in the literature. We give brief comments on how they affect the phase space.

\begin{itemize}
\item{\bf{Trimming:}}  The trimming procedure~\cite{Krohn:2009th} uses a  $k_t$ algorithm to divide 
a merged jet
into subjets with a size $R_{\mathrm{sub}}$, and then removes the subjets with  $P_{T(j_i)}/P_{T(\MJ)} < f_{\mathrm{cut}}$, where $f_{\mathrm{cut}}$ is a parameter.
The remaining subjets are then reclustered as a trimmed jet. 
Similar to the MDT, the $f_{\mathrm{cut}}$ applies a cut on the $P_T$ fraction of a subjet $z$ such that $z>f_{\mathrm{cut}}$. 
Like eq.~\eqref{eq:MDT_R}, it effectively reduces the cone size of MJs.
 \item{\bf{Pruning:}}  The pruning method~\cite{Ellis:2009su} uses the C/A or $k_t$ algorithm to cluster the jets. At each recombination step $j_1 j_2 \rightarrow \MJ$, either $\min(P_{T(j_1)},P_{T(j_2)})/P_{T(\MJ)}> z_{\mathrm{cut}}$ or $\Delta R_{j_1j_2} <  m_{\MJ}/P_{T(\MJ)}$ needs to be satisfied. Interestingly, both cuts set some upper limit on the cone size of MJs.
 \item{\bf{ N-subjettness:}} N-subjettiness~\cite{Thaler:2010tr,Thaler:2011gf} is defined as 
\beq
\tau_N=  \frac{1}{d_0} \sum_k P_{T,k} \min(\Delta R_{j_1k},...,\Delta R_{j_Nk})
\eeq
where $d_0=\sum_i P_{T,k} R_0$ with $R_0$ being the characteristic jet radius used in the original jet finding algorithm.
Here $j_i$ denotes the usual $i$th subjet, while $k$ runs over all constituent particles in a given MJ. 
For a two-prong MJ, usually $\tau_{21}\equiv\tau_2/\tau_1$ is computed with its upper limit/cut. Since $\tau_2=0$ at the parton level, it is interesting to see how a non-zero $\tau_{21}$ cut affects the relevant phase space at the reconstruction level.
In Figure~\ref{fig:Nsubjetcut} we show distributions in $\Delta R_{q\bar q}$ between two partons from a $Z$ boson decay for BP1 (left panel) and BP2 (right panel), after applying $\tau_{21}$ cuts to the corresponding reconstructed jets.
We see that a $\tau_{21}$ cut reduces the associated selection rate over the entire $\Delta R_{q\bar q}$ region, so it could be taken as an independent cut after a jet grooming procedure (e.g., MDT, trimming, or pruning).
\begin{figure}[t]
\begin{center}
 \includegraphics[width=0.97\textwidth]{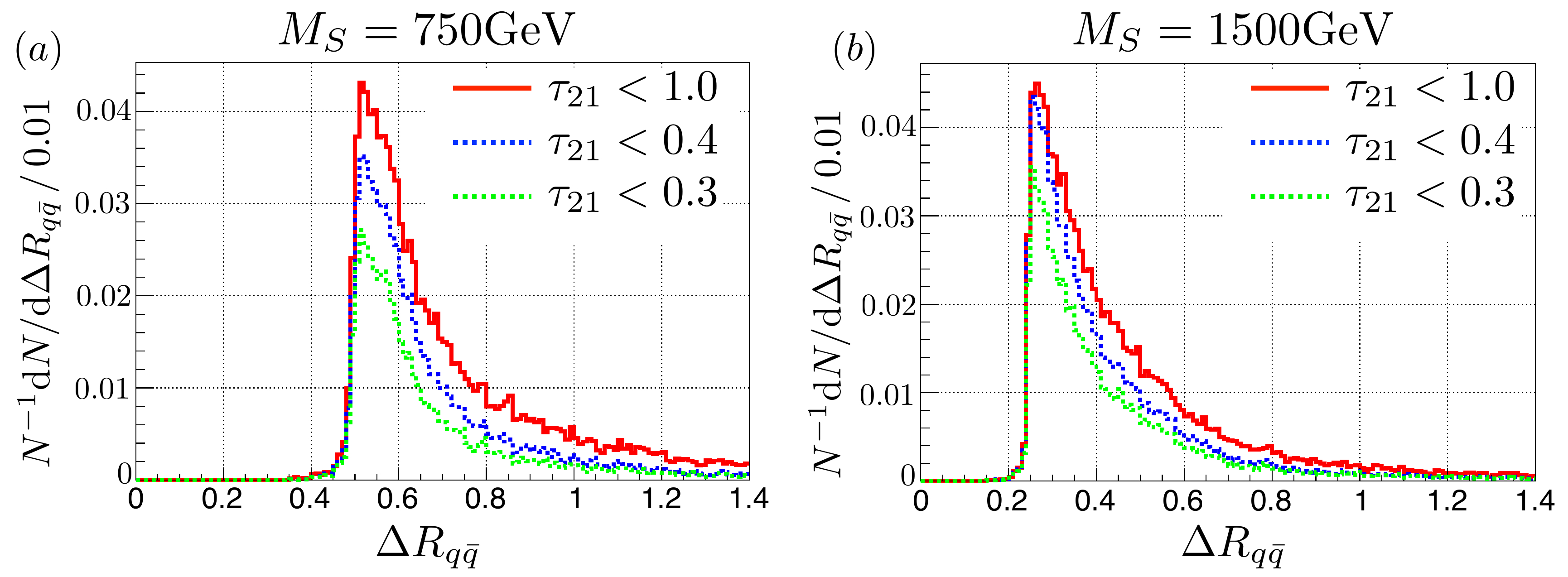}
 \end{center}
   \caption{Distributions of $\Delta R_{q\bar q}$ between two partons from a $Z$ boson decay for (a) $M_S =750\GeV$ and (b) $M_S=1500\GeV$, after we apply
$\tau_{21}$ cuts in the N-subjettiness onto the corresponding reconstructed jets.  } 
\label{fig:Nsubjetcut}
\end{figure}
\end{itemize}

\bibliographystyle{JHEP}
\bibliography{JetSubstructure_Diboson}

\end{document}